\documentclass[aps,prl,twocolumn,superscriptaddress,showpacs,preprintnumbers,longbibliography]{revtex4-1}
\usepackage[colorlinks,bookmarks=false,citecolor=blue,linkcolor=red,urlcolor=blue]{hyperref}
\usepackage{physics}
\input pdfcolor.tex
\usepackage{colortbl,amsthm,txfonts}
\usepackage{verbatim}
\usepackage{graphicx}
\usepackage{epsfig}
\usepackage{dcolumn}
\usepackage{bm}

\usepackage{amsmath,amssymb}

\makeatletter
\renewcommand*\env@matrix[1][*\c@MaxMatrixCols c]{%

\hskip -\arraycolsep
\let\@ifnextchar\new@ifnextchar
\array{#1}}
\makeatother
\usepackage{appendix}

\begin{document}

\title{Thermal stability of pair density wave in a $d$-wave altermagnetic superconductor}
\author{Amrutha N Madhusuthanan}
\author{Madhuparna Karmakar}
\email{madhuparna.k@gmail.com}
\affiliation{Department of Physics and Nanotechnology, 
SRM Institute of Science and Technology, 
Kattankulathur, Chennai 603203, India}

\date{\today}

\begin{abstract}
We study finite-momentum superconductivity in a two-dimensional $d$-wave altermagnetic superconductor using a non-perturbative Monte 
Carlo approach beyond mean-field theory. We show that altermagnetism stabilizes a pair density wave (PDW) state without external magnetic fields and enables its survival at finite temperatures with robust phase coherence. Our results establish altermagnetism as a promising route to realizing thermally stable PDW superconductivity and identify clear thermodynamic and spectroscopic signatures.
\end{abstract}

\maketitle

\textit{Introduction:}  Altermagnetism (ALM), arising purely out of the magnetic space group symmetry and considered to be the magnetic counterpart of unconventional superconductivity promises to bring in the best of both ferromagnetism and antiferromagnetism \cite{jungwirth_prx2022,pan_natrevmat2025,mason_annrev2025}. Characterized by a net zero magnetic polarization and non-relativistic momentum-dependent spin-splitting arising out of the ${\cal PT}$ symmetry breaking ALM is considered to be important not just for its potential application in spintronics but rather in its promise to bring forth esoteric quantum systems, phases and phase transitions \cite{pan_natrevmat2025,mason_annrev2025,jungwirth_prx2022,pereira_prb2024,smejkal_pnas2021,brink_mattodayphys2023,lu_natscirev2025,smejkal_arxiv2023,sinova_prb2024,haule_prl2025,olsen_apl2024,mazin_scipost2024,yang_chemsci2024,yang_jamchemsoc2025,rhone_prm2025,stroppa_prl2025,zhou_prl2025,smejkal_arxiv2024,brink_comphys2025,shen_prl2024,liu_natcom2025,ma_nanolet2025,qian_natphys2025,chen_natphys2025,mazin_prbl2023,kim_prl2024,sato_prb2024,jungwirth_nature2024,ji_prm2025,valenti_arxiv2025,sasaki_prr2025,cano_jap2025,seo_npjspintronics2025,valenti_prb2024,scheurer_prr2024,capone_prbl2025,lu_arxiv2025,fernandes_arxiv2025,fernandes2_arxiv2025,franz_prl2025,thomale_prl2025,fernandes_arxiv2025}. One such promising quantum system is the altermagnetic superconductor (ALM-SC) with finite momentum superconducting (SC) pairing viz. the pair density wave (PDW) \cite{sumita_prb2025,neupert_natcom2024,ohashi_arxiv2026,knoll_prbl2025,schaffer_prbl204,schaffer_prl2025,paramekanti_prb2024,fradkin_arxiv2026,zegrodnik_npjquantmat2025}.

Finite momentum pairing has been largely explored in the context of Pauli limited superconductors and Fulde-Ferell-Larkin-Ovchinnikov (FFLO) phase wherein an applied Zeeman field induced imbalance in the fermionic population promotes the pairing instability at finite momentum \cite{ff,lo}. FFLO and Pauli limited superconductors have been extensively investigated both for continuum and lattice fermion based models using a host of analytic and numerical approaches \cite{karmakar_pra2016,karmakar_pra2018,torma_prl2007,sarma_jap1963,zoller_pra2004,graf_prb2005,graf_prb2006,spalek_prb2011,vekhter_prb2010,yanase_jpsj2008,ting_prb2009,yang_prb2012,karmakar_jpcm2020,karmakar_jpcm2024}. Further, experimental realization of such a SC state has been reported in the context of ultracold atomic gases \cite{ketterle_nature2008,ketterle_science2007,ketterle_prl2006,mueller_nature2010} and solid state materials such as, heavy fermion superconductors \cite{sarro_prl2003,onuki_prb2002,flouquet_prl2008,kenzelman_science2008,sarro_prb2004,sarro_prb2005,gerber_natphys2013,matsuda_prl2006,movshovich_prx2016,movshovic_prl2020}, layered organic materials \cite{wosnitza_ltp2013,wosnitza_prl2007,mitrovic_natphys2014,wright_prl2011,montegomery_prb2011,lortz_prb2011,agosta_prb2012,schlueter_prb2009} and iron arsenide superconductors \cite{lohneysen_prl2013,prozorov_prb2011,kim_prb2011}.

The finite momentum paired PDW phase, though has been theoretically explored time and again \cite{agterberg_annrevcmp2020,tranquada_njp2009,kivelson_prl2010,raghu_prbl2023,tsunetsugu_natphys2008,kivelson_natphys2009,tranquada_rmp2015,kampf_prb2010} and experimentally observed in materials such as, La-based underdoped high-$T_{c}$ cuprates \cite{zimmermann_prb2008}, Kagome metals \cite{hasan_natmat2021,gao_nature2021}, UTe$_{2}$ \cite{liu_nature2023,madhavan_nature2023}, EuRbFe$_{4}$As$_{4}$ \cite{fujita_nature2023} etc. are found to be extremely fragile against thermal fluctuations and other perturbations.  Unlike FFLO, the finite momentum pairing in a PDW state is dictated purely by the Fermi surface topology and doesn't require an external magnetic field \cite{agterberg_annrevcmp2020,kampf_prb2010}.

The recently discovered ALM have put forward the precise platform which fulfils the required criteria for realizing the PDW state and expectedly there has been a flurry of theoretical and computational research both on the bulk and SC-ALM heterostructures to investigate a prospective PDW state \cite{sumita_prb2025,neupert_natcom2024,ohashi_arxiv2026,knoll_prbl2025,schaffer_prbl204,schaffer_prl2025,liu_prb2025,paramekanti_prb2024,fradkin_arxiv2026,zegrodnik_npjquantmat2025,kim_prb2025,wang_prbl2024,linder_prl2023,papaj_prbl2023}. The studies have however remained confined to the premises of the mean field theory (MFT) and phenomenological Ginzburg-Landau formalism, investigating the role of ALM towards a stable PDW ground state and its related phenomena \cite{sumita_prb2025,neupert_natcom2024,ohashi_arxiv2026,knoll_prbl2025,schaffer_prbl204,schaffer_prl2025,liu_prb2025,paramekanti_prb2024,fradkin_arxiv2026,zegrodnik_npjquantmat2025,kim_prb2025,wang_prbl2024,linder_prl2023,papaj_prbl2023}. It is noteworthy that the actualization of the PDW phase in real materials is crucially dependent on its stability against thermal fluctuations and perturbation; an issue that can't be addressed within the purview of the MFT.

In this letter we attempt to discourse on this issue within the framework of the non perturbative static path approximated (SPA) Monte Carlo technique that takes into account the spatial fluctuations of the fermionic correlators at all orders and bring forth quantitative signatures of the 
finite temperature PDW phase in two-dimensional (2D) $d$-wave ALM-SC.  Based on the thermodynamic and spectroscopic signatures of this system we for the first time exhibit the stability and provide the estimate of the thermal scales for a PDW state. 
Our primary results from this work entails: (i) we map out the ground state phase diagram of a $d$-wave ALM-SC comprising of a uniform Bardeen-Cooper-Schrieffer (BCS), a pair density wave (PDW), a quantum breached pair (QBP), amplitude modulated Larkin-Ovchnnikov (LO) and a polarized Fermi liquid (PFL) phases, (ii) based on our analysis of the thermodynamic and spectroscopic signatures we provide the first evidence of the stability of the PDW phase against thermal fluctuations and provide sharp quantitative estimates of the thermal scales.

\textit{Theoretical model:} We model the system based on the 2D attractive Hubbard Hamiltonian with inter-site 
interaction and spin-dependent anisotropic hopping on a square lattice (see supplementary materials (SM) for the details), that reads as, 
\begin{eqnarray}
\hat H & =& \sum_{\langle ij\rangle, \sigma}(-t_{ij}+\sigma t_{am}\eta_{ij})(\hat c_{i,\sigma}^{\dagger}\hat c_{j,\sigma}+h.c.) \nonumber \\ &&
- \sum_{i,\sigma}(\mu+\sigma_{i}^{z} h)\hat n_{i,\sigma} -\vert U\vert \sum_{\langle ij\rangle }\hat n_{i}\hat n_{j}  
\end{eqnarray}
where, $t_{ij}=t=0.5$ is the nearest neighbor hopping and sets the reference energy scale of the system. The second term depicts the 
$d_{x^{2}-y^{2}}$ ALM interaction such that, $t_{am}$ quantifies the strength of the interaction and $\eta_{ij}$ is the $d$-wave form factor, 
leading to $t_{\hat x} = t-\frac{\sigma t_{am}}{2}$ and $t_{\hat y} = t+\frac{\sigma t_{am}}{2}$; $\sigma=+(-)$ for the $\uparrow$($\downarrow$) 
spin species.  SC pairing is brought in via the attractive interaction $\vert U\vert > 0$, the chemical potential $\mu$ dictates the fermionic number density in the system and the Zeeman field $h$ allows for a population imbalance between the fermionic species, leading to a finite magnetic polarization, $m$. The model is made numerically tractable via Hubbard-Stratonovich (HS) \cite{hs1,hs2} decomposition of the interaction term, introducing the randomly fluctuating complex bosonic auxiliary field $\Delta_{ij} = \vert \Delta_{ij}\vert e^{\phi_{ij}} $ which couples to the $d$-wave pairing singlet, $(c_{i\uparrow}c_{j\downarrow}+c_{j\uparrow}c_{i\downarrow})$. The SC pairing amplitude is defined as, $\Delta_{ij} \propto \vert \Delta_{ij}\vert \cos({\bf q}.{\bf r}_{i})$ where ${\bf q} \in \{q_{x}, q_{y}\}$ is the pairing momentum and the anisotropic SC phase is  $\phi_{ij} \in \{\phi_{ij}^{x}, \phi_{ij}^{y}\}$ with the relative phase being $\phi_{ij}^{rel} = \phi_{ij}^{x}-\phi_{ij}^{y}$.

\textit{Numerical methods:} Our primary numerical approach is SPA, based on the adiabatic approximation of the slow (thermal) bosonic fields serving as a random, static disordered,  fluctuating background to the fast moving fermions \cite{ciuchi_scipost2021,fratini_prb2023,kivelson_pans2023,karmakar_prml2025}. The approximation allows one to treat the bosonic field as classical variable and provide access to real frequency dependent quantities without requiring an analytic continuation (see SM for details). In addition, for accessing the ground state of the system we use an alternate approach of the variational MFT. 
\begin{figure}
\begin{center}
\includegraphics[height=5.5cm,width=6.5cm,angle=0]{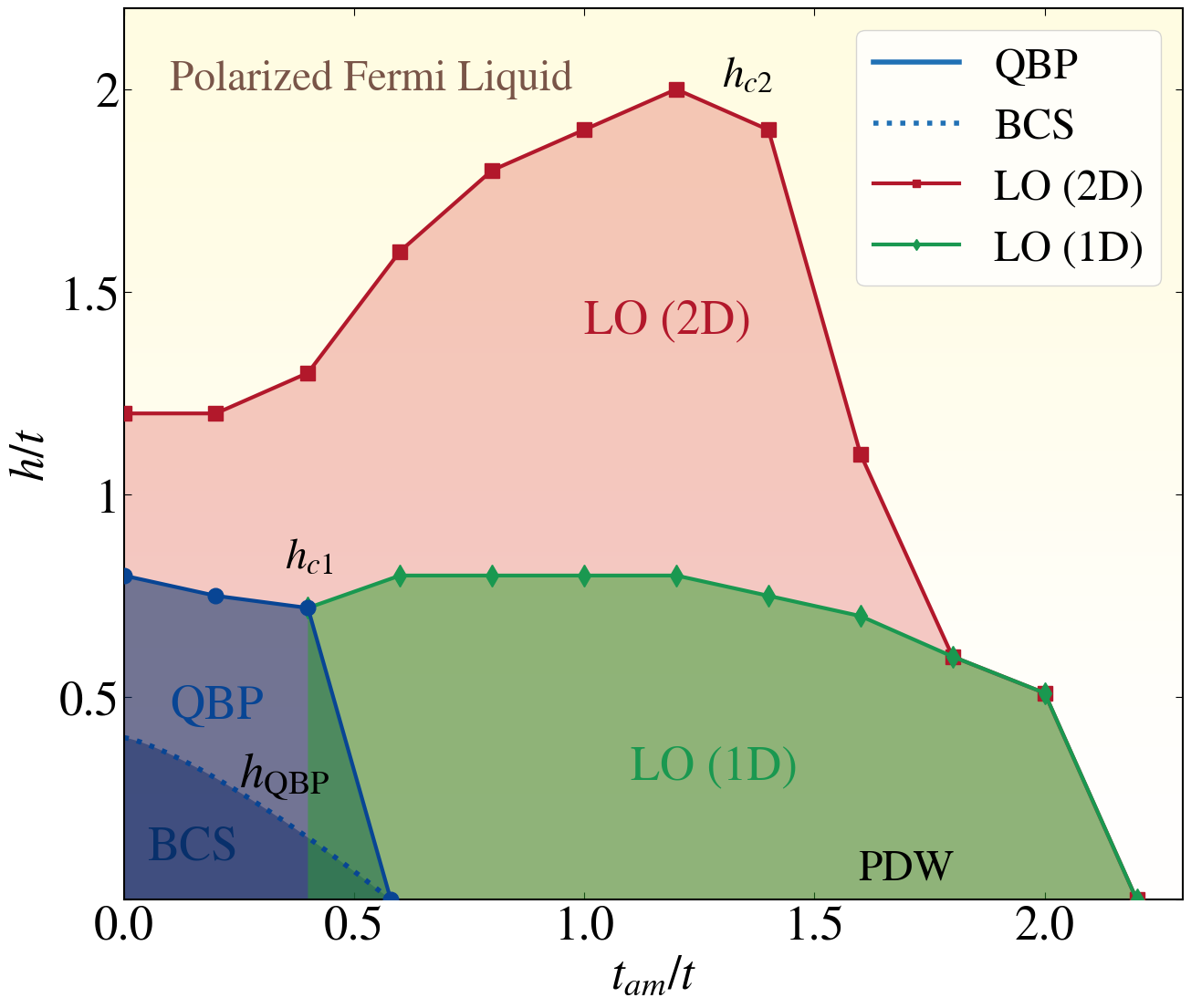}
\caption{Ground state phase diagram of $d$-wave altermagnetic superconductor in the $t_{am}-h$-plane showing the BCS, QBP, PDW, 
LO and PFL phases along with the corresponding transition scales (solid lines with points).}
\label{fig1}
\end{center}
\end{figure}

As $T \rightarrow 0$,  the thermal fluctuations die off, $\Delta_{ij}$ and $\{q_{x}, q_{y}\}$ can thence be used as variational parameters to minimize the energy. In the spirit of MFT we treat $\Delta_{ij} = \vert \Delta \vert$ as a real number and fix $\phi_{ij}^{rel} = \phi_{ij}^{x}-\phi_{ij}^{y} = \pi$. To capture the spatially modulated SC phases periodic configurations of the order parameter as, (i) $\Delta_{ij} \sim \vert \Delta\vert \cos(qx_{i})$, (ii) $\Delta_{ij} \sim \vert \Delta\vert (\cos(qx_{i})+\cos(qy_{i}))$ are optimized over the $t_{am}-h$ plane.

The thermodynamic phases and transition scales are quantified in terms of: (i) real space maps of the pairing field amplitude ($\vert \Delta_{ij}\vert$) and phase correlation ($\cos(\phi_{0}^{x}-\phi_{ij}^{x})$), (ii) single particle density of states (DOS) ($N(\omega)$), (iii) spin-resolved Fermi surface topology ($A_{\sigma}({\bf k}, 0)$) and (iv) magnetic polarization ($m$) (see SM for details). The results presented in this letter corresponds to a system size of $L=24$ unless specified otherwise and are found to be robust against finite system size effects. The calculations are carried out in the grand canonical ensemble at $\vert U\vert = 4t$ and $\mu=-0.2t$,  tantamount to a fermionic number density of $n \approx 0.9$. The ground state variational results are computed at a temperature of $T =10^{-3}t$.

\begin{figure}
\begin{center}
\includegraphics[height=10cm,width=8.5cm,angle=0]{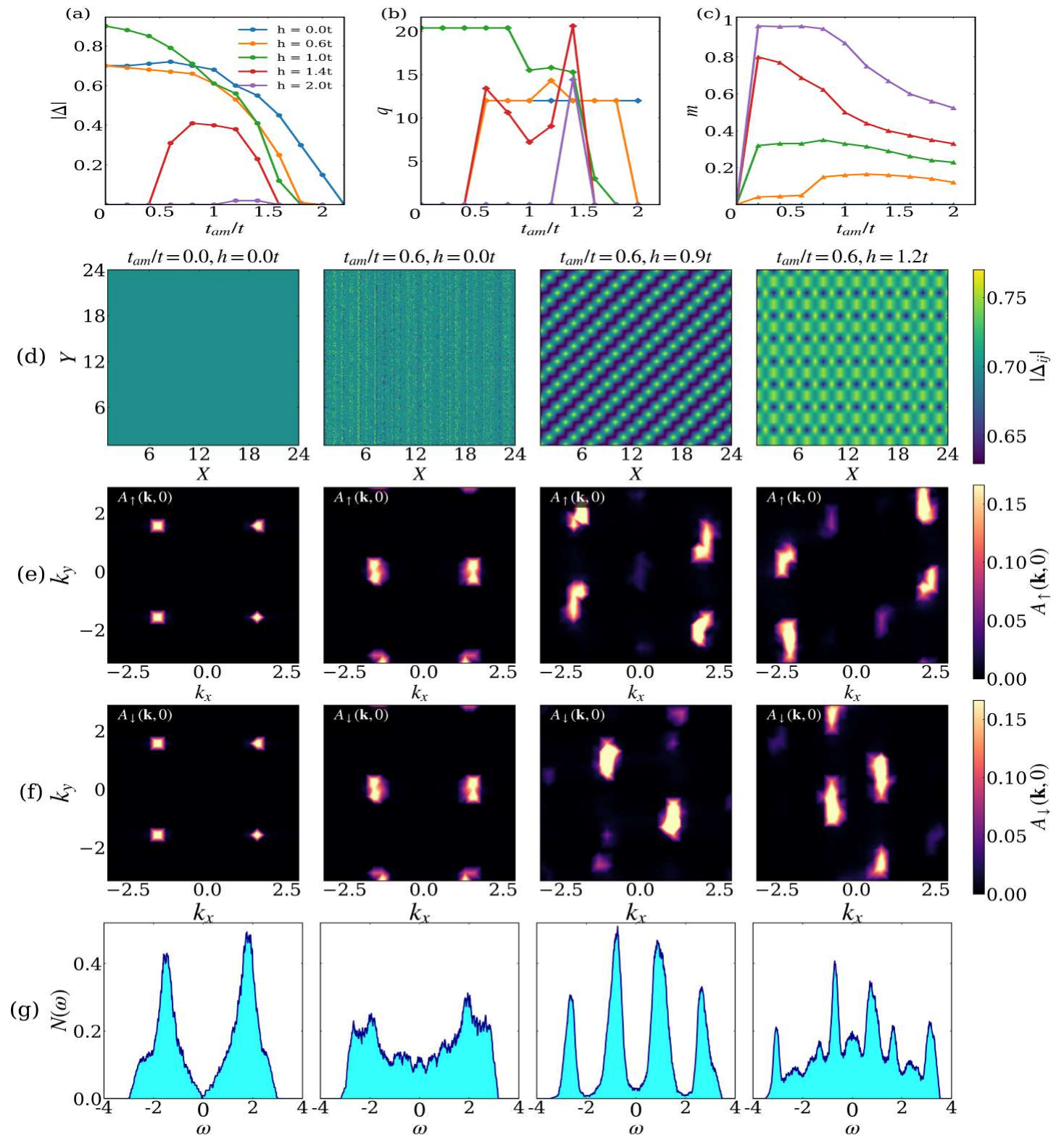}
\caption{Mean field estimates of the thermodynamic and spectroscopic quantities at selected $t_{am}-h$ cross sections.
(a) Pairing field amplitude ($\vert \Delta\vert$), (b) pairing momentum ($q=\sqrt{q_{x}^{2}+q_{y}^{2}}$), (c) magnetic polarization ($m$), 
(d) real space maps of the pairing field amplitude at representative fields corresponding to BCS, PDW, 1D-LO and 2D-LO phases, 
(e)-(f) spin resolved low energy spectral weight distribution ($A_{\sigma}({\bf k}, 0)$) mapping out the underlying Fermi surface, 
(g) single particle DOS ($N(\omega)$) at the representative $t_{am}-h$ cross sections.}
\label{fig2}
\end{center}
\end{figure}

\textit{Ground state:} Fig.\ref{fig1} sums up the ground state of the system in the $t_{am}-h$ plane, typified in terms of the thermodynamic and spectroscopic signatures shown in Fig.\ref{fig2}. The thermodynamic phases are mapped out based on the variational MFT calculation and are classified in terms of the pairing field amplitude ($\vert \Delta \vert$), pairing momentum (${\bf q}$)  and magnetic polarization ($m$) as: (i) BCS ($\vert \Delta\vert \neq 0$, ${\bf q} = 0$, $m=0$), (ii) PDW ($\vert \Delta \vert \neq 0$, ${\bf q} \neq 0$, $m=0$), (iii) LO (1D) ($\vert \Delta \vert \neq 0$, $q_{x} = 0$, $q_{y} \neq 0$, $m \neq 0$), (iv) LO (2D) ($\vert \Delta \vert \neq 0$, $q_{x} \neq 0$, $q_{y} \neq 0$, $m \neq 0$) and (v) PFL ($\vert \Delta \vert = 0$, ${\bf q} = 0$, $m \neq 0$).  Note that at $t_{am}=0$ the regime $0 < h \lesssim h_{c1}$ can further be sub-divided into an uniform (BCS) $d$-wave superconductor ($0 < h \lesssim h_{QBP}$) and a QBP phase ($h_{QBP} < h \lesssim h_{c1}$). The later is a {\it gapless} SC phase quantified in terms of $\vert \Delta\vert \neq 0$, ${\bf q} = 0$, $m \neq 0$ and characterized by a finite spectral weight at the Fermi level in the single particle DOS \cite{karmakar_jpcm2020,karmakar_jpcm2024}. It is a phase cohered, spatially inhomogeneous SC state with complimentary magnetic polarization and is expected to survive over the regime $0 < t_{am} \lesssim t_{am1}$, as shown in Fig.\ref{fig1}.  Since the focus of this work is on the ${\bf q} \neq 0$ SC states in the $t_{am}-h-T$ space we don't analyze the QBP phase any further in this manuscript and refer to all the ${\bf q} = 0$ phases as BCS.

In Fig.\ref{fig2}(a)-(c) we show the optimized $\vert \Delta \vert$, ${\bf q}$ and $m$, respectively, across the $t_{am}-h$ cross sections. While the $t_{am}=0$ limit essentially corresponds to a BCS superconductor, ALM ($t_{am} \neq 0$) weakly suppresses $\vert \Delta\vert$ (Fig.\ref{fig2}(a)) 
and simultaneously promotes a ${\bf q} \neq 0$ pairing (Fig.\ref{fig2}(b)); leading to a collinear (${\bf q} = \{0, \pi\}$) paired state at $h=0$.  The finite momentum paired state at $h=0$, $t_{am} \neq 0$ represents the illusive PDW, which unlike the standard FFLO state is devoid of 
any magnetic polarization, as observed from Fig.\ref{fig2}(c). Based on the pairing field amplitude and momentum we define the critical ALM scales at the ground state as, (i) $t_{am1} \sim 0.6t$ and (ii) $t_{am2} \sim 2.4t$, demarcating the second order transitions between the BCS-PDW and PDW-PFL phases, respectively. In a similar spirit, the applied Zeeman field regime $0 < h \lesssim h_{c1} \sim 0.8t$ corresponds to the BCS state for $t_{am} \lesssim t_{am1}$, with ${\bf q}=0$ and to a 1D modulated LO phase with ${\bf q} \neq 0$ and $m \neq 0$ for $t_{am1} < t_{am} \lesssim t_{am2}$. The Zeeman field promotes the FFLO pairing, such that, a first order transition takes place across $h_{c1}$ and a non-collinear ($q, q$) 2D modulated LO state with $m \neq 0$ is realized over the regime $h_{c1} < h \lesssim h_{c2}$, as is evident from Fig.\ref{fig2}(a-c).  The LO phase via a second order transition gives way to the PFL at $h \gtrsim h_{c2}$, which lacks any superconducting correlations.

The real space maps for the pairing field amplitude at the selected $t_{am}-h$ cross sections, representative of the various phases are shown in 
Fig.\ref{fig2}(d), while the corresponding spin resolved Fermi surface topology as determined from the low energy spectral weight distribution $A({\bf k}, 0)$ are presented in Fig.\ref{fig2}(e) and Fig.\ref{fig2}(f). In Fig.\ref{fig2}(g) we show the associated single particle DOS. The BCS state at $t_{am}=0.0, h=0.0$ is an uniform, long range phase cohered $d_{x^{2}-y^{2}}$ state with a nodal Fermi surface comprising of hot spots along the diagonals of the Brillouin zone (Fig.\ref{fig2}(e)-(f)). The corresponding single particle DOS with the characteristic ``V''-shape at the Fermi level is shown in Fig.\ref{fig2}(g). At $t_{am}=0.6t, h=0.0$ the system hosts a finite momentum ($0, \pi$) PDW phase, which though maps out a nearly uniform, phase cohered SC phase in real space, significantly alters the nodal Fermi surface topology of the $d_{x^{2}-y^{2}}$ uniform superconductor (Fig.\ref{fig2}(e)-(f)). Note that such segmentation of the Fermi surface is reported over a large class of SC systems such as, magnetic superconductor, magnet-superconductor hybrid, helical superconductor etc. and arises due to the finite momentum scattering of the quasiparticles. Unlike the standard BCS pairing which couples the $\vert {\bf k}_{\uparrow}\rangle$ to the $\vert {-\bf k}_{\downarrow}\rangle$ states, a finite momentum pairing connects the $\vert {\bf k}_{\uparrow}\rangle$ state to $\vert {\bf {-k + q}}_{\downarrow}\rangle$ and $\vert {\bf {-k - q}}_{\downarrow}\rangle$ states as well, giving rise to a dispersion spectra with multiple branches and non trivial van Hove singularities. Applied Zeeman field leads to spin dependent splitting of the dispersion spectra bringing forth additional dispersion branches. Real space modulated paired states are realized at $h = 0.9t$ and $h=1.2t$, representative of the 1D and 2D LO phases, respectively. The Zeeman field induced imbalance in the population of the fermionic species (and therefore $m \neq 0$) shows up in the form of the Fermi surface mismatch, as depicted in Fig.\ref{fig2}(e) and Fig.\ref{fig2}(f). The spectroscopic signature of ${\bf q} \neq 0$ pairing is in the form of gapless single particle DOS with finite spectral weight at the Fermi level, as shown in Fig.\ref{fig2}(g) for the PDW and LO phases.  
\begin{figure}
\begin{center}
\includegraphics[height=15cm,width=8.5cm,angle=0]{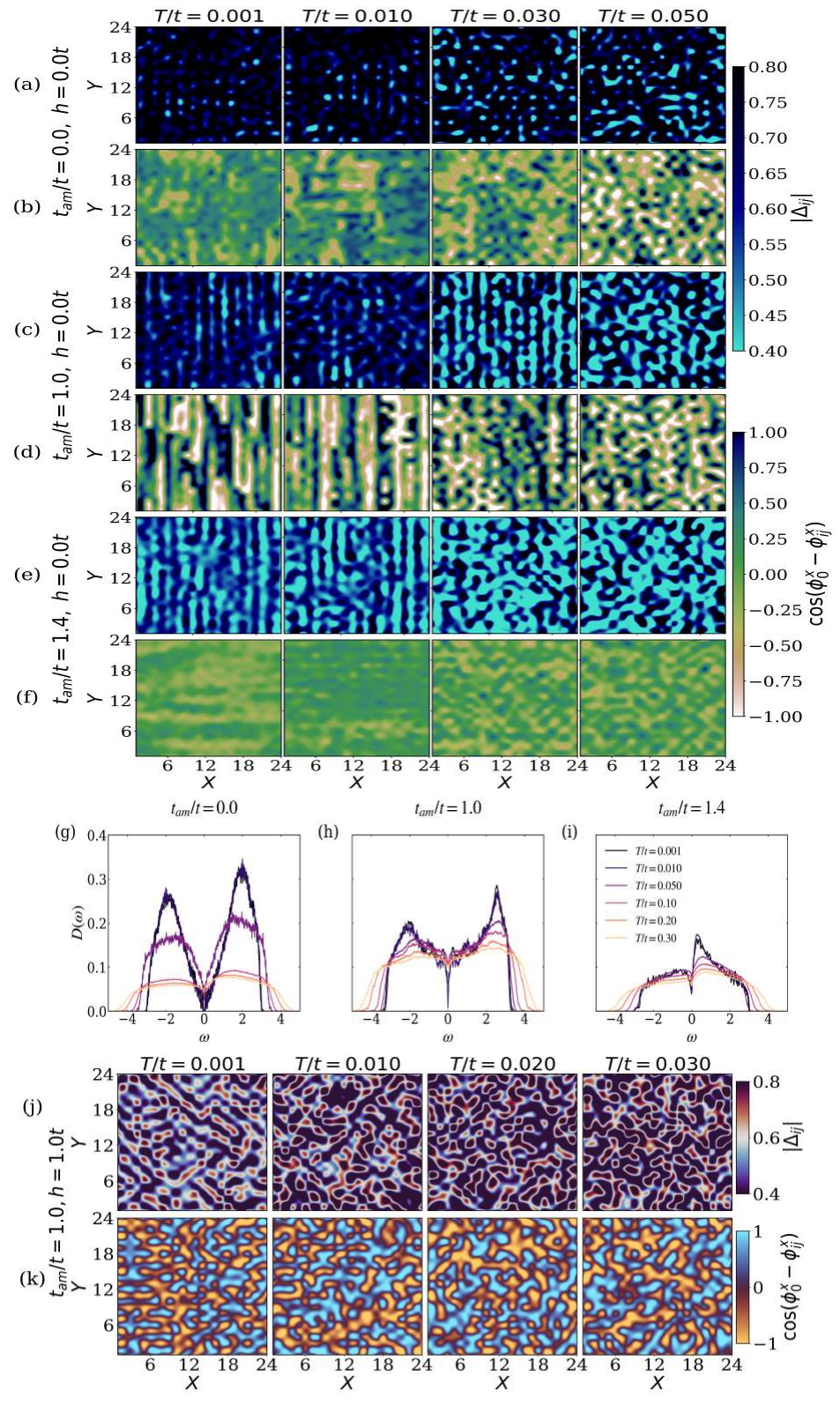}
\caption{Thermodynamic and spectroscopic signatures in the BCS and PDW phases at representative $t_{am}-T$ cross sections and $h=0$. 
(a), (c), (e) real space maps corresponding to the thermal evolution of the pairing field amplitude $\vert \Delta_{ij}\vert$ at the selected values 
of $t_{am}=0$, $t_{am}=1t$ and $t_{am}=1.4t$, respectively. (b), (d), (f) thermal evolution of the phase correlation ($\cos(\phi_{0}^{x}-\phi_{ij}^{x})$) at  $t_{am}=0$, $t_{am}=1t$ and $t_{am}=1.4t$, respectively. (g)-(i) corresponding single particle DOS as a function of temperature. (j)-(k) real space maps for the thermal evolution of $\vert \Delta_{ij}\vert$ and $\cos(\phi_{0}^{x}-\phi_{ij}^{x})$ at a Zeeman field of $h=1.0t$ and $t_{am}=1.0t$.}
\label{fig3}
\end{center}
\end{figure}

\textit{Finite temperature ${\bf q} \neq 0$ phases:} In Fig.\ref{fig3} we present the main result of this work highlighting the thermal stability of the PDW phase. While the possible ground state realizations of the PDW phase have been discussed in significant detail across literature even in the context of ALM-SC, the phase is known to be extremely fragile to thermal fluctuations, thereby evading its finite temperature actualization \cite{agterberg_annrevcmp2020}. We provide unambiguous evidence of the finite temperature PDW in Fig.\ref{fig3} as determined using SPA in terms of the stability of the pairing field amplitude and (quasi) long range phase coherence against thermal fluctuations, at selected $t_{am}$ and $h=0$. In agreement with the real space signatures the underlying Fermi surface is segmented and exhibit hot spots for quasiparticle scattering at isolated points of the Brillouin zone which undergoes thermal broadening with temperature (see SM).

The uniform, phase cohered BCS state at $t_{am}=0.0$ undergoes thermal fluctuation induced spatial fragmentation with temperature, such that 
the global phase coherence is lost at $T_{c} \sim 0.05t$ conforming to the standard $d_{x^{2}-y^{2}}$ paired superconductor \cite{karmakar_jpcm2020}. The loss of the global phase coherence is spatially quantified in terms of short range correlated isolated islands separated by non SC regimes as shown in Fig.\ref{fig3}(a) and Fig.\ref{fig3}(b). The PDW phase at $t_{am}=1.0t$ is typified by real space modulation in pairing field amplitude and phase correlation, as shown in Fig.\ref{fig3}(c)-(d). The regime of stability of the PDW phase at low temperatures as obtained via SPA is in agreement with the MFT estimates shown in Fig.\ref{fig1}. Thermal fluctuations lead to rapid disordering of the PDW state with the corresponding $T_{c} \sim 0.03t$. At $t_{am}=1.4t$ the PDW phase is fragile and is destroyed for $T \gtrsim 0$, see Fig.\ref{fig3}(e)-(f).

The spectroscopic fingerprint of the finite temperature PDW phase is analyzed in terms of the thermal evolution of the single particle DOS shown in Fig.\ref{fig3}(g)-(i), at selected $t_{am}$. At $t_{am}=0.0$ the single particle DOS at low temperatures behaves as $N(\omega) \propto \omega$ as $\omega \rightarrow 0$, characteristic to the $d_{x^{2}-y^{2}}$ pairing symmetry. Thermal fluctuations close the gap and broaden the spectra via large transfer of spectral weight away from the Fermi level. At $t_{am}=1.0t$ the single particle spectra is gapless and undergoes progressive temperature induced broadening. The $t_{am}=1.4t$ cross section is largely devoid of SC pair correlations and the corresponding single particle spectra maps out the underlying non interacting band structure which undergoes thermal fluctuations induced broadening. Finally, Fig.\ref{fig3}(j)-(k) shows the thermal fluctuation induced melting of the LO phase in terms of the pairing field amplitude and phase coherence at $t_{am}=1.0t$ and $h=1.0t$.

\begin{figure}
\begin{center}
\includegraphics[height=5cm,width=6.5cm,angle=0]{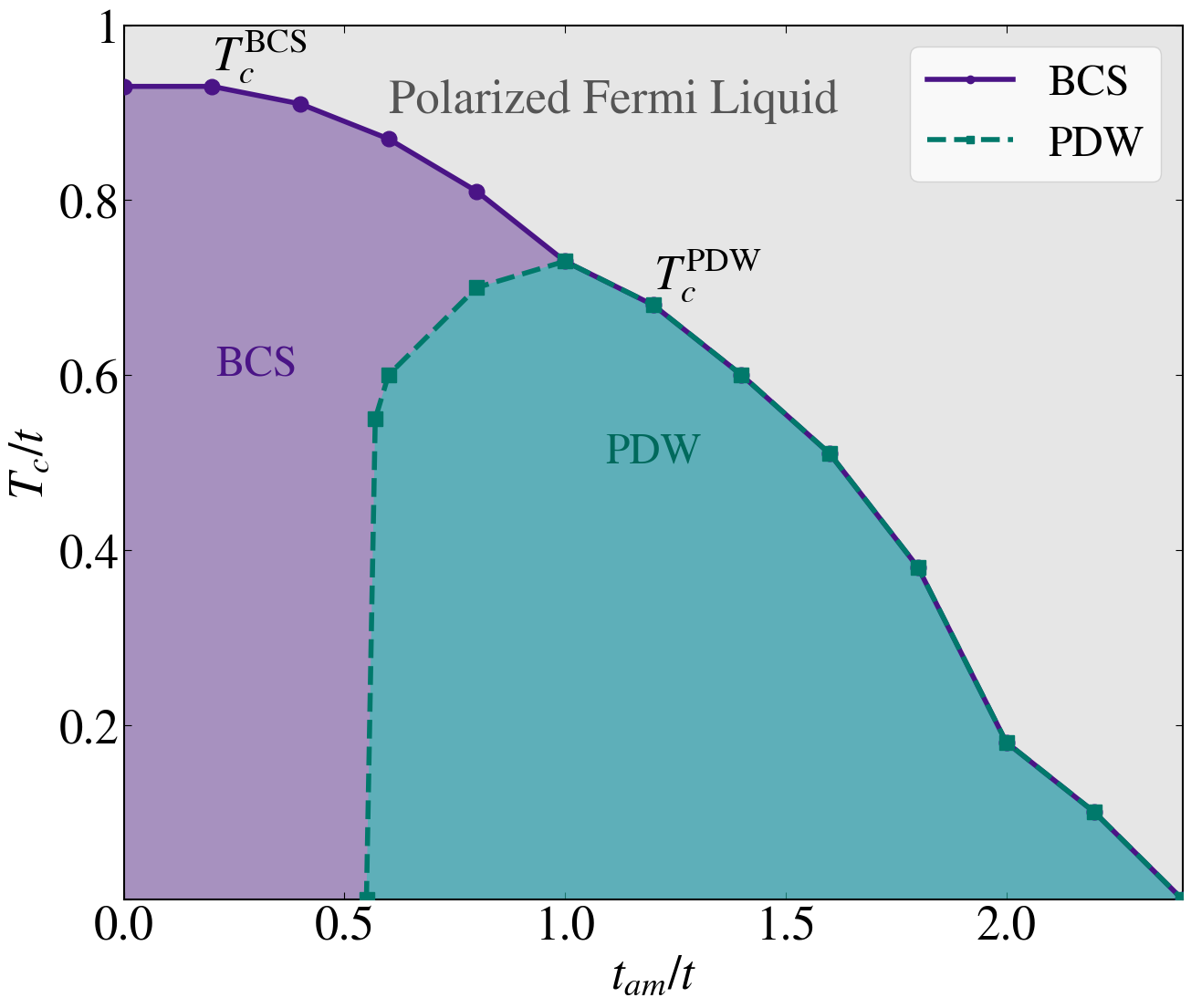}
\caption{Thermal phase diagram as obtained from the variational mean field theory, showing the BCS and PDW phases and the corresponding 
thermal scales $T_{c}^{BCS}$ and $T_{c}^{PDW}$. Note that while the thermodynamic phases and the order of phase transitions are well 
captured by MFT, the thermal scales are overestimated.}
\label{fig4}
\end{center}
\end{figure}

\textit{Discussion and conclusions:} The $d$-wave superconductivity is prone to thermal fluctuations induced disordering owing to the nodal 
order parameter and low energy excitations \cite{karmakar_jpcm2020,karmakar_jpcm2024}.  The phase coherence decay sharply with temperature, making the thermal transition scales non trivial to access, particularly in case of ${\bf q} \neq 0$ pairing which tends to suppress the phase correlation even further. In the above sections we have shown that ALM can stabilize a finite temperature PDW against thermal fluctuations and have provided the ballpark estimates of the corresponding $T_{c}$. These estimates can further be concretized in terms of the temperature dependence of the superfluid stiffness  by determining the current-current correlator which is beyond the scope of the present work. A thermal phase diagram can however be mapped out based on the variational MFT calculations as shown in Fig.\ref{fig4} which provides a qualitative estimate of the transitions in the $t_{am}-T$ plane at $h=0$. It must however be noted that MFT though captures the thermodynamic phases with reasonable accuracy, grossly over estimates the transition scales owing to its neglect of the phase fluctuations. Nevertheless, Fig.\ref{fig4} provides the estimate of the regimes over which the various SC correlations dominate. Over the regime $0 < t_{am} \lesssim t_{am1}$ the uniform $d$-wave SC state is realized which gives way to the PDW phase over the regime $t_{am1} < t_{am} \lesssim t_{am2}$. Thermal transitions to the PFL from both these phases are of second order. 

The experimental evidence of the Pauli limited superconductivity and FFLO phase in solid state materials largely stems from the specific heat, 
magnetic torque, muon spin rotation ($\mu$SR), nuclear magnetic resonance (NMR) and magnetic neutron scattering measurements \cite{sarro_prl2003,onuki_prb2002,flouquet_prl2008,kenzelman_science2008,sarro_prb2004,sarro_prb2005,gerber_natphys2013,matsuda_prl2006,movshovich_prx2016,movshovic_prl2020,wosnitza_ltp2013,wosnitza_prl2007,mitrovic_natphys2014,wright_prl2011,montegomery_prb2011,lortz_prb2011,agosta_prb2012,schlueter_prb2009,lohneysen_prl2013,prozorov_prb2011,kim_prb2011}. Among the candidate materials, heavy fermion superconductor CeCoIn$_{5}$ with its large effective mass \cite{sarro_prl2003,onuki_prb2002,flouquet_prl2008,kenzelman_science2008,sarro_prb2004,sarro_prb2005,gerber_natphys2013,matsuda_prl2006,movshovich_prx2016,movshovic_prl2020}, 2D layered organic superconductor $\kappa$-BEDT \cite{wosnitza_ltp2013,wosnitza_prl2007,mitrovic_natphys2014,wright_prl2011,montegomery_prb2011,lortz_prb2011,agosta_prb2012,schlueter_prb2009}, iron pnictide KFe$_{2}$As$_{2}$ \cite{lohneysen_prl2013,prozorov_prb2011,kim_prb2011} and iron chalcogenide FeSe \cite{ok_prb2020,kasahara_prl2021}, stands out. The low temperature, high magnetic field regime of these systems are reported to host finite momentum paired states. On the other hand, inspite of the several theoretical proposals so far there has not been an experimental evidence either for an ALM-SC or for the PDW phase therein. However, there are promising candidates which can potentially host such a state. For example, spatially modulated superconducting gap has been observed in Kagome superconductors AV$_{3}$Sb$_{5}$ through spectroscopic imaging \cite{hasan_natmat2021}. Similarly, scanning tunnelling microscopy and spectroscopy (STM/STS) and Josephson STS revealed a state akin to PDW in CsV$_{3}$Sb$_{5}$ \cite{gao_nature2021,goh_nanolett2023}, intertwined with a charge density wave (CDW) order and exhibiting pseudogap like features \cite{gao_nature2021,goh_nanolett2023}. Owing to the absence of the depairing magnetic field ALM provides a highly plausible route to actualize the PDW state in real materials.

The results presented in this manuscript are determined using SPA Monte Carlo simulation based on the adiabatic approximation of the slow bosonic fields \cite{ciuchi_scipost2021,fratini_prb2023,kivelson_pans2023,karmakar_prml2025}.  The method is well suited to address quantum phases and has been suitably used to study a wide class of many body systems such as, unconventional superconductivity \cite{karmakar_pra2016,karmakar_pra2018}, geometrically frustrated lattices and Mott insulator-metal transitions (MIT) \cite{karmakar_spinliq,karmakar_tri}, flat and multiband systems \cite{lieb_strain,shashi_kagome2024} and most recently MIT in altermagnetic metals \cite{santhosh_dalm2026}. While the method is accurate to capture the finite temperature physics of the system, it doesn't take into account the quantum fluctuations at $T \rightarrow 0$ and effectively boils down to the MFT.  The $T \neq 0$ results discussed in this work are however accurate within the purview of SPA for $T > T_{FL}$ where, $T_{FL}$ is the Fermi liquid coherence temperature.

\textit{Acknowledgment:} M.K. would like to acknowledge the use of the high performance computing facility (AQUA) at the Indian Institute of Technology, Madras, India.

\section{Supplementary Information}

\textit{Model Hamiltonian and $d$-wave pairing:} An attractive fermionic interaction doesn't lead to an inter-site pairing by itself. If we start with a repulsive Hubbard model the $s$-wave pairing is inhibited by the formation of the local moments. In the strong coupling regime a $t-J$ model obtains from the repulsive Hubbard model if the double occupancy is projected out as,
\begin{eqnarray}
\hat H & = & {\cal P}[\sum_{\langle ij\rangle, \sigma}t_{ij}(\hat c_{i,\sigma}^{\dagger}\hat c_{j,\sigma}+h.c.) + \sum_{ij}J_{ij}(\vec \sigma_{i}.\vec \sigma_{j}-\frac{1}{4}\hat n_{i}\hat n_{j}) \nonumber \\ && - \mu N]{\cal P}
\end{eqnarray} 
where, ${\cal P}$ is the projection operator that eliminates the double occupancy, $J_{ij}=4t_{ij}^{2}/U$ and $\vec \sigma$ is the electron spin operator. 
The Hamiltonian now contains spin-spin and density-density coupling but the projection needs to be retained for further calculation. An alternate approach can be implemented wherein both the Hubbard and the inter-site interactions are retained. The resulting $t-J-V$ model doesn't require 
explicit projection and the corresponding Hamiltonian reads as, 
\begin{eqnarray}
\hat H & = & \sum_{\langle ij\rangle, \sigma}(-t_{ij}+\sigma t_{am}\eta_{ij})(\hat c_{i,\sigma}^{\dagger}\hat c_{j,\sigma} - U \sum_{\langle ij\rangle}\hat n_{i}\hat n_{j} \nonumber \\ && + V\sum_{i}\hat n_{i\uparrow}\hat n_{i\downarrow} - \mu N)
\end{eqnarray}
where, we have taken into account the spin-dependent ALM interaction such that, $t_{am}$ quantifies the strength of the interaction and $\eta_{ij}$ is 
the $d$-wave form factor, leading to $t_{\hat x} = t-\frac{\sigma t_{am}}{2}$ and $t_{\hat y} = t+\frac{\sigma t_{am}}{2}$; $\sigma=+(-)$ for the $\uparrow$($\downarrow$) spin species. We treat $U$ and $V$ independently and consider the $V=0$ limit such that, the local moment formation is completely suppressed. The resulting Hamiltonian includes density-density coupling which can be decomposed into bosonic auxiliary pairing field $\Delta_{ij}$ as 
follows:

The partition function of the system is written in the functional integral form in terms of the Grassmann fields $\psi_{i\sigma}(\tau)$ and $\bar \psi_{i\sigma}(\tau)$,
 
 {\begin{eqnarray}
Z & = & \int \cal{D}\psi \cal{D}\bar \psi \exp{-\int_{\mathrm{0}}^{\beta} {\mathrm {d}\tau} \cal{L}(\tau)}  \\ 
\cal{L}(\tau) & = & \cal{L}_{\mathrm{0}}(\tau) + \cal{L}_{\mathrm{U}}(\tau) \\ 
\cal{L}_{\mathrm{0}}(\tau) & = & \sum_{\langle ij\rangle, \sigma}\{\bar \psi_{i\sigma}((\partial_{\tau}-\mu-\sigma_{i}^{z}h)\delta_{ij} 
+ t_{ij}-\sigma t_{am}\eta_{ij})\psi_{j\sigma}\} \\ 
\cal{L}_{\mathrm{U}}(\tau) & = & -U\sum_{\langle ij\rangle, \sigma \sigma^{\prime}}\bar \psi_{i\sigma}\psi_{i\sigma}
\bar \psi_{j\sigma^{\prime}}\psi_{j\sigma^{\prime}} 
\end{eqnarray}
where, $t_{ij}=t=0.5$ is the nearest neighbor hopping and sets the reference energy scale of the system. Superconducting pairing is brought in via the attractive interaction $\vert U\vert > 0$, the chemical potential $\mu$ dictates the fermionic number density in the system and the Zeeman field $h$ allows for a population imbalance between the fermionic species, leading to a finite magnetic polarization, $m$. $\beta$ is the inverse temperature.

We decompose the interaction term using Hubbard Stratonovich (HS) decomposition \cite{hs1,hs2} introducing the bosonic auxiliary $d$-wave pairing singlet $\Delta_{ij}(\tau)$. Here $ij$ 
and $\tau$ refers to the spatial and imaginary time dependence of the pairing field, respectively. In terms of the 
Matsubara frequency $\Omega_{n}=2\pi nT$ the pairing field reads as, $\Delta_{ijn}$, where $T$ is temperature. 
The resulting partition function is given as, 
{\begin{eqnarray}
Z & = & \int {\cal D}\psi {\cal D} \bar \psi {\cal D} \Delta {\cal D} \Delta^{*}e^{-\int_{\mathrm 0}^{\beta}
{\mathrm d}\tau {\cal L}(\tau)} \\ 
{\cal L}(\tau) & = & {\cal L}_{\mathrm 0}(\tau) + {\cal L}_{\mathrm U}(\tau) + {\cal L}_{cl}(\tau) \\ 
{\cal L}_{\mathrm 0}(\tau) & = & \sum_{\langle ij\rangle, \sigma} 
\{\bar \psi_{i\sigma}((\partial_{\tau}-\mu -\sigma_{i}^{z}h)\delta_{ij} + t_{ij}-\sigma t_{am}\eta_{ij})\psi_{j\sigma}\} \\ 
{\cal L}_{\mathrm U}(\tau) & = & -\sum_{i\neq j}\Delta_{ij}(\bar \psi_{i\uparrow}\bar \psi_{j\downarrow} + 
\bar \psi_{j\uparrow}\bar \psi_{i\downarrow}) + h. c. \\
{\cal L}_{\mathrm cl}(\tau) & = & 4\sum_{i\neq j}\frac{\vert \Delta_{ij}\vert^{2}}{\vert U\vert}
\end{eqnarray}

The fermions are now quadratic but at the cost of an extra integral over $\Delta$ and $\Delta^{*}$. The 
$\int \cal{D}\psi \cal{D}\bar \psi$ integral can now be performed to generate the effective action for the random 
background fields $\{\Delta\}$,
\begin{eqnarray}
Z & = & \int {\cal D}\Delta {\cal D} \Delta^{*} e^{-S_{eff}\{\Delta, \Delta^{*}\}} \\ 
S_{eff} & = & {\mathrm {ln}}{\mathrm {Det}}[{\cal G}^{-1}\{\Delta, \Delta^{*}\}] + 
\int_{\mathrm 0}^{\beta} {\mathrm d}\tau {\cal L}_{\mathrm cl}(\tau)
\end{eqnarray}
here, $\cal{G}$ is the electronic Green's function in the $\{\Delta\}$ background.

\textit{SPA Monte Carlo:}
Within the framework of SPA, the auxiliary field retains the spatial fluctuations at all orders but retains only the $\Omega_{n}=0$ mode 
of the Matsubara frequency, such that, $\Delta_{ij}(\tau) \rightarrow \Delta_{ij}$. The system can be thought of as fermions moving on a random correlated background of classical $\Delta_{ij}$. The resulting effective Hamiltonian reads as,

\begin{eqnarray}
  H_{eff} & = & \sum_{\langle ij\rangle, \sigma}(-t_{ij}+\sigma t_{am}\eta_{ij})(c_{i\sigma}^{\dagger}c_{j\sigma} + h. c.) \nonumber \\ && 
  + \sum_{i\neq j}\Delta_{ij}(c_{i\uparrow}^{\dagger}c_{j\downarrow}^{\dagger} +
  c_{j\uparrow}^{\dagger}c_{i\downarrow}^{\dagger}) + h. c. -\mu\sum_{i,\sigma}\hat n_{i,\sigma} \nonumber \\ && 
  -h\sum_{i,\sigma}\sigma_{i}^{z}\hat n_{i,\sigma} + 4\sum_{i\neq j} \frac{\vert \Delta_{ij}\vert^{2}}{\vert U\vert}
\end{eqnarray}  
where, the last term corresponds to the stiffness cost associated with the now classical auxiliary field.

The $\{\Delta_{ij}\}$ background obeys the Boltzmann distribution, $P\{\Delta_{ij}\} \propto Tr_{c, c^{\dagger}}e^{-\beta H_{eff}}$, related to the free energy of the system. For large, random background the trace is taken numerically. The background configurations are generated by Monte Carlo simulation, diagonalizing $H_{eff}$ for each attempted update of $\Delta_{ij}$. The computation cost is brought down by implementing traveling cluster 
approximation (TCA) scheme \cite{karmakar_jpcm2020,karmakar_jpcm2024}. The required fermionic correlators are then computed on the optimized background configurations. 
\begin{figure}
\begin{center}
\includegraphics[height=4.5cm,width=8.5cm,angle=0]{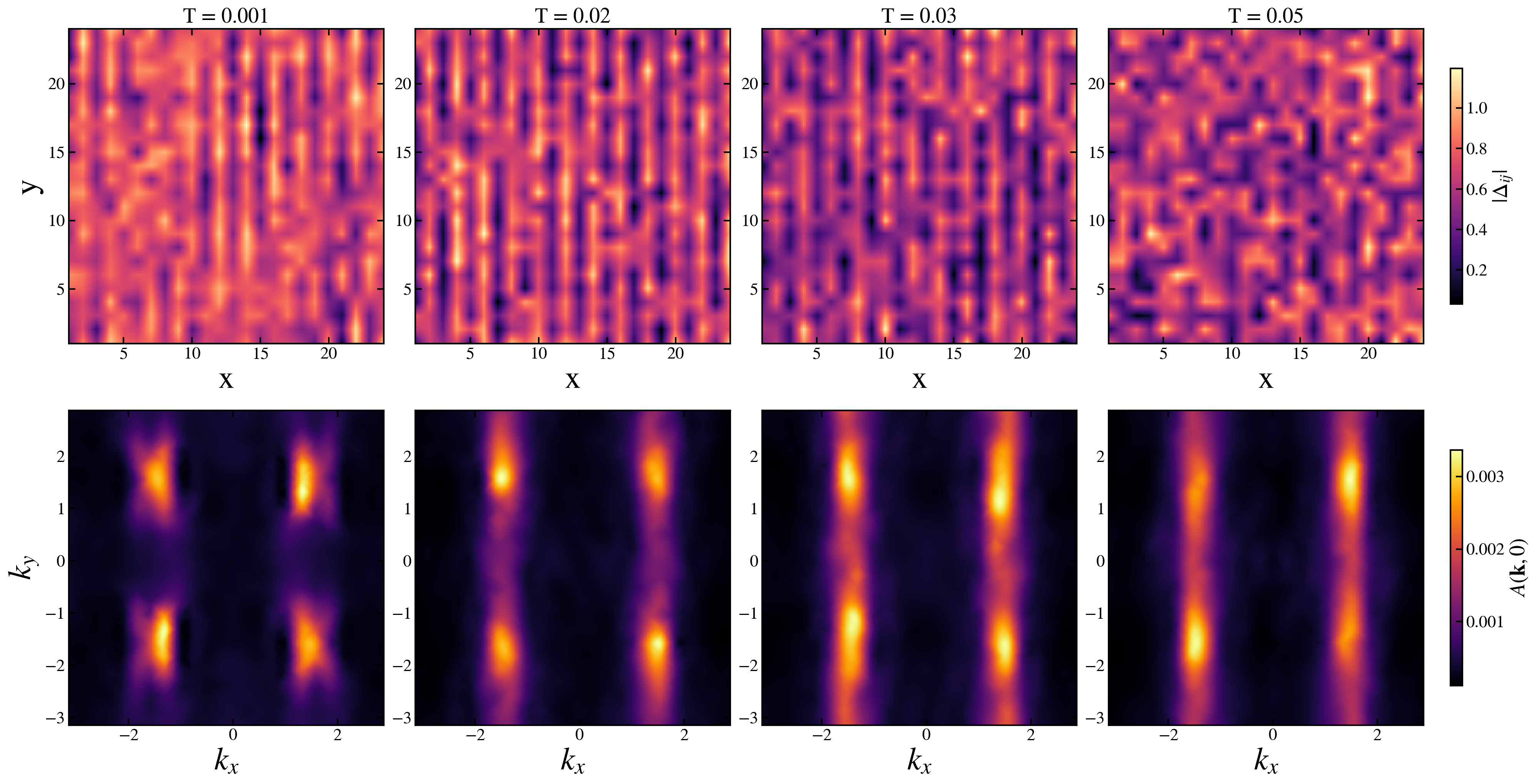}
\caption{The thermal evolution of the PDW state at $t_{am} = 1.0t$ and $h = 0.0$. Top row: Spatial maps corresponding to the 
pairing amplitude $\vert \Delta_{ij}\vert$. Bottom row: Zero-energy spectral weight distributions $A({\bf k}, 0)$ that demonstrate 
the reconstruction of the Fermi surface.}
\label{fig1_suppl}
\end{center}
\end{figure} 

\textit{Observables:}
The ground state and finite temperature phases are characterized based on the following thermodynamic and spectroscopic signatures, 
\begin{itemize}
\item{Magnetic polarization, 
\begin{eqnarray}
m & = & \frac{1}{N} \langle n_{\uparrow}-n_{\downarrow}\rangle
\end{eqnarray}
}
\item{Single particle DOS, 
\begin{eqnarray}
N(\omega) & = & \frac{1}{N}\langle \sum_{i}(\vert u_{n}^{i}\vert^{2}\delta(\omega-E_{n})+\vert v_{n}^{i}\vert^{2}\delta(\omega+E_{n})\rangle
\end{eqnarray}
}
\item{Spin resolved low energy spectral weight distribution, 
\begin{eqnarray}
A_{\sigma}({\bf k}, 0) & = & -(1/\pi) \Im G_{\sigma}({\bf k}, \omega \rightarrow 0)
\end{eqnarray}
}
\end{itemize}
where, $i$ and $j$ correspond to two different sites on the lattice. $\langle \cdot\rangle$ correspond to thermal average and $\sigma$ is the 
spin label. $n_{i\sigma}$ are the number of the individual fermionic species, while $u_{n}^{i}$ and $v_{n}^{i}$ are Bogoliubov eigenfunctions corresponding to the eigenvalue $E_{n}$. The single particle Green's function reads as, $G({\bf k}, \omega) = \lim_{\delta \rightarrow 0} G({\bf k}, i\omega_{n})\vert_{i\omega_{n} \rightarrow \omega + i\delta}$, where $G({\bf k}, \omega)$ is the imaginary frequency transform of $\langle c_{\bf k}(\tau)c_{\bf k}^{\dagger}(0)\rangle$.

\textit{Thermal evolution of the Fermi surface:}
We show the temperature dependence of the Fermi surface in conjugation with the corresponding real space signatures in the PDW phase in 
Fig.\ref{fig1_suppl} at $t_{am}=1.0t$ and $h=0$. The low temperature phase exhibit 1D modulations in the pairing field amplitude with the 
underlying Fermi surface being nodal. Thermal fluctuations lead to progressive fragmentation of the SC phase and the associated segmented 
Fermi surface undergoes broadening.

\bibliography{dscalm}

\begin{thebibliography}{124}%
\makeatletter
\providecommand \@ifxundefined [1]{%
 \@ifx{#1\undefined}
}%
\providecommand \@ifnum [1]{%
 \ifnum #1\expandafter \@firstoftwo
 \else \expandafter \@secondoftwo
 \fi
}%
\providecommand \@ifx [1]{%
 \ifx #1\expandafter \@firstoftwo
 \else \expandafter \@secondoftwo
 \fi
}%
\providecommand \natexlab [1]{#1}%
\providecommand \enquote  [1]{``#1''}%
\providecommand \bibnamefont  [1]{#1}%
\providecommand \bibfnamefont [1]{#1}%
\providecommand \citenamefont [1]{#1}%
\providecommand \href@noop [0]{\@secondoftwo}%
\providecommand \href [0]{\begingroup \@sanitize@url \@href}%
\providecommand \@href[1]{\@@startlink{#1}\@@href}%
\providecommand \@@href[1]{\endgroup#1\@@endlink}%
\providecommand \@sanitize@url [0]{\catcode `\\12\catcode `\$12\catcode
  `\&12\catcode `\#12\catcode `\^12\catcode `\_12\catcode `\%12\relax}%
\providecommand \@@startlink[1]{}%
\providecommand \@@endlink[0]{}%
\providecommand \url  [0]{\begingroup\@sanitize@url \@url }%
\providecommand \@url [1]{\endgroup\@href {#1}{\urlprefix }}%
\providecommand \urlprefix  [0]{URL }%
\providecommand \Eprint [0]{\href }%
\providecommand \doibase [0]{http://dx.doi.org/}%
\providecommand \selectlanguage [0]{\@gobble}%
\providecommand \bibinfo  [0]{\@secondoftwo}%
\providecommand \bibfield  [0]{\@secondoftwo}%
\providecommand \translation [1]{[#1]}%
\providecommand \BibitemOpen [0]{}%
\providecommand \bibitemStop [0]{}%
\providecommand \bibitemNoStop [0]{.\EOS\space}%
\providecommand \EOS [0]{\spacefactor3000\relax}%
\providecommand \BibitemShut  [1]{\csname bibitem#1\endcsname}%
\let\auto@bib@innerbib\@empty
\bibitem [{\citenamefont {\ifmmode~\check{S}\else \v{S}\fi{}mejkal}\ \emph
  {et~al.}(2022)\citenamefont {\ifmmode~\check{S}\else \v{S}\fi{}mejkal},
  \citenamefont {Sinova},\ and\ \citenamefont {Jungwirth}}]{jungwirth_prx2022}%
  \BibitemOpen
  \bibfield  {author} {\bibinfo {author} {\bibfnamefont {Libor}\ \bibnamefont
  {\ifmmode~\check{S}\else \v{S}\fi{}mejkal}}, \bibinfo {author} {\bibfnamefont
  {Jairo}\ \bibnamefont {Sinova}}, \ and\ \bibinfo {author} {\bibfnamefont
  {Tomas}\ \bibnamefont {Jungwirth}},\ }\bibfield  {title} {\enquote {\bibinfo
  {title} {Beyond conventional ferromagnetism and antiferromagnetism: A phase
  with nonrelativistic spin and crystal rotation symmetry},}\ }\href {\doibase
  10.1103/PhysRevX.12.031042} {\bibfield  {journal} {\bibinfo  {journal} {Phys.
  Rev. X}\ }\textbf {\bibinfo {volume} {12}},\ \bibinfo {pages} {031042}
  (\bibinfo {year} {2022})}\BibitemShut {NoStop}%
\bibitem [{\citenamefont {Song}\ \emph {et~al.}(2025)\citenamefont {Song},
  \citenamefont {Bai}, \citenamefont {Zhou}, \citenamefont {Han}, \citenamefont
  {Reichlova}, \citenamefont {Dil}, \citenamefont {Liu}, \citenamefont {Chen},\
  and\ \citenamefont {Pan}}]{pan_natrevmat2025}%
  \BibitemOpen
  \bibfield  {author} {\bibinfo {author} {\bibfnamefont {Cheng}\ \bibnamefont
  {Song}}, \bibinfo {author} {\bibfnamefont {Hua}\ \bibnamefont {Bai}},
  \bibinfo {author} {\bibfnamefont {Zhiyuan}\ \bibnamefont {Zhou}}, \bibinfo
  {author} {\bibfnamefont {Lei}\ \bibnamefont {Han}}, \bibinfo {author}
  {\bibfnamefont {Helena}\ \bibnamefont {Reichlova}}, \bibinfo {author}
  {\bibfnamefont {J.~Hugo}\ \bibnamefont {Dil}}, \bibinfo {author}
  {\bibfnamefont {Junwei}\ \bibnamefont {Liu}}, \bibinfo {author}
  {\bibfnamefont {Xianzhe}\ \bibnamefont {Chen}}, \ and\ \bibinfo {author}
  {\bibfnamefont {Feng}\ \bibnamefont {Pan}},\ }\bibfield  {title} {\enquote
  {\bibinfo {title} {Altermagnets as a new class of functional materials},}\
  }\href {\doibase 10.1038/s41578-025-00779-1} {\bibfield  {journal} {\bibinfo
  {journal} {Nature Reviews Materials}\ }\textbf {\bibinfo {volume} {10}},\
  \bibinfo {pages} {473} (\bibinfo {year} {2025})}\BibitemShut {NoStop}%
\bibitem [{\citenamefont {Shim}\ \emph {et~al.}(2025)\citenamefont {Shim},
  \citenamefont {Mehraeen}, \citenamefont {Sklenar}, \citenamefont {Zhang},
  \citenamefont {Hoffmann},\ and\ \citenamefont {Mason}}]{mason_annrev2025}%
  \BibitemOpen
  \bibfield  {author} {\bibinfo {author} {\bibfnamefont {Soho}\ \bibnamefont
  {Shim}}, \bibinfo {author} {\bibfnamefont {M.}~\bibnamefont {Mehraeen}},
  \bibinfo {author} {\bibfnamefont {Joseph}\ \bibnamefont {Sklenar}}, \bibinfo
  {author} {\bibfnamefont {Steven S.-L.}\ \bibnamefont {Zhang}}, \bibinfo
  {author} {\bibfnamefont {Axel}\ \bibnamefont {Hoffmann}}, \ and\ \bibinfo
  {author} {\bibfnamefont {Nadya}\ \bibnamefont {Mason}},\ }\bibfield  {title}
  {\enquote {\bibinfo {title} {Spin-polarized antiferromagnetic metals},}\
  }\href {\doibase https://doi.org/10.1146/annurev-conmatphys-042924-123620}
  {\bibfield  {journal} {\bibinfo  {journal} {Annual Review of Condensed Matter
  Physics}\ }\textbf {\bibinfo {volume} {16}},\ \bibinfo {pages} {103}
  (\bibinfo {year} {2025})}\BibitemShut {NoStop}%
\bibitem [{\citenamefont {Fernandes}\ \emph {et~al.}(2024)\citenamefont
  {Fernandes}, \citenamefont {de~Carvalho}, \citenamefont {Birol},\ and\
  \citenamefont {Pereira}}]{pereira_prb2024}%
  \BibitemOpen
  \bibfield  {author} {\bibinfo {author} {\bibfnamefont {Rafael~M.}\
  \bibnamefont {Fernandes}}, \bibinfo {author} {\bibfnamefont {Vanuildo~S.}\
  \bibnamefont {de~Carvalho}}, \bibinfo {author} {\bibfnamefont {Turan}\
  \bibnamefont {Birol}}, \ and\ \bibinfo {author} {\bibfnamefont {Rodrigo~G.}\
  \bibnamefont {Pereira}},\ }\bibfield  {title} {\enquote {\bibinfo {title}
  {Topological transition from nodal to nodeless zeeman splitting in
  altermagnets},}\ }\href {\doibase 10.1103/PhysRevB.109.024404} {\bibfield
  {journal} {\bibinfo  {journal} {Phys. Rev. B}\ }\textbf {\bibinfo {volume}
  {109}},\ \bibinfo {pages} {024404} (\bibinfo {year} {2024})}\BibitemShut
  {NoStop}%
\bibitem [{\citenamefont {Mazin}\ \emph {et~al.}(2021)\citenamefont {Mazin},
  \citenamefont {Koepernik}, \citenamefont {Johannes}, \citenamefont
  {González-Hernández},\ and\ \citenamefont {Šmejkal}}]{smejkal_pnas2021}%
  \BibitemOpen
  \bibfield  {author} {\bibinfo {author} {\bibfnamefont {Igor~I.}\ \bibnamefont
  {Mazin}}, \bibinfo {author} {\bibfnamefont {Klaus}\ \bibnamefont
  {Koepernik}}, \bibinfo {author} {\bibfnamefont {Michelle~D.}\ \bibnamefont
  {Johannes}}, \bibinfo {author} {\bibfnamefont {Rafael}\ \bibnamefont
  {González-Hernández}}, \ and\ \bibinfo {author} {\bibfnamefont {Libor}\
  \bibnamefont {Šmejkal}},\ }\bibfield  {title} {\enquote {\bibinfo {title}
  {Prediction of unconventional magnetism in doped fesb<sub>2</sub>},}\ }\href
  {\doibase 10.1073/pnas.2108924118} {\bibfield  {journal} {\bibinfo  {journal}
  {Proceedings of the National Academy of Sciences}\ }\textbf {\bibinfo
  {volume} {118}},\ \bibinfo {pages} {e2108924118} (\bibinfo {year}
  {2021})}\BibitemShut {NoStop}%
\bibitem [{bri(2023)}]{brink_mattodayphys2023}%
  \BibitemOpen
  \bibfield  {title} {\enquote {\bibinfo {title} {Spin-split collinear
  antiferromagnets: A large-scale ab-initio study},}\ }\href {\doibase
  https://doi.org/10.1016/j.mtphys.2023.100991} {\bibfield  {journal} {\bibinfo
   {journal} {Materials Today Physics}\ }\textbf {\bibinfo {volume} {32}},\
  \bibinfo {pages} {100991} (\bibinfo {year} {2023})}\BibitemShut {NoStop}%
\bibitem [{\citenamefont {Gao}\ \emph {et~al.}(2025)\citenamefont {Gao},
  \citenamefont {Qu}, \citenamefont {Zeng}, \citenamefont {Liu}, \citenamefont
  {Wen}, \citenamefont {Sun}, \citenamefont {Guo},\ and\ \citenamefont
  {Lu}}]{lu_natscirev2025}%
  \BibitemOpen
  \bibfield  {author} {\bibinfo {author} {\bibfnamefont {Ze-Feng}\ \bibnamefont
  {Gao}}, \bibinfo {author} {\bibfnamefont {Shuai}\ \bibnamefont {Qu}},
  \bibinfo {author} {\bibfnamefont {Bocheng}\ \bibnamefont {Zeng}}, \bibinfo
  {author} {\bibfnamefont {Yang}\ \bibnamefont {Liu}}, \bibinfo {author}
  {\bibfnamefont {Ji-Rong}\ \bibnamefont {Wen}}, \bibinfo {author}
  {\bibfnamefont {Hao}\ \bibnamefont {Sun}}, \bibinfo {author} {\bibfnamefont
  {Peng-Jie}\ \bibnamefont {Guo}}, \ and\ \bibinfo {author} {\bibfnamefont
  {Zhong-Yi}\ \bibnamefont {Lu}},\ }\bibfield  {title} {\enquote {\bibinfo
  {title} {Ai-accelerated discovery of altermagnetic materials},}\ }\href
  {\doibase 10.1093/nsr/nwaf066} {\bibfield  {journal} {\bibinfo  {journal}
  {National Science Review}\ }\textbf {\bibinfo {volume} {12}},\ \bibinfo
  {pages} {nwaf066} (\bibinfo {year} {2025})}\BibitemShut {NoStop}%
\bibitem [{\citenamefont {Mazin}\ \emph {et~al.}(2023)\citenamefont {Mazin},
  \citenamefont {González-Hernández},\ and\ \citenamefont
  {Šmejkal}}]{smejkal_arxiv2023}%
  \BibitemOpen
  \bibfield  {author} {\bibinfo {author} {\bibfnamefont {Igor}\ \bibnamefont
  {Mazin}}, \bibinfo {author} {\bibfnamefont {Rafael}\ \bibnamefont
  {González-Hernández}}, \ and\ \bibinfo {author} {\bibfnamefont {Libor}\
  \bibnamefont {Šmejkal}},\ }\bibfield  {title} {\enquote {\bibinfo {title}
  {Induced monolayer altermagnetism in mnp(s,se)$_3$ and fese},}\ }\href
  {https://arxiv.org/abs/2309.02355} {\  (\bibinfo {year} {2023})},\ \Eprint
  {http://arxiv.org/abs/2309.02355} {arXiv:2309.02355 [cond-mat.mes-hall]}
  \BibitemShut {NoStop}%
\bibitem [{\citenamefont {Jaeschke-Ubiergo}\ \emph {et~al.}(2024)\citenamefont
  {Jaeschke-Ubiergo}, \citenamefont {Bharadwaj}, \citenamefont {Jungwirth},
  \citenamefont {\ifmmode~\check{S}\else \v{S}\fi{}mejkal},\ and\ \citenamefont
  {Sinova}}]{sinova_prb2024}%
  \BibitemOpen
  \bibfield  {author} {\bibinfo {author} {\bibfnamefont {Rodrigo}\ \bibnamefont
  {Jaeschke-Ubiergo}}, \bibinfo {author} {\bibfnamefont {Venkata~Krishna}\
  \bibnamefont {Bharadwaj}}, \bibinfo {author} {\bibfnamefont {Tomas}\
  \bibnamefont {Jungwirth}}, \bibinfo {author} {\bibfnamefont {Libor}\
  \bibnamefont {\ifmmode~\check{S}\else \v{S}\fi{}mejkal}}, \ and\ \bibinfo
  {author} {\bibfnamefont {Jairo}\ \bibnamefont {Sinova}},\ }\bibfield  {title}
  {\enquote {\bibinfo {title} {Supercell altermagnets},}\ }\href {\doibase
  10.1103/PhysRevB.109.094425} {\bibfield  {journal} {\bibinfo  {journal}
  {Phys. Rev. B}\ }\textbf {\bibinfo {volume} {109}},\ \bibinfo {pages}
  {094425} (\bibinfo {year} {2024})}\BibitemShut {NoStop}%
\bibitem [{\citenamefont {Wan}\ \emph {et~al.}(2025)\citenamefont {Wan},
  \citenamefont {Mandal}, \citenamefont {Guo},\ and\ \citenamefont
  {Haule}}]{haule_prl2025}%
  \BibitemOpen
  \bibfield  {author} {\bibinfo {author} {\bibfnamefont {Xuhao}\ \bibnamefont
  {Wan}}, \bibinfo {author} {\bibfnamefont {Subhasish}\ \bibnamefont {Mandal}},
  \bibinfo {author} {\bibfnamefont {Yuzheng}\ \bibnamefont {Guo}}, \ and\
  \bibinfo {author} {\bibfnamefont {Kristjan}\ \bibnamefont {Haule}},\
  }\bibfield  {title} {\enquote {\bibinfo {title} {High-throughput search for
  metallic altermagnets by embedded dynamical mean field theory},}\ }\href
  {\doibase 10.1103/k47t-23gp} {\bibfield  {journal} {\bibinfo  {journal}
  {Phys. Rev. Lett.}\ }\textbf {\bibinfo {volume} {135}},\ \bibinfo {pages}
  {106501} (\bibinfo {year} {2025})}\BibitemShut {NoStop}%
\bibitem [{\citenamefont {Sødequist}\ and\ \citenamefont
  {Olsen}()}]{olsen_apl2024}%
  \BibitemOpen
  \bibfield  {author} {\bibinfo {author} {\bibfnamefont {Joachim}\ \bibnamefont
  {Sødequist}}\ and\ \bibinfo {author} {\bibfnamefont {Thomas}\ \bibnamefont
  {Olsen}},\ }\bibfield  {title} {\enquote {\bibinfo {title} {Two-dimensional
  altermagnets from high throughput computational screening: Symmetry
  requirements, chiral magnons, and spin-orbit effects},}\ }\href {\doibase
  10.1063/5.0198285} {\bibfield  {journal} {\bibinfo  {journal} {Applied
  Physics Letters}\ }\textbf {\bibinfo {volume} {124}},\ \bibinfo {pages}
  {182409}}\BibitemShut {NoStop}%
\bibitem [{\citenamefont {Smolyanyuk}\ \emph {et~al.}(2024)\citenamefont
  {Smolyanyuk}, \citenamefont {Šmejkal},\ and\ \citenamefont
  {Mazin}}]{mazin_scipost2024}%
  \BibitemOpen
  \bibfield  {author} {\bibinfo {author} {\bibfnamefont {Andriy}\ \bibnamefont
  {Smolyanyuk}}, \bibinfo {author} {\bibfnamefont {Libor}\ \bibnamefont
  {Šmejkal}}, \ and\ \bibinfo {author} {\bibfnamefont {Igor~I.}\ \bibnamefont
  {Mazin}},\ }\bibfield  {title} {\enquote {\bibinfo {title} {A tool to check
  whether a symmetry-compensated collinear magnetic material is antiferro- or
  altermagnetic},}\ }\href {https://arxiv.org/abs/2401.08784} {\  (\bibinfo
  {year} {2024})},\ \Eprint {http://arxiv.org/abs/2401.08784} {arXiv:2401.08784
  [cond-mat.mtrl-sci]} \BibitemShut {NoStop}%
\bibitem [{\citenamefont {Che}\ \emph {et~al.}()\citenamefont {Che},
  \citenamefont {Lv}, \citenamefont {Wu},\ and\ \citenamefont
  {Yang}}]{yang_chemsci2024}%
  \BibitemOpen
  \bibfield  {author} {\bibinfo {author} {\bibfnamefont {Yixuan}\ \bibnamefont
  {Che}}, \bibinfo {author} {\bibfnamefont {Haifeng}\ \bibnamefont {Lv}},
  \bibinfo {author} {\bibfnamefont {Xiaojun}\ \bibnamefont {Wu}}, \ and\
  \bibinfo {author} {\bibfnamefont {Jinlong}\ \bibnamefont {Yang}},\ }\bibfield
   {title} {\enquote {\bibinfo {title} {Realizing altermagnetism in
  two-dimensional metal–organic framework semiconductors with
  electric-field-controlled anisotropic spin current},}\ }\href {\doibase
  10.1039/D4SC04125A} {\bibfield  {journal} {\bibinfo  {journal} {Chem. Sci.}\
  }\textbf {\bibinfo {volume} {15}},\ 10.1039/D4SC04125A}\BibitemShut {NoStop}%
\bibitem [{\citenamefont {Che}\ \emph {et~al.}(2025)\citenamefont {Che},
  \citenamefont {Lv}, \citenamefont {Wu},\ and\ \citenamefont
  {Yang}}]{yang_jamchemsoc2025}%
  \BibitemOpen
  \bibfield  {author} {\bibinfo {author} {\bibfnamefont {Yixuan}\ \bibnamefont
  {Che}}, \bibinfo {author} {\bibfnamefont {Haifeng}\ \bibnamefont {Lv}},
  \bibinfo {author} {\bibfnamefont {Xiaojun}\ \bibnamefont {Wu}}, \ and\
  \bibinfo {author} {\bibfnamefont {Jinlong}\ \bibnamefont {Yang}},\ }\bibfield
   {title} {\enquote {\bibinfo {title} {Bilayer metal–organic framework
  altermagnets with electrically tunable spin-split valleys},}\ }\href
  {\doibase 10.1021/jacs.5c04106} {\bibfield  {journal} {\bibinfo  {journal}
  {Journal of the American Chemical Society}\ }\textbf {\bibinfo {volume}
  {147}},\ \bibinfo {pages} {14806} (\bibinfo {year} {2025})}\BibitemShut
  {NoStop}%
\bibitem [{\citenamefont {Bhattarai}\ \emph {et~al.}(2025)\citenamefont
  {Bhattarai}, \citenamefont {Minch},\ and\ \citenamefont
  {Rhone}}]{rhone_prm2025}%
  \BibitemOpen
  \bibfield  {author} {\bibinfo {author} {\bibfnamefont {Romakanta}\
  \bibnamefont {Bhattarai}}, \bibinfo {author} {\bibfnamefont {Peter}\
  \bibnamefont {Minch}}, \ and\ \bibinfo {author} {\bibfnamefont
  {Trevor~David}\ \bibnamefont {Rhone}},\ }\bibfield  {title} {\enquote
  {\bibinfo {title} {High-throughput screening of altermagnetic materials},}\
  }\href {\doibase 10.1103/PhysRevMaterials.9.064403} {\bibfield  {journal}
  {\bibinfo  {journal} {Phys. Rev. Mater.}\ }\textbf {\bibinfo {volume} {9}},\
  \bibinfo {pages} {064403} (\bibinfo {year} {2025})}\BibitemShut {NoStop}%
\bibitem [{\citenamefont {Gu}\ \emph {et~al.}(2025)\citenamefont {Gu},
  \citenamefont {Liu}, \citenamefont {Zhu}, \citenamefont {Yananose},
  \citenamefont {Chen}, \citenamefont {Hu}, \citenamefont {Stroppa},\ and\
  \citenamefont {Liu}}]{stroppa_prl2025}%
  \BibitemOpen
  \bibfield  {author} {\bibinfo {author} {\bibfnamefont {Mingqiang}\
  \bibnamefont {Gu}}, \bibinfo {author} {\bibfnamefont {Yuntian}\ \bibnamefont
  {Liu}}, \bibinfo {author} {\bibfnamefont {Haiyuan}\ \bibnamefont {Zhu}},
  \bibinfo {author} {\bibfnamefont {Kunihiro}\ \bibnamefont {Yananose}},
  \bibinfo {author} {\bibfnamefont {Xiaobing}\ \bibnamefont {Chen}}, \bibinfo
  {author} {\bibfnamefont {Yongkang}\ \bibnamefont {Hu}}, \bibinfo {author}
  {\bibfnamefont {Alessandro}\ \bibnamefont {Stroppa}}, \ and\ \bibinfo
  {author} {\bibfnamefont {Qihang}\ \bibnamefont {Liu}},\ }\bibfield  {title}
  {\enquote {\bibinfo {title} {Ferroelectric switchable altermagnetism},}\
  }\href {\doibase 10.1103/PhysRevLett.134.106802} {\bibfield  {journal}
  {\bibinfo  {journal} {Phys. Rev. Lett.}\ }\textbf {\bibinfo {volume} {134}},\
  \bibinfo {pages} {106802} (\bibinfo {year} {2025})}\BibitemShut {NoStop}%
\bibitem [{\citenamefont {Duan}\ \emph {et~al.}(2025)\citenamefont {Duan},
  \citenamefont {Zhang}, \citenamefont {Zhu}, \citenamefont {Liu},
  \citenamefont {Zhang}, \citenamefont {\ifmmode \check{Z}\else
  \v{Z}\fi{}uti\ifmmode~\acute{c}\else \'{c}\fi{}},\ and\ \citenamefont
  {Zhou}}]{zhou_prl2025}%
  \BibitemOpen
  \bibfield  {author} {\bibinfo {author} {\bibfnamefont {Xunkai}\ \bibnamefont
  {Duan}}, \bibinfo {author} {\bibfnamefont {Jiayong}\ \bibnamefont {Zhang}},
  \bibinfo {author} {\bibfnamefont {Ziye}\ \bibnamefont {Zhu}}, \bibinfo
  {author} {\bibfnamefont {Yuntian}\ \bibnamefont {Liu}}, \bibinfo {author}
  {\bibfnamefont {Zhenyu}\ \bibnamefont {Zhang}}, \bibinfo {author}
  {\bibfnamefont {Igor}\ \bibnamefont {\ifmmode \check{Z}\else
  \v{Z}\fi{}uti\ifmmode~\acute{c}\else \'{c}\fi{}}}, \ and\ \bibinfo {author}
  {\bibfnamefont {Tong}\ \bibnamefont {Zhou}},\ }\bibfield  {title} {\enquote
  {\bibinfo {title} {Antiferroelectric altermagnets: Antiferroelectricity
  alters magnets},}\ }\href {\doibase 10.1103/PhysRevLett.134.106801}
  {\bibfield  {journal} {\bibinfo  {journal} {Phys. Rev. Lett.}\ }\textbf
  {\bibinfo {volume} {134}},\ \bibinfo {pages} {106801} (\bibinfo {year}
  {2025})}\BibitemShut {NoStop}%
\bibitem [{\citenamefont {Šmejkal}(2024)}]{smejkal_arxiv2024}%
  \BibitemOpen
  \bibfield  {author} {\bibinfo {author} {\bibfnamefont {Libor}\ \bibnamefont
  {Šmejkal}},\ }\bibfield  {title} {\enquote {\bibinfo {title} {Altermagnetic
  multiferroics and altermagnetoelectric effect},}\ }\href
  {https://arxiv.org/abs/2411.19928} {\  (\bibinfo {year} {2024})},\ \Eprint
  {http://arxiv.org/abs/2411.19928} {arXiv:2411.19928 [cond-mat.mtrl-sci]}
  \BibitemShut {NoStop}%
\bibitem [{\citenamefont {Li}\ \emph {et~al.}(2025)\citenamefont {Li},
  \citenamefont {Hu}, \citenamefont {Li}, \citenamefont {Wang}, \citenamefont
  {Chen}, \citenamefont {Thiagarajan}, \citenamefont {Leandersson},
  \citenamefont {Polley}, \citenamefont {Kim}, \citenamefont {Liu},
  \citenamefont {Fulga}, \citenamefont {Vergniory}, \citenamefont {Janson},
  \citenamefont {Tjernberg},\ and\ \citenamefont {van~den
  Brink}}]{brink_comphys2025}%
  \BibitemOpen
  \bibfield  {author} {\bibinfo {author} {\bibfnamefont {Cong}\ \bibnamefont
  {Li}}, \bibinfo {author} {\bibfnamefont {Mengli}\ \bibnamefont {Hu}},
  \bibinfo {author} {\bibfnamefont {Zhilin}\ \bibnamefont {Li}}, \bibinfo
  {author} {\bibfnamefont {Yang}\ \bibnamefont {Wang}}, \bibinfo {author}
  {\bibfnamefont {Wanyu}\ \bibnamefont {Chen}}, \bibinfo {author}
  {\bibfnamefont {Balasubramanian}\ \bibnamefont {Thiagarajan}}, \bibinfo
  {author} {\bibfnamefont {Mats}\ \bibnamefont {Leandersson}}, \bibinfo
  {author} {\bibfnamefont {Craig}\ \bibnamefont {Polley}}, \bibinfo {author}
  {\bibfnamefont {Timur}\ \bibnamefont {Kim}}, \bibinfo {author} {\bibfnamefont
  {Hui}\ \bibnamefont {Liu}}, \bibinfo {author} {\bibfnamefont {Cosma}\
  \bibnamefont {Fulga}}, \bibinfo {author} {\bibfnamefont {Maia~G.}\
  \bibnamefont {Vergniory}}, \bibinfo {author} {\bibfnamefont {Oleg}\
  \bibnamefont {Janson}}, \bibinfo {author} {\bibfnamefont {Oscar}\
  \bibnamefont {Tjernberg}}, \ and\ \bibinfo {author} {\bibfnamefont {Jeroen}\
  \bibnamefont {van~den Brink}},\ }\bibfield  {title} {\enquote {\bibinfo
  {title} {Topological weyl altermagnetism in crsb},}\ }\href {\doibase
  10.1038/s42005-025-02232-9} {\bibfield  {journal} {\bibinfo  {journal}
  {Communications Physics}\ }\textbf {\bibinfo {volume} {8}},\ \bibinfo {pages}
  {311} (\bibinfo {year} {2025})}\BibitemShut {NoStop}%
\bibitem [{\citenamefont {Ding}\ \emph {et~al.}(2024)\citenamefont {Ding},
  \citenamefont {Jiang}, \citenamefont {Chen}, \citenamefont {Tao},
  \citenamefont {Liu}, \citenamefont {Li}, \citenamefont {Liu}, \citenamefont
  {Sun}, \citenamefont {Cheng}, \citenamefont {Liu}, \citenamefont {Yang},
  \citenamefont {Zhang}, \citenamefont {Deng}, \citenamefont {Jing},
  \citenamefont {Huang}, \citenamefont {Shi}, \citenamefont {Ye}, \citenamefont
  {Qiao}, \citenamefont {Wang}, \citenamefont {Guo}, \citenamefont {Feng},\
  and\ \citenamefont {Shen}}]{shen_prl2024}%
  \BibitemOpen
  \bibfield  {author} {\bibinfo {author} {\bibfnamefont {Jianyang}\
  \bibnamefont {Ding}}, \bibinfo {author} {\bibfnamefont {Zhicheng}\
  \bibnamefont {Jiang}}, \bibinfo {author} {\bibfnamefont {Xiuhua}\
  \bibnamefont {Chen}}, \bibinfo {author} {\bibfnamefont {Zicheng}\
  \bibnamefont {Tao}}, \bibinfo {author} {\bibfnamefont {Zhengtai}\
  \bibnamefont {Liu}}, \bibinfo {author} {\bibfnamefont {Tongrui}\ \bibnamefont
  {Li}}, \bibinfo {author} {\bibfnamefont {Jishan}\ \bibnamefont {Liu}},
  \bibinfo {author} {\bibfnamefont {Jianping}\ \bibnamefont {Sun}}, \bibinfo
  {author} {\bibfnamefont {Jinguang}\ \bibnamefont {Cheng}}, \bibinfo {author}
  {\bibfnamefont {Jiayu}\ \bibnamefont {Liu}}, \bibinfo {author} {\bibfnamefont
  {Yichen}\ \bibnamefont {Yang}}, \bibinfo {author} {\bibfnamefont {Runfeng}\
  \bibnamefont {Zhang}}, \bibinfo {author} {\bibfnamefont {Liwei}\ \bibnamefont
  {Deng}}, \bibinfo {author} {\bibfnamefont {Wenchuan}\ \bibnamefont {Jing}},
  \bibinfo {author} {\bibfnamefont {Yu}~\bibnamefont {Huang}}, \bibinfo
  {author} {\bibfnamefont {Yuming}\ \bibnamefont {Shi}}, \bibinfo {author}
  {\bibfnamefont {Mao}\ \bibnamefont {Ye}}, \bibinfo {author} {\bibfnamefont
  {Shan}\ \bibnamefont {Qiao}}, \bibinfo {author} {\bibfnamefont {Yilin}\
  \bibnamefont {Wang}}, \bibinfo {author} {\bibfnamefont {Yanfeng}\
  \bibnamefont {Guo}}, \bibinfo {author} {\bibfnamefont {Donglai}\ \bibnamefont
  {Feng}}, \ and\ \bibinfo {author} {\bibfnamefont {Dawei}\ \bibnamefont
  {Shen}},\ }\bibfield  {title} {\enquote {\bibinfo {title} {Large band
  splitting in $g$-wave altermagnet crsb},}\ }\href {\doibase
  10.1103/PhysRevLett.133.206401} {\bibfield  {journal} {\bibinfo  {journal}
  {Phys. Rev. Lett.}\ }\textbf {\bibinfo {volume} {133}},\ \bibinfo {pages}
  {206401} (\bibinfo {year} {2024})}\BibitemShut {NoStop}%
\bibitem [{\citenamefont {Yang}\ \emph {et~al.}(2025)\citenamefont {Yang},
  \citenamefont {Li}, \citenamefont {Yang}, \citenamefont {Li}, \citenamefont
  {Zheng}, \citenamefont {Zhu}, \citenamefont {Pan}, \citenamefont {Xu},
  \citenamefont {Cao}, \citenamefont {Zhao}, \citenamefont {Jana},
  \citenamefont {Zhang}, \citenamefont {Ye}, \citenamefont {Song},
  \citenamefont {Hu}, \citenamefont {Yang}, \citenamefont {Fujii},
  \citenamefont {Vobornik}, \citenamefont {Shi}, \citenamefont {Yuan},
  \citenamefont {Zhang}, \citenamefont {Xu},\ and\ \citenamefont
  {Liu}}]{liu_natcom2025}%
  \BibitemOpen
  \bibfield  {author} {\bibinfo {author} {\bibfnamefont {Guowei}\ \bibnamefont
  {Yang}}, \bibinfo {author} {\bibfnamefont {Zhanghuan}\ \bibnamefont {Li}},
  \bibinfo {author} {\bibfnamefont {Sai}\ \bibnamefont {Yang}}, \bibinfo
  {author} {\bibfnamefont {Jiyuan}\ \bibnamefont {Li}}, \bibinfo {author}
  {\bibfnamefont {Hao}\ \bibnamefont {Zheng}}, \bibinfo {author} {\bibfnamefont
  {Weifan}\ \bibnamefont {Zhu}}, \bibinfo {author} {\bibfnamefont
  {Ze}~\bibnamefont {Pan}}, \bibinfo {author} {\bibfnamefont {Yifu}\
  \bibnamefont {Xu}}, \bibinfo {author} {\bibfnamefont {Saizheng}\ \bibnamefont
  {Cao}}, \bibinfo {author} {\bibfnamefont {Wenxuan}\ \bibnamefont {Zhao}},
  \bibinfo {author} {\bibfnamefont {Anupam}\ \bibnamefont {Jana}}, \bibinfo
  {author} {\bibfnamefont {Jiawen}\ \bibnamefont {Zhang}}, \bibinfo {author}
  {\bibfnamefont {Mao}\ \bibnamefont {Ye}}, \bibinfo {author} {\bibfnamefont
  {Yu}~\bibnamefont {Song}}, \bibinfo {author} {\bibfnamefont {Lun-Hui}\
  \bibnamefont {Hu}}, \bibinfo {author} {\bibfnamefont {Lexian}\ \bibnamefont
  {Yang}}, \bibinfo {author} {\bibfnamefont {Jun}\ \bibnamefont {Fujii}},
  \bibinfo {author} {\bibfnamefont {Ivana}\ \bibnamefont {Vobornik}}, \bibinfo
  {author} {\bibfnamefont {Ming}\ \bibnamefont {Shi}}, \bibinfo {author}
  {\bibfnamefont {Huiqiu}\ \bibnamefont {Yuan}}, \bibinfo {author}
  {\bibfnamefont {Yongjun}\ \bibnamefont {Zhang}}, \bibinfo {author}
  {\bibfnamefont {Yuanfeng}\ \bibnamefont {Xu}}, \ and\ \bibinfo {author}
  {\bibfnamefont {Yang}\ \bibnamefont {Liu}},\ }\bibfield  {title} {\enquote
  {\bibinfo {title} {Three-dimensional mapping of the altermagnetic spin
  splitting in crsb},}\ }\href {\doibase 10.1038/s41467-025-56647-7} {\bibfield
   {journal} {\bibinfo  {journal} {Nature Communications}\ }\textbf {\bibinfo
  {volume} {16}},\ \bibinfo {pages} {1442} (\bibinfo {year}
  {2025})}\BibitemShut {NoStop}%
\bibitem [{\citenamefont {Lu}\ \emph {et~al.}(2025)\citenamefont {Lu},
  \citenamefont {Feng}, \citenamefont {Wang}, \citenamefont {Chen},
  \citenamefont {Lin}, \citenamefont {Liang}, \citenamefont {Liu},
  \citenamefont {Feng}, \citenamefont {Yamagami}, \citenamefont {Liu},
  \citenamefont {Felser}, \citenamefont {Wu},\ and\ \citenamefont
  {Ma}}]{ma_nanolet2025}%
  \BibitemOpen
  \bibfield  {author} {\bibinfo {author} {\bibfnamefont {Wenlong}\ \bibnamefont
  {Lu}}, \bibinfo {author} {\bibfnamefont {Shiyu}\ \bibnamefont {Feng}},
  \bibinfo {author} {\bibfnamefont {Yuzhi}\ \bibnamefont {Wang}}, \bibinfo
  {author} {\bibfnamefont {Dong}\ \bibnamefont {Chen}}, \bibinfo {author}
  {\bibfnamefont {Zihan}\ \bibnamefont {Lin}}, \bibinfo {author} {\bibfnamefont
  {Xin}\ \bibnamefont {Liang}}, \bibinfo {author} {\bibfnamefont {Siyuan}\
  \bibnamefont {Liu}}, \bibinfo {author} {\bibfnamefont {Wanxiang}\
  \bibnamefont {Feng}}, \bibinfo {author} {\bibfnamefont {Kohei}\ \bibnamefont
  {Yamagami}}, \bibinfo {author} {\bibfnamefont {Junwei}\ \bibnamefont {Liu}},
  \bibinfo {author} {\bibfnamefont {Claudia}\ \bibnamefont {Felser}}, \bibinfo
  {author} {\bibfnamefont {Quansheng}\ \bibnamefont {Wu}}, \ and\ \bibinfo
  {author} {\bibfnamefont {Junzhang}\ \bibnamefont {Ma}},\ }\bibfield  {title}
  {\enquote {\bibinfo {title} {Signature of topological surface bands in
  altermagnetic weyl semimetal crsb},}\ }\href {\doibase
  10.1021/acs.nanolett.5c00482} {\bibfield  {journal} {\bibinfo  {journal}
  {Nano Letters}\ }\textbf {\bibinfo {volume} {25}},\ \bibinfo {pages} {7343}
  (\bibinfo {year} {2025})}\BibitemShut {NoStop}%
\bibitem [{\citenamefont {iang}\ \emph {et~al.}(2025)\citenamefont {iang},
  \citenamefont {Hu}, \citenamefont {Bai}, \citenamefont {Song}, \citenamefont
  {Mu}, \citenamefont {Qu}, \citenamefont {Li}, \citenamefont {Zhu},
  \citenamefont {Pi}, \citenamefont {Wei}, \citenamefont {Sun}, \citenamefont
  {Huang}, \citenamefont {Zheng}, \citenamefont {Peng}, \citenamefont {He},
  \citenamefont {Li}, \citenamefont {Luo}, \citenamefont {Li}, \citenamefont
  {Chen}, \citenamefont {Li}, \citenamefont {Weng},\ and\ \citenamefont
  {Qian}}]{qian_natphys2025}%
  \BibitemOpen
  \bibfield  {author} {\bibinfo {author} {\bibfnamefont {Bei}\ \bibnamefont
  {iang}}, \bibinfo {author} {\bibfnamefont {Mingzhe}\ \bibnamefont {Hu}},
  \bibinfo {author} {\bibfnamefont {Jianli}\ \bibnamefont {Bai}}, \bibinfo
  {author} {\bibfnamefont {Ziyin}\ \bibnamefont {Song}}, \bibinfo {author}
  {\bibfnamefont {Chao}\ \bibnamefont {Mu}}, \bibinfo {author} {\bibfnamefont
  {Gexing}\ \bibnamefont {Qu}}, \bibinfo {author} {\bibfnamefont {Wan}\
  \bibnamefont {Li}}, \bibinfo {author} {\bibfnamefont {Wenliang}\ \bibnamefont
  {Zhu}}, \bibinfo {author} {\bibfnamefont {Hanqi}\ \bibnamefont {Pi}},
  \bibinfo {author} {\bibfnamefont {Zhongxu}\ \bibnamefont {Wei}}, \bibinfo
  {author} {\bibfnamefont {Yu-Jie}\ \bibnamefont {Sun}}, \bibinfo {author}
  {\bibfnamefont {Yaobo}\ \bibnamefont {Huang}}, \bibinfo {author}
  {\bibfnamefont {Xiquan}\ \bibnamefont {Zheng}}, \bibinfo {author}
  {\bibfnamefont {Yingying}\ \bibnamefont {Peng}}, \bibinfo {author}
  {\bibfnamefont {Lunhua}\ \bibnamefont {He}}, \bibinfo {author} {\bibfnamefont
  {Shiliang}\ \bibnamefont {Li}}, \bibinfo {author} {\bibfnamefont {Jianlin}\
  \bibnamefont {Luo}}, \bibinfo {author} {\bibfnamefont {Zheng}\ \bibnamefont
  {Li}}, \bibinfo {author} {\bibfnamefont {Genfu}\ \bibnamefont {Chen}},
  \bibinfo {author} {\bibfnamefont {Hang}\ \bibnamefont {Li}}, \bibinfo
  {author} {\bibfnamefont {Hongming}\ \bibnamefont {Weng}}, \ and\ \bibinfo
  {author} {\bibfnamefont {Tian}\ \bibnamefont {Qian}},\ }\bibfield  {title}
  {\enquote {\bibinfo {title} {A metallic room-temperature d-wave
  altermagnet},}\ }\href {\doibase 10.1038/s41567-025-02822-y} {\bibfield
  {journal} {\bibinfo  {journal} {Nature Physics}\ }\textbf {\bibinfo {volume}
  {21}},\ \bibinfo {pages} {754} (\bibinfo {year} {2025})}\BibitemShut
  {NoStop}%
\bibitem [{\citenamefont {Zhang}\ \emph {et~al.}(2025)\citenamefont {Zhang},
  \citenamefont {Cheng}, \citenamefont {Yin}, \citenamefont {Liu},
  \citenamefont {Deng}, \citenamefont {Qiao}, \citenamefont {Shi},
  \citenamefont {Zhang}, \citenamefont {Lin}, \citenamefont {Liu},
  \citenamefont {Ye}, \citenamefont {Huang}, \citenamefont {Meng},
  \citenamefont {Zhang}, \citenamefont {Okuda}, \citenamefont {Shimada},
  \citenamefont {Cui}, \citenamefont {Zhao}, \citenamefont {Cao}, \citenamefont
  {Qiao}, \citenamefont {Liu},\ and\ \citenamefont {Chen}}]{chen_natphys2025}%
  \BibitemOpen
  \bibfield  {author} {\bibinfo {author} {\bibfnamefont {Fayuan}\ \bibnamefont
  {Zhang}}, \bibinfo {author} {\bibfnamefont {Xingkai}\ \bibnamefont {Cheng}},
  \bibinfo {author} {\bibfnamefont {Zhouyi}\ \bibnamefont {Yin}}, \bibinfo
  {author} {\bibfnamefont {Changchao}\ \bibnamefont {Liu}}, \bibinfo {author}
  {\bibfnamefont {Liwei}\ \bibnamefont {Deng}}, \bibinfo {author}
  {\bibfnamefont {Yuxi}\ \bibnamefont {Qiao}}, \bibinfo {author} {\bibfnamefont
  {Zheng}\ \bibnamefont {Shi}}, \bibinfo {author} {\bibfnamefont {Shuxuan}\
  \bibnamefont {Zhang}}, \bibinfo {author} {\bibfnamefont {Junhao}\
  \bibnamefont {Lin}}, \bibinfo {author} {\bibfnamefont {Zhengtai}\
  \bibnamefont {Liu}}, \bibinfo {author} {\bibfnamefont {Mao}\ \bibnamefont
  {Ye}}, \bibinfo {author} {\bibfnamefont {Yaobo}\ \bibnamefont {Huang}},
  \bibinfo {author} {\bibfnamefont {Xiangyu}\ \bibnamefont {Meng}}, \bibinfo
  {author} {\bibfnamefont {Cheng}\ \bibnamefont {Zhang}}, \bibinfo {author}
  {\bibfnamefont {Taichi}\ \bibnamefont {Okuda}}, \bibinfo {author}
  {\bibfnamefont {Kenya}\ \bibnamefont {Shimada}}, \bibinfo {author}
  {\bibfnamefont {Shengtao}\ \bibnamefont {Cui}}, \bibinfo {author}
  {\bibfnamefont {Yue}\ \bibnamefont {Zhao}}, \bibinfo {author} {\bibfnamefont
  {Guang-Han}\ \bibnamefont {Cao}}, \bibinfo {author} {\bibfnamefont {Shan}\
  \bibnamefont {Qiao}}, \bibinfo {author} {\bibfnamefont {Junwei}\ \bibnamefont
  {Liu}}, \ and\ \bibinfo {author} {\bibfnamefont {Chaoyu}\ \bibnamefont
  {Chen}},\ }\bibfield  {title} {\enquote {\bibinfo {title}
  {Crystal-symmetry-paired spin–valley locking in a layered room-temperature
  metallic altermagnet candidate},}\ }\href {\doibase
  10.1038/s41567-025-02864-2} {\bibfield  {journal} {\bibinfo  {journal}
  {Nature Physics}\ }\textbf {\bibinfo {volume} {21}},\ \bibinfo {pages} {760}
  (\bibinfo {year} {2025})}\BibitemShut {NoStop}%
\bibitem [{\citenamefont {Mazin}(2023)}]{mazin_prbl2023}%
  \BibitemOpen
  \bibfield  {author} {\bibinfo {author} {\bibfnamefont {I.~I.}\ \bibnamefont
  {Mazin}},\ }\bibfield  {title} {\enquote {\bibinfo {title} {Altermagnetism in
  mnte: Origin, predicted manifestations, and routes to detwinning},}\ }\href
  {\doibase 10.1103/PhysRevB.107.L100418} {\bibfield  {journal} {\bibinfo
  {journal} {Phys. Rev. B}\ }\textbf {\bibinfo {volume} {107}},\ \bibinfo
  {pages} {L100418} (\bibinfo {year} {2023})}\BibitemShut {NoStop}%
\bibitem [{\citenamefont {Lee}\ \emph {et~al.}(2024)\citenamefont {Lee},
  \citenamefont {Lee}, \citenamefont {Jung}, \citenamefont {Jung},
  \citenamefont {Kim}, \citenamefont {Lee}, \citenamefont {Seok}, \citenamefont
  {Kim}, \citenamefont {Park}, \citenamefont {\ifmmode~\check{S}\else
  \v{S}\fi{}mejkal}, \citenamefont {Kang},\ and\ \citenamefont
  {Kim}}]{kim_prl2024}%
  \BibitemOpen
  \bibfield  {author} {\bibinfo {author} {\bibfnamefont {Suyoung}\ \bibnamefont
  {Lee}}, \bibinfo {author} {\bibfnamefont {Sangjae}\ \bibnamefont {Lee}},
  \bibinfo {author} {\bibfnamefont {Saegyeol}\ \bibnamefont {Jung}}, \bibinfo
  {author} {\bibfnamefont {Jiwon}\ \bibnamefont {Jung}}, \bibinfo {author}
  {\bibfnamefont {Donghan}\ \bibnamefont {Kim}}, \bibinfo {author}
  {\bibfnamefont {Yeonjae}\ \bibnamefont {Lee}}, \bibinfo {author}
  {\bibfnamefont {Byeongjun}\ \bibnamefont {Seok}}, \bibinfo {author}
  {\bibfnamefont {Jaeyoung}\ \bibnamefont {Kim}}, \bibinfo {author}
  {\bibfnamefont {Byeong~Gyu}\ \bibnamefont {Park}}, \bibinfo {author}
  {\bibfnamefont {Libor}\ \bibnamefont {\ifmmode~\check{S}\else
  \v{S}\fi{}mejkal}}, \bibinfo {author} {\bibfnamefont {Chang-Jong}\
  \bibnamefont {Kang}}, \ and\ \bibinfo {author} {\bibfnamefont {Changyoung}\
  \bibnamefont {Kim}},\ }\bibfield  {title} {\enquote {\bibinfo {title} {Broken
  kramers degeneracy in altermagnetic mnte},}\ }\href {\doibase
  10.1103/PhysRevLett.132.036702} {\bibfield  {journal} {\bibinfo  {journal}
  {Phys. Rev. Lett.}\ }\textbf {\bibinfo {volume} {132}},\ \bibinfo {pages}
  {036702} (\bibinfo {year} {2024})}\BibitemShut {NoStop}%
\bibitem [{\citenamefont {Osumi}\ \emph {et~al.}(2024)\citenamefont {Osumi},
  \citenamefont {Souma}, \citenamefont {Aoyama}, \citenamefont {Yamauchi},
  \citenamefont {Honma}, \citenamefont {Nakayama}, \citenamefont {Takahashi},
  \citenamefont {Ohgushi},\ and\ \citenamefont {Sato}}]{sato_prb2024}%
  \BibitemOpen
  \bibfield  {author} {\bibinfo {author} {\bibfnamefont {T.}~\bibnamefont
  {Osumi}}, \bibinfo {author} {\bibfnamefont {S.}~\bibnamefont {Souma}},
  \bibinfo {author} {\bibfnamefont {T.}~\bibnamefont {Aoyama}}, \bibinfo
  {author} {\bibfnamefont {K.}~\bibnamefont {Yamauchi}}, \bibinfo {author}
  {\bibfnamefont {A.}~\bibnamefont {Honma}}, \bibinfo {author} {\bibfnamefont
  {K.}~\bibnamefont {Nakayama}}, \bibinfo {author} {\bibfnamefont
  {T.}~\bibnamefont {Takahashi}}, \bibinfo {author} {\bibfnamefont
  {K.}~\bibnamefont {Ohgushi}}, \ and\ \bibinfo {author} {\bibfnamefont
  {T.}~\bibnamefont {Sato}},\ }\bibfield  {title} {\enquote {\bibinfo {title}
  {Observation of a giant band splitting in altermagnetic mnte},}\ }\href
  {\doibase 10.1103/PhysRevB.109.115102} {\bibfield  {journal} {\bibinfo
  {journal} {Phys. Rev. B}\ }\textbf {\bibinfo {volume} {109}},\ \bibinfo
  {pages} {115102} (\bibinfo {year} {2024})}\BibitemShut {NoStop}%
\bibitem [{\citenamefont {Krempaský}\ \emph {et~al.}(2024)\citenamefont
  {Krempaský}, \citenamefont {Šmejkal}, \citenamefont {D’Souza},
  \citenamefont {Hajlaoui}, \citenamefont {Springholz}, \citenamefont
  {Uhlířová}, \citenamefont {Alarab}, \citenamefont {Constantinou},
  \citenamefont {Strocov}, \citenamefont {Usanov}, \citenamefont {Pudelko},
  \citenamefont {González-Hernández}, \citenamefont {Birk~Hellenes},
  \citenamefont {Jansa}, \citenamefont {Reichlová}, \citenamefont {Šobáň},
  \citenamefont {Gonzalez~Betancourt}, \citenamefont {Wadley}, \citenamefont
  {Sinova}, \citenamefont {Kriegner}, \citenamefont {Minár}, \citenamefont
  {Dil},\ and\ \citenamefont {Jungwirth}}]{jungwirth_nature2024}%
  \BibitemOpen
  \bibfield  {author} {\bibinfo {author} {\bibfnamefont {J.}~\bibnamefont
  {Krempaský}}, \bibinfo {author} {\bibfnamefont {L.}~\bibnamefont
  {Šmejkal}}, \bibinfo {author} {\bibfnamefont {S.~W.}\ \bibnamefont
  {D’Souza}}, \bibinfo {author} {\bibfnamefont {M.}~\bibnamefont {Hajlaoui}},
  \bibinfo {author} {\bibfnamefont {G.}~\bibnamefont {Springholz}}, \bibinfo
  {author} {\bibfnamefont {K.}~\bibnamefont {Uhlířová}}, \bibinfo {author}
  {\bibfnamefont {F.}~\bibnamefont {Alarab}}, \bibinfo {author} {\bibfnamefont
  {P.~C.}\ \bibnamefont {Constantinou}}, \bibinfo {author} {\bibfnamefont
  {V.}~\bibnamefont {Strocov}}, \bibinfo {author} {\bibfnamefont
  {D.}~\bibnamefont {Usanov}}, \bibinfo {author} {\bibfnamefont {W.~R.}\
  \bibnamefont {Pudelko}}, \bibinfo {author} {\bibfnamefont {R.}~\bibnamefont
  {González-Hernández}}, \bibinfo {author} {\bibfnamefont {A.}~\bibnamefont
  {Birk~Hellenes}}, \bibinfo {author} {\bibfnamefont {Z.}~\bibnamefont
  {Jansa}}, \bibinfo {author} {\bibfnamefont {H.}~\bibnamefont {Reichlová}},
  \bibinfo {author} {\bibfnamefont {Z.}~\bibnamefont {Šobáň}}, \bibinfo
  {author} {\bibfnamefont {R.~D.}\ \bibnamefont {Gonzalez~Betancourt}},
  \bibinfo {author} {\bibfnamefont {P.}~\bibnamefont {Wadley}}, \bibinfo
  {author} {\bibfnamefont {J.}~\bibnamefont {Sinova}}, \bibinfo {author}
  {\bibfnamefont {D.}~\bibnamefont {Kriegner}}, \bibinfo {author}
  {\bibfnamefont {J.}~\bibnamefont {Minár}}, \bibinfo {author} {\bibfnamefont
  {J.~H.}\ \bibnamefont {Dil}}, \ and\ \bibinfo {author} {\bibfnamefont
  {T.}~\bibnamefont {Jungwirth}},\ }\bibfield  {title} {\enquote {\bibinfo
  {title} {Altermagnetic lifting of kramers spin degeneracy},}\ }\href
  {\doibase 10.1038/s41586-023-06907-7} {\bibfield  {journal} {\bibinfo
  {journal} {Nature}\ }\textbf {\bibinfo {volume} {626}},\ \bibinfo {pages}
  {517} (\bibinfo {year} {2024})}\BibitemShut {NoStop}%
\bibitem [{\citenamefont {Wei}\ \emph {et~al.}(2025)\citenamefont {Wei},
  \citenamefont {Li}, \citenamefont {Hatt}, \citenamefont {Huai}, \citenamefont
  {Liu}, \citenamefont {Singh}, \citenamefont {Kim}, \citenamefont {Fernandes},
  \citenamefont {Cardon}, \citenamefont {Zhao}, \citenamefont {Tran},
  \citenamefont {Frandsen}, \citenamefont {Burch}, \citenamefont {Liu},\ and\
  \citenamefont {Ji}}]{ji_prm2025}%
  \BibitemOpen
  \bibfield  {author} {\bibinfo {author} {\bibfnamefont {Chao-Chun}\
  \bibnamefont {Wei}}, \bibinfo {author} {\bibfnamefont {Xiaoyin}\ \bibnamefont
  {Li}}, \bibinfo {author} {\bibfnamefont {Sabrina}\ \bibnamefont {Hatt}},
  \bibinfo {author} {\bibfnamefont {Xudong}\ \bibnamefont {Huai}}, \bibinfo
  {author} {\bibfnamefont {Jue}\ \bibnamefont {Liu}}, \bibinfo {author}
  {\bibfnamefont {Birender}\ \bibnamefont {Singh}}, \bibinfo {author}
  {\bibfnamefont {Kyung-Mo}\ \bibnamefont {Kim}}, \bibinfo {author}
  {\bibfnamefont {Rafael~M.}\ \bibnamefont {Fernandes}}, \bibinfo {author}
  {\bibfnamefont {Paul}\ \bibnamefont {Cardon}}, \bibinfo {author}
  {\bibfnamefont {Liuyan}\ \bibnamefont {Zhao}}, \bibinfo {author}
  {\bibfnamefont {Thao~T.}\ \bibnamefont {Tran}}, \bibinfo {author}
  {\bibfnamefont {Benjamin~A.}\ \bibnamefont {Frandsen}}, \bibinfo {author}
  {\bibfnamefont {Kenneth~S.}\ \bibnamefont {Burch}}, \bibinfo {author}
  {\bibfnamefont {Feng}\ \bibnamefont {Liu}}, \ and\ \bibinfo {author}
  {\bibfnamefont {Huiwen}\ \bibnamefont {Ji}},\ }\bibfield  {title} {\enquote
  {\bibinfo {title}
  {${\mathrm{la}}_{2}{\mathrm{o}}_{3}{\mathrm{mn}}_{2}{\mathrm{se}}_{2}$: A
  correlated insulating layered d-wave altermagnet},}\ }\href {\doibase
  10.1103/PhysRevMaterials.9.024402} {\bibfield  {journal} {\bibinfo  {journal}
  {Phys. Rev. Mater.}\ }\textbf {\bibinfo {volume} {9}},\ \bibinfo {pages}
  {024402} (\bibinfo {year} {2025})}\BibitemShut {NoStop}%
\bibitem [{\citenamefont {Garcia-Gassull}\ \emph {et~al.}(2025)\citenamefont
  {Garcia-Gassull}, \citenamefont {Razpopov}, \citenamefont {Stavropoulos},
  \citenamefont {Mazin},\ and\ \citenamefont {Valentí}}]{valenti_arxiv2025}%
  \BibitemOpen
  \bibfield  {author} {\bibinfo {author} {\bibfnamefont {Laura}\ \bibnamefont
  {Garcia-Gassull}}, \bibinfo {author} {\bibfnamefont {Aleksandar}\
  \bibnamefont {Razpopov}}, \bibinfo {author} {\bibfnamefont
  {Panagiotis~Peter}\ \bibnamefont {Stavropoulos}}, \bibinfo {author}
  {\bibfnamefont {Igor~I}\ \bibnamefont {Mazin}}, \ and\ \bibinfo {author}
  {\bibfnamefont {Roser}\ \bibnamefont {Valentí}},\ }\bibfield  {title}
  {\enquote {\bibinfo {title} {Microscopic origin of the magnetic interactions
  and their experimental signatures in altermagnetic
  la$_2$o$_3$mn$_2$se$_2$},}\ }\href {https://arxiv.org/abs/2506.21661} {\
  (\bibinfo {year} {2025})},\ \Eprint {http://arxiv.org/abs/2506.21661}
  {arXiv:2506.21661 [cond-mat.str-el]} \BibitemShut {NoStop}%
\bibitem [{\citenamefont {Iguchi}\ \emph {et~al.}(2025)\citenamefont {Iguchi},
  \citenamefont {Kobayashi}, \citenamefont {Ikemoto}, \citenamefont {Furukawa},
  \citenamefont {Itoh}, \citenamefont {Iwai}, \citenamefont {Moriwaki},\ and\
  \citenamefont {Sasaki}}]{sasaki_prr2025}%
  \BibitemOpen
  \bibfield  {author} {\bibinfo {author} {\bibfnamefont {Satoshi}\ \bibnamefont
  {Iguchi}}, \bibinfo {author} {\bibfnamefont {Hiroki}\ \bibnamefont
  {Kobayashi}}, \bibinfo {author} {\bibfnamefont {Yuka}\ \bibnamefont
  {Ikemoto}}, \bibinfo {author} {\bibfnamefont {Tetsuya}\ \bibnamefont
  {Furukawa}}, \bibinfo {author} {\bibfnamefont {Hirotake}\ \bibnamefont
  {Itoh}}, \bibinfo {author} {\bibfnamefont {Shinichiro}\ \bibnamefont {Iwai}},
  \bibinfo {author} {\bibfnamefont {Taro}\ \bibnamefont {Moriwaki}}, \ and\
  \bibinfo {author} {\bibfnamefont {Takahiko}\ \bibnamefont {Sasaki}},\
  }\bibfield  {title} {\enquote {\bibinfo {title} {Magneto-optical spectra of
  an organic antiferromagnet as a candidate for an altermagnet},}\ }\href
  {\doibase 10.1103/nnz3-tq7y} {\bibfield  {journal} {\bibinfo  {journal}
  {Phys. Rev. Res.}\ }\textbf {\bibinfo {volume} {7}},\ \bibinfo {pages}
  {033026} (\bibinfo {year} {2025})}\BibitemShut {NoStop}%
\bibitem [{\citenamefont {Bernardini}\ \emph {et~al.}(2025)\citenamefont
  {Bernardini}, \citenamefont {Fiebig},\ and\ \citenamefont
  {Cano}}]{cano_jap2025}%
  \BibitemOpen
  \bibfield  {author} {\bibinfo {author} {\bibfnamefont {Fabio}\ \bibnamefont
  {Bernardini}}, \bibinfo {author} {\bibfnamefont {Manfred}\ \bibnamefont
  {Fiebig}}, \ and\ \bibinfo {author} {\bibfnamefont {Andrés}\ \bibnamefont
  {Cano}},\ }\bibfield  {title} {\enquote {\bibinfo {title}
  {Ruddlesden–popper and perovskite phases as a material platform for
  altermagnetism},}\ }\href {\doibase 10.1063/5.0252836} {\bibfield  {journal}
  {\bibinfo  {journal} {Journal of Applied Physics}\ }\textbf {\bibinfo
  {volume} {137}},\ \bibinfo {pages} {103903} (\bibinfo {year}
  {2025})}\BibitemShut {NoStop}%
\bibitem [{\citenamefont {Naka}\ \emph {et~al.}(2025)\citenamefont {Naka},
  \citenamefont {Motome},\ and\ \citenamefont {Seo}}]{seo_npjspintronics2025}%
  \BibitemOpen
  \bibfield  {author} {\bibinfo {author} {\bibfnamefont {Makoto}\ \bibnamefont
  {Naka}}, \bibinfo {author} {\bibfnamefont {Yukitoshi}\ \bibnamefont
  {Motome}}, \ and\ \bibinfo {author} {\bibfnamefont {Hitoshi}\ \bibnamefont
  {Seo}},\ }\bibfield  {title} {\enquote {\bibinfo {title} {Altermagnetic
  perovskites},}\ }\href {\doibase 10.1038/s44306-024-00066-9} {\bibfield
  {journal} {\bibinfo  {journal} {npj Spintronics}\ }\textbf {\bibinfo {volume}
  {3}},\ \bibinfo {pages} {1} (\bibinfo {year} {2025})}\BibitemShut {NoStop}%
\bibitem [{\citenamefont {Ferrari}\ and\ \citenamefont
  {Valent\'{\i}}(2024)}]{valenti_prb2024}%
  \BibitemOpen
  \bibfield  {author} {\bibinfo {author} {\bibfnamefont {Francesco}\
  \bibnamefont {Ferrari}}\ and\ \bibinfo {author} {\bibfnamefont {Roser}\
  \bibnamefont {Valent\'{\i}}},\ }\bibfield  {title} {\enquote {\bibinfo
  {title} {Altermagnetism on the shastry-sutherland lattice},}\ }\href
  {\doibase 10.1103/PhysRevB.110.205140} {\bibfield  {journal} {\bibinfo
  {journal} {Phys. Rev. B}\ }\textbf {\bibinfo {volume} {110}},\ \bibinfo
  {pages} {205140} (\bibinfo {year} {2024})}\BibitemShut {NoStop}%
\bibitem [{\citenamefont {Sobral}\ \emph {et~al.}(2025)\citenamefont {Sobral},
  \citenamefont {Mandal},\ and\ \citenamefont {Scheurer}}]{scheurer_prr2024}%
  \BibitemOpen
  \bibfield  {author} {\bibinfo {author} {\bibfnamefont {Jo\~ao~Augusto}\
  \bibnamefont {Sobral}}, \bibinfo {author} {\bibfnamefont {Subrata}\
  \bibnamefont {Mandal}}, \ and\ \bibinfo {author} {\bibfnamefont {Mathias~S.}\
  \bibnamefont {Scheurer}},\ }\bibfield  {title} {\enquote {\bibinfo {title}
  {Fractionalized altermagnets: From neighboring and altermagnetic spin liquids
  to spin-symmetric band splitting},}\ }\href {\doibase
  10.1103/PhysRevResearch.7.023152} {\bibfield  {journal} {\bibinfo  {journal}
  {Phys. Rev. Res.}\ }\textbf {\bibinfo {volume} {7}},\ \bibinfo {pages}
  {023152} (\bibinfo {year} {2025})}\BibitemShut {NoStop}%
\bibitem [{\citenamefont {Giuli}\ \emph {et~al.}(2025)\citenamefont {Giuli},
  \citenamefont {Mejuto-Zaera},\ and\ \citenamefont
  {Capone}}]{capone_prbl2025}%
  \BibitemOpen
  \bibfield  {author} {\bibinfo {author} {\bibfnamefont {Samuele}\ \bibnamefont
  {Giuli}}, \bibinfo {author} {\bibfnamefont {Carlos}\ \bibnamefont
  {Mejuto-Zaera}}, \ and\ \bibinfo {author} {\bibfnamefont {Massimo}\
  \bibnamefont {Capone}},\ }\bibfield  {title} {\enquote {\bibinfo {title}
  {Altermagnetism from interaction-driven itinerant magnetism},}\ }\href
  {\doibase 10.1103/PhysRevB.111.L020401} {\bibfield  {journal} {\bibinfo
  {journal} {Phys. Rev. B}\ }\textbf {\bibinfo {volume} {111}},\ \bibinfo
  {pages} {L020401} (\bibinfo {year} {2025})}\BibitemShut {NoStop}%
\bibitem [{\citenamefont {Ouyang}\ \emph {et~al.}(2025)\citenamefont {Ouyang},
  \citenamefont {Guo}, \citenamefont {He},\ and\ \citenamefont
  {Lu}}]{lu_arxiv2025}%
  \BibitemOpen
  \bibfield  {author} {\bibinfo {author} {\bibfnamefont {Zhenfeng}\
  \bibnamefont {Ouyang}}, \bibinfo {author} {\bibfnamefont {Peng-Jie}\
  \bibnamefont {Guo}}, \bibinfo {author} {\bibfnamefont {Rong-Qiang}\
  \bibnamefont {He}}, \ and\ \bibinfo {author} {\bibfnamefont {Zhong-Yi}\
  \bibnamefont {Lu}},\ }\bibfield  {title} {\enquote {\bibinfo {title}
  {Strongly correlated altermagnet cacro$_3$},}\ }\href
  {https://arxiv.org/abs/2507.14081} {\  (\bibinfo {year} {2025})},\ \Eprint
  {http://arxiv.org/abs/2507.14081} {arXiv:2507.14081 [cond-mat.str-el]}
  \BibitemShut {NoStop}%
\bibitem [{\citenamefont {Jungwirth}\ \emph {et~al.}(2025)\citenamefont
  {Jungwirth}, \citenamefont {Sinova}, \citenamefont {Fernandes}, \citenamefont
  {Liu}, \citenamefont {Watanabe}, \citenamefont {Murakami}, \citenamefont
  {Nakatsuji},\ and\ \citenamefont {Smejkal}}]{fernandes_arxiv2025}%
  \BibitemOpen
  \bibfield  {author} {\bibinfo {author} {\bibfnamefont {Tomas}\ \bibnamefont
  {Jungwirth}}, \bibinfo {author} {\bibfnamefont {Jairo}\ \bibnamefont
  {Sinova}}, \bibinfo {author} {\bibfnamefont {Rafael~M.}\ \bibnamefont
  {Fernandes}}, \bibinfo {author} {\bibfnamefont {Qihang}\ \bibnamefont {Liu}},
  \bibinfo {author} {\bibfnamefont {Hikaru}\ \bibnamefont {Watanabe}}, \bibinfo
  {author} {\bibfnamefont {Shuichi}\ \bibnamefont {Murakami}}, \bibinfo
  {author} {\bibfnamefont {Satoru}\ \bibnamefont {Nakatsuji}}, \ and\ \bibinfo
  {author} {\bibfnamefont {Libor}\ \bibnamefont {Smejkal}},\ }\bibfield
  {title} {\enquote {\bibinfo {title} {Symmetry, microscopy and spectroscopy
  signatures of altermagnetism},}\ }\href {https://arxiv.org/abs/2506.22860} {\
   (\bibinfo {year} {2025})},\ \Eprint {http://arxiv.org/abs/2506.22860}
  {arXiv:2506.22860 [cond-mat.mtrl-sci]} \BibitemShut {NoStop}%
\bibitem [{\citenamefont {Park}\ \emph {et~al.}(2025)\citenamefont {Park},
  \citenamefont {Birol}, \citenamefont {Georges},\ and\ \citenamefont
  {Fernandes}}]{fernandes2_arxiv2025}%
  \BibitemOpen
  \bibfield  {author} {\bibinfo {author} {\bibfnamefont {Ina}\ \bibnamefont
  {Park}}, \bibinfo {author} {\bibfnamefont {Turan}\ \bibnamefont {Birol}},
  \bibinfo {author} {\bibfnamefont {Antoine}\ \bibnamefont {Georges}}, \ and\
  \bibinfo {author} {\bibfnamefont {Rafael~M.}\ \bibnamefont {Fernandes}},\
  }\bibfield  {title} {\enquote {\bibinfo {title} {Impact of strong electronic
  correlations on altermagnets: the case of nis2},}\ }\href
  {https://arxiv.org/abs/2512.17059} {\  (\bibinfo {year} {2025})},\ \Eprint
  {http://arxiv.org/abs/2512.17059} {arXiv:2512.17059 [cond-mat.str-el]}
  \BibitemShut {NoStop}%
\bibitem [{\citenamefont {Kaushal}\ and\ \citenamefont
  {Franz}(2025)}]{franz_prl2025}%
  \BibitemOpen
  \bibfield  {author} {\bibinfo {author} {\bibfnamefont {Nitin}\ \bibnamefont
  {Kaushal}}\ and\ \bibinfo {author} {\bibfnamefont {Marcel}\ \bibnamefont
  {Franz}},\ }\href {\doibase 10.1103/s31h-hk2v} {\enquote {\bibinfo {title}
  {Altermagnetism in modified lieb lattice hubbard model},}\ } (\bibinfo {year}
  {2025})\BibitemShut {NoStop}%
\bibitem [{\citenamefont {D\"urrnagel}\ \emph {et~al.}(2025)\citenamefont
  {D\"urrnagel}, \citenamefont {Hohmann}, \citenamefont {Maity}, \citenamefont
  {Seufert}, \citenamefont {Klett}, \citenamefont {Klebl},\ and\ \citenamefont
  {Thomale}}]{thomale_prl2025}%
  \BibitemOpen
  \bibfield  {author} {\bibinfo {author} {\bibfnamefont {Matteo}\ \bibnamefont
  {D\"urrnagel}}, \bibinfo {author} {\bibfnamefont {Hendrik}\ \bibnamefont
  {Hohmann}}, \bibinfo {author} {\bibfnamefont {Atanu}\ \bibnamefont {Maity}},
  \bibinfo {author} {\bibfnamefont {Jannis}\ \bibnamefont {Seufert}}, \bibinfo
  {author} {\bibfnamefont {Michael}\ \bibnamefont {Klett}}, \bibinfo {author}
  {\bibfnamefont {Lennart}\ \bibnamefont {Klebl}}, \ and\ \bibinfo {author}
  {\bibfnamefont {Ronny}\ \bibnamefont {Thomale}},\ }\bibfield  {title}
  {\enquote {\bibinfo {title} {Altermagnetic phase transition in a lieb
  metal},}\ }\href {\doibase 10.1103/2g3v-z76q} {\bibfield  {journal} {\bibinfo
   {journal} {Phys. Rev. Lett.}\ }\textbf {\bibinfo {volume} {135}},\ \bibinfo
  {pages} {036502} (\bibinfo {year} {2025})}\BibitemShut {NoStop}%
\bibitem [{\citenamefont {Sumita}\ \emph {et~al.}(2025)\citenamefont {Sumita},
  \citenamefont {Naka},\ and\ \citenamefont {Seo}}]{sumita_prb2025}%
  \BibitemOpen
  \bibfield  {author} {\bibinfo {author} {\bibfnamefont {Shuntaro}\
  \bibnamefont {Sumita}}, \bibinfo {author} {\bibfnamefont {Makoto}\
  \bibnamefont {Naka}}, \ and\ \bibinfo {author} {\bibfnamefont {Hitoshi}\
  \bibnamefont {Seo}},\ }\bibfield  {title} {\enquote {\bibinfo {title}
  {Phase-modulated superconductivity via altermagnetism},}\ }\href {\doibase
  10.1103/3k12-2467} {\bibfield  {journal} {\bibinfo  {journal} {Phys. Rev. B}\
  }\textbf {\bibinfo {volume} {112}},\ \bibinfo {pages} {144510} (\bibinfo
  {year} {2025})}\BibitemShut {NoStop}%
\bibitem [{\citenamefont {Zhang}\ \emph {et~al.}(2024)\citenamefont {Zhang},
  \citenamefont {Hu},\ and\ \citenamefont {Neupert}}]{neupert_natcom2024}%
  \BibitemOpen
  \bibfield  {author} {\bibinfo {author} {\bibfnamefont {Song-Bo}\ \bibnamefont
  {Zhang}}, \bibinfo {author} {\bibfnamefont {Lun-Hui}\ \bibnamefont {Hu}}, \
  and\ \bibinfo {author} {\bibfnamefont {Titus}\ \bibnamefont {Neupert}},\
  }\bibfield  {title} {\enquote {\bibinfo {title} {Finite-momentum cooper
  pairing in proximitized altermagnets},}\ }\href {\doibase
  10.1038/s41467-024-45951-3} {\bibfield  {journal} {\bibinfo  {journal}
  {Nature Communications}\ }\textbf {\bibinfo {volume} {15}},\ \bibinfo {pages}
  {1801} (\bibinfo {year} {2024})}\BibitemShut {NoStop}%
\bibitem [{\citenamefont {Hu}\ \emph {et~al.}(2026)\citenamefont {Hu},
  \citenamefont {Liu}, \citenamefont {Wang}, \citenamefont {Liu},\ and\
  \citenamefont {Ohashi}}]{ohashi_arxiv2026}%
  \BibitemOpen
  \bibfield  {author} {\bibinfo {author} {\bibfnamefont {Hui}\ \bibnamefont
  {Hu}}, \bibinfo {author} {\bibfnamefont {Zhao}\ \bibnamefont {Liu}}, \bibinfo
  {author} {\bibfnamefont {Jia}\ \bibnamefont {Wang}}, \bibinfo {author}
  {\bibfnamefont {Xia-Ji}\ \bibnamefont {Liu}}, \ and\ \bibinfo {author}
  {\bibfnamefont {Yoji}\ \bibnamefont {Ohashi}},\ }\bibfield  {title} {\enquote
  {\bibinfo {title} {Finite-momentum superconductivity with singlet-triplet
  mixing in an altermagnetic metal: A pairing instability analysis},}\ }\href
  {https://arxiv.org/abs/2603.12897} {\  (\bibinfo {year} {2026})},\ \Eprint
  {http://arxiv.org/abs/2603.12897} {arXiv:2603.12897 [cond-mat.supr-con]}
  \BibitemShut {NoStop}%
\bibitem [{\citenamefont {Sim}\ and\ \citenamefont
  {Knolle}(2025)}]{knoll_prbl2025}%
  \BibitemOpen
  \bibfield  {author} {\bibinfo {author} {\bibfnamefont {GiBaik}\ \bibnamefont
  {Sim}}\ and\ \bibinfo {author} {\bibfnamefont {Johannes}\ \bibnamefont
  {Knolle}},\ }\bibfield  {title} {\enquote {\bibinfo {title} {Pair density
  waves and supercurrent diode effect in altermagnets},}\ }\href {\doibase
  10.1103/b7rh-v7nq} {\bibfield  {journal} {\bibinfo  {journal} {Phys. Rev. B}\
  }\textbf {\bibinfo {volume} {112}},\ \bibinfo {pages} {L020502} (\bibinfo
  {year} {2025})}\BibitemShut {NoStop}%
\bibitem [{\citenamefont {Chakraborty}\ and\ \citenamefont
  {Black-Schaffer}(2024)}]{schaffer_prbl204}%
  \BibitemOpen
  \bibfield  {author} {\bibinfo {author} {\bibfnamefont {Debmalya}\
  \bibnamefont {Chakraborty}}\ and\ \bibinfo {author} {\bibfnamefont
  {Annica~M.}\ \bibnamefont {Black-Schaffer}},\ }\bibfield  {title} {\enquote
  {\bibinfo {title} {Zero-field finite-momentum and field-induced
  superconductivity in altermagnets},}\ }\href {\doibase
  10.1103/PhysRevB.110.L060508} {\bibfield  {journal} {\bibinfo  {journal}
  {Phys. Rev. B}\ }\textbf {\bibinfo {volume} {110}},\ \bibinfo {pages}
  {L060508} (\bibinfo {year} {2024})}\BibitemShut {NoStop}%
\bibitem [{\citenamefont {Chakraborty}\ and\ \citenamefont
  {Black-Schaffer}(2025)}]{schaffer_prl2025}%
  \BibitemOpen
  \bibfield  {author} {\bibinfo {author} {\bibfnamefont {Debmalya}\
  \bibnamefont {Chakraborty}}\ and\ \bibinfo {author} {\bibfnamefont
  {Annica~M.}\ \bibnamefont {Black-Schaffer}},\ }\bibfield  {title} {\enquote
  {\bibinfo {title} {Perfect superconducting diode effect in altermagnets},}\
  }\href {\doibase 10.1103/cv8s-tk4c} {\bibfield  {journal} {\bibinfo
  {journal} {Phys. Rev. Lett.}\ }\textbf {\bibinfo {volume} {135}},\ \bibinfo
  {pages} {026001} (\bibinfo {year} {2025})}\BibitemShut {NoStop}%
\bibitem [{\citenamefont {Bose}\ \emph {et~al.}(2024)\citenamefont {Bose},
  \citenamefont {Vadnais},\ and\ \citenamefont
  {Paramekanti}}]{paramekanti_prb2024}%
  \BibitemOpen
  \bibfield  {author} {\bibinfo {author} {\bibfnamefont {Anjishnu}\
  \bibnamefont {Bose}}, \bibinfo {author} {\bibfnamefont {Samuel}\ \bibnamefont
  {Vadnais}}, \ and\ \bibinfo {author} {\bibfnamefont {Arun}\ \bibnamefont
  {Paramekanti}},\ }\bibfield  {title} {\enquote {\bibinfo {title}
  {Altermagnetism and superconductivity in a multiorbital $t\ensuremath{-}j$
  model},}\ }\href {\doibase 10.1103/PhysRevB.110.205120} {\bibfield  {journal}
  {\bibinfo  {journal} {Phys. Rev. B}\ }\textbf {\bibinfo {volume} {110}},\
  \bibinfo {pages} {205120} (\bibinfo {year} {2024})}\BibitemShut {NoStop}%
\bibitem [{\citenamefont {Zou}\ \emph {et~al.}(2026)\citenamefont {Zou},
  \citenamefont {Fernandes},\ and\ \citenamefont
  {Fradkin}}]{fradkin_arxiv2026}%
  \BibitemOpen
  \bibfield  {author} {\bibinfo {author} {\bibfnamefont {Xuan}\ \bibnamefont
  {Zou}}, \bibinfo {author} {\bibfnamefont {Rafael~M.}\ \bibnamefont
  {Fernandes}}, \ and\ \bibinfo {author} {\bibfnamefont {Eduardo}\ \bibnamefont
  {Fradkin}},\ }\bibfield  {title} {\enquote {\bibinfo {title} {Superconducting
  states and intertwined orders in metallic altermagnets},}\ }\href
  {https://arxiv.org/abs/2603.04503} {\  (\bibinfo {year} {2026})},\ \Eprint
  {http://arxiv.org/abs/2603.04503} {arXiv:2603.04503 [cond-mat.supr-con]}
  \BibitemShut {NoStop}%
\bibitem [{\citenamefont {Jasiewicz}\ \emph {et~al.}(2025)\citenamefont
  {Jasiewicz}, \citenamefont {Wójcik}, \citenamefont {Nowak},\ and\
  \citenamefont {Zegrodnik}}]{zegrodnik_npjquantmat2025}%
  \BibitemOpen
  \bibfield  {author} {\bibinfo {author} {\bibfnamefont {Kinga}\ \bibnamefont
  {Jasiewicz}}, \bibinfo {author} {\bibfnamefont {Paweł}\ \bibnamefont
  {Wójcik}}, \bibinfo {author} {\bibfnamefont {Michał~P.}\ \bibnamefont
  {Nowak}}, \ and\ \bibinfo {author} {\bibfnamefont {Michał}\ \bibnamefont
  {Zegrodnik}},\ }\bibfield  {title} {\enquote {\bibinfo {title} {Interplay
  between altermagnetism and superconductivity in two dimensions: intertwined
  symmetries and singlet-triplet mixing},}\ }\href {\doibase
  10.1038/s41535-025-00840-w} {\bibfield  {journal} {\bibinfo  {journal} {npj
  Quantum Materials}\ } (\bibinfo {year} {2025}),\
  10.1038/s41535-025-00840-w}\BibitemShut {NoStop}%
\bibitem [{\citenamefont {Fulde}\ and\ \citenamefont {Ferrell}(1964)}]{ff}%
  \BibitemOpen
  \bibfield  {author} {\bibinfo {author} {\bibfnamefont {Peter}\ \bibnamefont
  {Fulde}}\ and\ \bibinfo {author} {\bibfnamefont {Richard~A.}\ \bibnamefont
  {Ferrell}},\ }\bibfield  {title} {\enquote {\bibinfo {title}
  {Superconductivity in a strong spin-exchange field},}\ }\href {\doibase
  10.1103/PhysRev.135.A550} {\bibfield  {journal} {\bibinfo  {journal} {Phys.
  Rev.}\ }\textbf {\bibinfo {volume} {135}},\ \bibinfo {pages} {A550--A563}
  (\bibinfo {year} {1964})}\BibitemShut {NoStop}%
\bibitem [{\citenamefont {Larkin}\ and\ \citenamefont
  {Ovchinnikov}(1964)}]{lo}%
  \BibitemOpen
  \bibfield  {author} {\bibinfo {author} {\bibfnamefont {A~I}\ \bibnamefont
  {Larkin}}\ and\ \bibinfo {author} {\bibfnamefont {Yu~N}\ \bibnamefont
  {Ovchinnikov}},\ }\bibfield  {title} {\enquote {\bibinfo {title} {Nonuniform
  state of superconductors},}\ }\href@noop {} {\bibfield  {journal} {\bibinfo
  {journal} {Zh. Eksp. Teor. Fiz.}\ }\textbf {\bibinfo {volume} {47}},\
  \bibinfo {pages} {1136} (\bibinfo {year} {1964})}\BibitemShut {NoStop}%
\bibitem [{\citenamefont {Karmakar}\ and\ \citenamefont
  {Majumdar}(2016)}]{karmakar_pra2016}%
  \BibitemOpen
  \bibfield  {author} {\bibinfo {author} {\bibfnamefont {Madhuparna}\
  \bibnamefont {Karmakar}}\ and\ \bibinfo {author} {\bibfnamefont {Pinaki}\
  \bibnamefont {Majumdar}},\ }\bibfield  {title} {\enquote {\bibinfo {title}
  {Population-imbalanced lattice fermions near the bcs-bec crossover: Thermal
  physics of the breached pair and fulde-ferrell-larkin-ovchinnikov phases},}\
  }\href {\doibase 10.1103/PhysRevA.93.053609} {\bibfield  {journal} {\bibinfo
  {journal} {Phys. Rev. A}\ }\textbf {\bibinfo {volume} {93}},\ \bibinfo
  {pages} {053609} (\bibinfo {year} {2016})}\BibitemShut {NoStop}%
\bibitem [{\citenamefont {Karmakar}(2018)}]{karmakar_pra2018}%
  \BibitemOpen
  \bibfield  {author} {\bibinfo {author} {\bibfnamefont {Madhuparna}\
  \bibnamefont {Karmakar}},\ }\bibfield  {title} {\enquote {\bibinfo {title}
  {Thermal transitions, pseudogap behavior, and bcs-bec crossover in
  fermi-fermi mixtures},}\ }\href {\doibase 10.1103/PhysRevA.97.033617}
  {\bibfield  {journal} {\bibinfo  {journal} {Phys. Rev. A}\ }\textbf {\bibinfo
  {volume} {97}},\ \bibinfo {pages} {033617} (\bibinfo {year}
  {2018})}\BibitemShut {NoStop}%
\bibitem [{\citenamefont {Koponen}\ \emph {et~al.}(2007)\citenamefont
  {Koponen}, \citenamefont {Paananen}, \citenamefont {Martikainen},\ and\
  \citenamefont {T\"orm\"a}}]{torma_prl2007}%
  \BibitemOpen
  \bibfield  {author} {\bibinfo {author} {\bibfnamefont {T.~K.}\ \bibnamefont
  {Koponen}}, \bibinfo {author} {\bibfnamefont {T.}~\bibnamefont {Paananen}},
  \bibinfo {author} {\bibfnamefont {J.-P.}\ \bibnamefont {Martikainen}}, \ and\
  \bibinfo {author} {\bibfnamefont {P.}~\bibnamefont {T\"orm\"a}},\ }\bibfield
  {title} {\enquote {\bibinfo {title} {Finite-temperature phase diagram of a
  polarized fermi gas in an optical lattice},}\ }\href {\doibase
  10.1103/PhysRevLett.99.120403} {\bibfield  {journal} {\bibinfo  {journal}
  {Phys. Rev. Lett.}\ }\textbf {\bibinfo {volume} {99}},\ \bibinfo {pages}
  {120403} (\bibinfo {year} {2007})}\BibitemShut {NoStop}%
\bibitem [{\citenamefont {De~Gennes}\ and\ \citenamefont
  {Sarma}(1963)}]{sarma_jap1963}%
  \BibitemOpen
  \bibfield  {author} {\bibinfo {author} {\bibfnamefont {P.~G.}\ \bibnamefont
  {De~Gennes}}\ and\ \bibinfo {author} {\bibfnamefont {G.}~\bibnamefont
  {Sarma}},\ }\bibfield  {title} {\enquote {\bibinfo {title} {{Some Relations
  Between Superconducting and Magnetic Properties}},}\ }\href {\doibase
  10.1063/1.1729520} {\bibfield  {journal} {\bibinfo  {journal} {Journal of
  Applied Physics}\ }\textbf {\bibinfo {volume} {34}},\ \bibinfo {pages}
  {1380--1385} (\bibinfo {year} {1963})}\BibitemShut {NoStop}%
\bibitem [{\citenamefont {Liu}\ \emph {et~al.}(2004)\citenamefont {Liu},
  \citenamefont {Wilczek},\ and\ \citenamefont {Zoller}}]{zoller_pra2004}%
  \BibitemOpen
  \bibfield  {author} {\bibinfo {author} {\bibfnamefont {W.~Vincent}\
  \bibnamefont {Liu}}, \bibinfo {author} {\bibfnamefont {Frank}\ \bibnamefont
  {Wilczek}}, \ and\ \bibinfo {author} {\bibfnamefont {Peter}\ \bibnamefont
  {Zoller}},\ }\bibfield  {title} {\enquote {\bibinfo {title} {Spin-dependent
  hubbard model and a quantum phase transition in cold atoms},}\ }\href
  {https://link.aps.org/doi/10.1103/PhysRevA.70.033603} {\bibfield  {journal}
  {\bibinfo  {journal} {Phys. Rev. A}\ }\textbf {\bibinfo {volume} {70}},\
  \bibinfo {pages} {033603} (\bibinfo {year} {2004})}\BibitemShut {NoStop}%
\bibitem [{\citenamefont {Vorontsov}\ \emph {et~al.}(2005)\citenamefont
  {Vorontsov}, \citenamefont {Sauls},\ and\ \citenamefont
  {Graf}}]{graf_prb2005}%
  \BibitemOpen
  \bibfield  {author} {\bibinfo {author} {\bibfnamefont {A.~B.}\ \bibnamefont
  {Vorontsov}}, \bibinfo {author} {\bibfnamefont {J.~A.}\ \bibnamefont
  {Sauls}}, \ and\ \bibinfo {author} {\bibfnamefont {M.~J.}\ \bibnamefont
  {Graf}},\ }\bibfield  {title} {\enquote {\bibinfo {title} {Phase diagram and
  spectroscopy of fulde-ferrell-larkin-ovchinnikov states of two-dimensional
  $d$-wave superconductors},}\ }\href {\doibase 10.1103/PhysRevB.72.184501}
  {\bibfield  {journal} {\bibinfo  {journal} {Phys. Rev. B}\ }\textbf {\bibinfo
  {volume} {72}},\ \bibinfo {pages} {184501} (\bibinfo {year}
  {2005})}\BibitemShut {NoStop}%
\bibitem [{\citenamefont {Vorontsov}\ and\ \citenamefont
  {Graf}(2006)}]{graf_prb2006}%
  \BibitemOpen
  \bibfield  {author} {\bibinfo {author} {\bibfnamefont {Anton~B.}\
  \bibnamefont {Vorontsov}}\ and\ \bibinfo {author} {\bibfnamefont
  {Matthias~J.}\ \bibnamefont {Graf}},\ }\bibfield  {title} {\enquote {\bibinfo
  {title} {Fermi-liquid effects in the fulde-ferrell-larkin-ovchinnikov state
  of two-dimensional $d$-wave superconductors},}\ }\href {\doibase
  10.1103/PhysRevB.74.172504} {\bibfield  {journal} {\bibinfo  {journal} {Phys.
  Rev. B}\ }\textbf {\bibinfo {volume} {74}},\ \bibinfo {pages} {172504}
  (\bibinfo {year} {2006})}\BibitemShut {NoStop}%
\bibitem [{\citenamefont {Kaczmarczyk}\ \emph {et~al.}(2011)\citenamefont
  {Kaczmarczyk}, \citenamefont {Sadzikowski},\ and\ \citenamefont
  {Spa\l{}ek}}]{spalek_prb2011}%
  \BibitemOpen
  \bibfield  {author} {\bibinfo {author} {\bibfnamefont {Jan}\ \bibnamefont
  {Kaczmarczyk}}, \bibinfo {author} {\bibfnamefont {Mariusz}\ \bibnamefont
  {Sadzikowski}}, \ and\ \bibinfo {author} {\bibfnamefont {Jozef}\ \bibnamefont
  {Spa\l{}ek}},\ }\bibfield  {title} {\enquote {\bibinfo {title} {Conductance
  spectroscopy of a correlated superconductor in a magnetic field in the pauli
  limit: Evidence for strong correlations},}\ }\href {\doibase
  10.1103/PhysRevB.84.094525} {\bibfield  {journal} {\bibinfo  {journal} {Phys.
  Rev. B}\ }\textbf {\bibinfo {volume} {84}},\ \bibinfo {pages} {094525}
  (\bibinfo {year} {2011})}\BibitemShut {NoStop}%
\bibitem [{\citenamefont {Beaird}\ \emph {et~al.}(2010)\citenamefont {Beaird},
  \citenamefont {Vorontsov},\ and\ \citenamefont {Vekhter}}]{vekhter_prb2010}%
  \BibitemOpen
  \bibfield  {author} {\bibinfo {author} {\bibfnamefont {Robert}\ \bibnamefont
  {Beaird}}, \bibinfo {author} {\bibfnamefont {Anton~B.}\ \bibnamefont
  {Vorontsov}}, \ and\ \bibinfo {author} {\bibfnamefont {Ilya}\ \bibnamefont
  {Vekhter}},\ }\bibfield  {title} {\enquote {\bibinfo {title} {Pauli-limited
  superconductivity with classical magnetic fluctuations},}\ }\href {\doibase
  10.1103/PhysRevB.81.224501} {\bibfield  {journal} {\bibinfo  {journal} {Phys.
  Rev. B}\ }\textbf {\bibinfo {volume} {81}},\ \bibinfo {pages} {224501}
  (\bibinfo {year} {2010})}\BibitemShut {NoStop}%
\bibitem [{\citenamefont {Yanase}(2008)}]{yanase_jpsj2008}%
  \BibitemOpen
  \bibfield  {author} {\bibinfo {author} {\bibfnamefont {Youichi}\ \bibnamefont
  {Yanase}},\ }\bibfield  {title} {\enquote {\bibinfo {title} {Fflo
  superconductivity near the antiferromagnetic quantum critical point},}\
  }\href {\doibase 10.1143/JPSJ.77.063705} {\bibfield  {journal} {\bibinfo
  {journal} {Journal of the Physical Society of Japan}\ }\textbf {\bibinfo
  {volume} {77}},\ \bibinfo {pages} {063705} (\bibinfo {year}
  {2008})}\BibitemShut {NoStop}%
\bibitem [{\citenamefont {Zhou}\ and\ \citenamefont
  {Ting}(2009)}]{ting_prb2009}%
  \BibitemOpen
  \bibfield  {author} {\bibinfo {author} {\bibfnamefont {Tao}\ \bibnamefont
  {Zhou}}\ and\ \bibinfo {author} {\bibfnamefont {C.~S.}\ \bibnamefont
  {Ting}},\ }\bibfield  {title} {\enquote {\bibinfo {title} {Phase diagram and
  local tunneling spectroscopy of the fulde-ferrell-larkin-ovchinnikov states
  of a two-dimensional square-lattice $d$-wave superconductor},}\ }\href
  {\doibase 10.1103/PhysRevB.80.224515} {\bibfield  {journal} {\bibinfo
  {journal} {Phys. Rev. B}\ }\textbf {\bibinfo {volume} {80}},\ \bibinfo
  {pages} {224515} (\bibinfo {year} {2009})}\BibitemShut {NoStop}%
\bibitem [{\citenamefont {Cui}\ \emph {et~al.}(2012)\citenamefont {Cui},
  \citenamefont {Hu}, \citenamefont {Wei},\ and\ \citenamefont
  {Yang}}]{yang_prb2012}%
  \BibitemOpen
  \bibfield  {author} {\bibinfo {author} {\bibfnamefont {Qinghong}\
  \bibnamefont {Cui}}, \bibinfo {author} {\bibfnamefont {C.-R.}\ \bibnamefont
  {Hu}}, \bibinfo {author} {\bibfnamefont {J.~Y.~T.}\ \bibnamefont {Wei}}, \
  and\ \bibinfo {author} {\bibfnamefont {Kun}\ \bibnamefont {Yang}},\
  }\bibfield  {title} {\enquote {\bibinfo {title} {Spectroscopic signatures of
  the larkin-ovchinnikov state in the conductance characteristics of a
  normal-metal/superconductor junction},}\ }\href {\doibase
  10.1103/PhysRevB.85.014503} {\bibfield  {journal} {\bibinfo  {journal} {Phys.
  Rev. B}\ }\textbf {\bibinfo {volume} {85}},\ \bibinfo {pages} {014503}
  (\bibinfo {year} {2012})}\BibitemShut {NoStop}%
\bibitem [{\citenamefont {Karmakar}(2020)}]{karmakar_jpcm2020}%
  \BibitemOpen
  \bibfield  {author} {\bibinfo {author} {\bibfnamefont {Madhuparna}\
  \bibnamefont {Karmakar}},\ }\bibfield  {title} {\enquote {\bibinfo {title}
  {Pauli limited d-wave superconductors: quantum breached pair phase and
  thermal transitions},}\ }\href {\doibase 10.1088/1361-648X/ab926a} {\bibfield
   {journal} {\bibinfo  {journal} {Journal of Physics: Condensed Matter}\
  }\textbf {\bibinfo {volume} {32}},\ \bibinfo {pages} {405604} (\bibinfo
  {year} {2020})}\BibitemShut {NoStop}%
\bibitem [{\citenamefont {Karmakar}(2024)}]{karmakar_jpcm2024}%
  \BibitemOpen
  \bibfield  {author} {\bibinfo {author} {\bibfnamefont {Madhuparna}\
  \bibnamefont {Karmakar}},\ }\bibfield  {title} {\enquote {\bibinfo {title}
  {Magnetotransport and fermi surface segmentation in pauli limited
  superconductors},}\ }\href {\doibase 10.1088/1361-648X/ad1bf6} {\bibfield
  {journal} {\bibinfo  {journal} {Journal of Physics: Condensed Matter}\
  }\textbf {\bibinfo {volume} {36}},\ \bibinfo {pages} {165601} (\bibinfo
  {year} {2024})}\BibitemShut {NoStop}%
\bibitem [{\citenamefont {Shin}\ \emph {et~al.}(2008)\citenamefont {Shin},
  \citenamefont {Schunck}, \citenamefont {Schirotzek},\ and\ \citenamefont
  {Ketterle}}]{ketterle_nature2008}%
  \BibitemOpen
  \bibfield  {author} {\bibinfo {author} {\bibfnamefont {Yong-il}\ \bibnamefont
  {Shin}}, \bibinfo {author} {\bibfnamefont {Christian~H.}\ \bibnamefont
  {Schunck}}, \bibinfo {author} {\bibfnamefont {André}\ \bibnamefont
  {Schirotzek}}, \ and\ \bibinfo {author} {\bibfnamefont {Wolfgang}\
  \bibnamefont {Ketterle}},\ }\bibfield  {title} {\enquote {\bibinfo {title}
  {Phase diagram of a two-component fermi gas with resonant interactions},}\
  }\href {\doibase 10.1038/nature06473} {\bibfield  {journal} {\bibinfo
  {journal} {Nature}\ }\textbf {\bibinfo {volume} {451}},\ \bibinfo {pages}
  {689} (\bibinfo {year} {2008})}\BibitemShut {NoStop}%
\bibitem [{\citenamefont {Schunck}\ \emph {et~al.}(2007)\citenamefont
  {Schunck}, \citenamefont {Shin}, \citenamefont {Schirotzek}, \citenamefont
  {Zwierlein},\ and\ \citenamefont {Ketterle}}]{ketterle_science2007}%
  \BibitemOpen
  \bibfield  {author} {\bibinfo {author} {\bibfnamefont {C.~H.}\ \bibnamefont
  {Schunck}}, \bibinfo {author} {\bibfnamefont {Y.}~\bibnamefont {Shin}},
  \bibinfo {author} {\bibfnamefont {A.}~\bibnamefont {Schirotzek}}, \bibinfo
  {author} {\bibfnamefont {M.~W.}\ \bibnamefont {Zwierlein}}, \ and\ \bibinfo
  {author} {\bibfnamefont {W.}~\bibnamefont {Ketterle}},\ }\bibfield  {title}
  {\enquote {\bibinfo {title} {Pairing without superfluidity: The ground state
  of an imbalanced fermi mixture},}\ }\href {\doibase 10.1126/science.1140749}
  {\bibfield  {journal} {\bibinfo  {journal} {Science}\ }\textbf {\bibinfo
  {volume} {316}},\ \bibinfo {pages} {867--870} (\bibinfo {year}
  {2007})}\BibitemShut {NoStop}%
\bibitem [{\citenamefont {Shin}\ \emph {et~al.}(2006)\citenamefont {Shin},
  \citenamefont {Zwierlein}, \citenamefont {Schunck}, \citenamefont
  {Schirotzek},\ and\ \citenamefont {Ketterle}}]{ketterle_prl2006}%
  \BibitemOpen
  \bibfield  {author} {\bibinfo {author} {\bibfnamefont {Y.}~\bibnamefont
  {Shin}}, \bibinfo {author} {\bibfnamefont {M.~W.}\ \bibnamefont {Zwierlein}},
  \bibinfo {author} {\bibfnamefont {C.~H.}\ \bibnamefont {Schunck}}, \bibinfo
  {author} {\bibfnamefont {A.}~\bibnamefont {Schirotzek}}, \ and\ \bibinfo
  {author} {\bibfnamefont {W.}~\bibnamefont {Ketterle}},\ }\bibfield  {title}
  {\enquote {\bibinfo {title} {Observation of phase separation in a strongly
  interacting imbalanced fermi gas},}\ }\href {\doibase
  10.1103/PhysRevLett.97.030401} {\bibfield  {journal} {\bibinfo  {journal}
  {Phys. Rev. Lett.}\ }\textbf {\bibinfo {volume} {97}},\ \bibinfo {pages}
  {030401} (\bibinfo {year} {2006})}\BibitemShut {NoStop}%
\bibitem [{\citenamefont {Liao}\ \emph {et~al.}(2010)\citenamefont {Liao},
  \citenamefont {Rittner}, \citenamefont {Paprotta}, \citenamefont {Li},
  \citenamefont {Partridge}, \citenamefont {Hulet}, \citenamefont {Baur},\ and\
  \citenamefont {Mueller}}]{mueller_nature2010}%
  \BibitemOpen
  \bibfield  {author} {\bibinfo {author} {\bibfnamefont {Yean-an}\ \bibnamefont
  {Liao}}, \bibinfo {author} {\bibfnamefont {Ann Sophie~C.}\ \bibnamefont
  {Rittner}}, \bibinfo {author} {\bibfnamefont {Tobias}\ \bibnamefont
  {Paprotta}}, \bibinfo {author} {\bibfnamefont {Wenhui}\ \bibnamefont {Li}},
  \bibinfo {author} {\bibfnamefont {Guthrie~B.}\ \bibnamefont {Partridge}},
  \bibinfo {author} {\bibfnamefont {Randall~G.}\ \bibnamefont {Hulet}},
  \bibinfo {author} {\bibfnamefont {Stefan~K.}\ \bibnamefont {Baur}}, \ and\
  \bibinfo {author} {\bibfnamefont {Erich~J.}\ \bibnamefont {Mueller}},\
  }\bibfield  {title} {\enquote {\bibinfo {title} {Spin-imbalance in a
  one-dimensional fermi gas},}\ }\href {\doibase 10.1038/nature09393}
  {\bibfield  {journal} {\bibinfo  {journal} {Nature}\ }\textbf {\bibinfo
  {volume} {467}},\ \bibinfo {pages} {567} (\bibinfo {year}
  {2010})}\BibitemShut {NoStop}%
\bibitem [{\citenamefont {Bianchi}\ \emph {et~al.}(2003)\citenamefont
  {Bianchi}, \citenamefont {Movshovich}, \citenamefont {Capan}, \citenamefont
  {Pagliuso},\ and\ \citenamefont {Sarrao}}]{sarro_prl2003}%
  \BibitemOpen
  \bibfield  {author} {\bibinfo {author} {\bibfnamefont {A.}~\bibnamefont
  {Bianchi}}, \bibinfo {author} {\bibfnamefont {R.}~\bibnamefont {Movshovich}},
  \bibinfo {author} {\bibfnamefont {C.}~\bibnamefont {Capan}}, \bibinfo
  {author} {\bibfnamefont {P.~G.}\ \bibnamefont {Pagliuso}}, \ and\ \bibinfo
  {author} {\bibfnamefont {J.~L.}\ \bibnamefont {Sarrao}},\ }\bibfield  {title}
  {\enquote {\bibinfo {title} {Possible fulde-ferrell-larkin-ovchinnikov
  superconducting state in
  ${\mathrm{c}\mathrm{e}\mathrm{c}\mathrm{o}\mathrm{i}\mathrm{n}}_{5}$},}\
  }\href {\doibase 10.1103/PhysRevLett.91.187004} {\bibfield  {journal}
  {\bibinfo  {journal} {Phys. Rev. Lett.}\ }\textbf {\bibinfo {volume} {91}},\
  \bibinfo {pages} {187004} (\bibinfo {year} {2003})}\BibitemShut {NoStop}%
\bibitem [{\citenamefont {Tayama}\ \emph {et~al.}(2002)\citenamefont {Tayama},
  \citenamefont {Harita}, \citenamefont {Sakakibara}, \citenamefont {Haga},
  \citenamefont {Shishido}, \citenamefont {Settai},\ and\ \citenamefont
  {Onuki}}]{onuki_prb2002}%
  \BibitemOpen
  \bibfield  {author} {\bibinfo {author} {\bibfnamefont {T.}~\bibnamefont
  {Tayama}}, \bibinfo {author} {\bibfnamefont {A.}~\bibnamefont {Harita}},
  \bibinfo {author} {\bibfnamefont {T.}~\bibnamefont {Sakakibara}}, \bibinfo
  {author} {\bibfnamefont {Y.}~\bibnamefont {Haga}}, \bibinfo {author}
  {\bibfnamefont {H.}~\bibnamefont {Shishido}}, \bibinfo {author}
  {\bibfnamefont {R.}~\bibnamefont {Settai}}, \ and\ \bibinfo {author}
  {\bibfnamefont {Y.}~\bibnamefont {Onuki}},\ }\bibfield  {title} {\enquote
  {\bibinfo {title} {Unconventional heavy-fermion superconductor
  ${\mathrm{cecoin}}_{5}:$ dc magnetization study at temperatures down to 50
  mk},}\ }\href {\doibase 10.1103/PhysRevB.65.180504} {\bibfield  {journal}
  {\bibinfo  {journal} {Phys. Rev. B}\ }\textbf {\bibinfo {volume} {65}},\
  \bibinfo {pages} {180504} (\bibinfo {year} {2002})}\BibitemShut {NoStop}%
\bibitem [{\citenamefont {Koutroulakis}\ \emph {et~al.}(2010)\citenamefont
  {Koutroulakis}, \citenamefont {Stewart}, \citenamefont
  {Mitrovi\ifmmode~\acute{c}\else \'{c}\fi{}}, \citenamefont
  {Horvati\ifmmode~\acute{c}\else \'{c}\fi{}}, \citenamefont {Berthier},
  \citenamefont {Lapertot},\ and\ \citenamefont {Flouquet}}]{flouquet_prl2008}%
  \BibitemOpen
  \bibfield  {author} {\bibinfo {author} {\bibfnamefont {G.}~\bibnamefont
  {Koutroulakis}}, \bibinfo {author} {\bibfnamefont {M.~D.}\ \bibnamefont
  {Stewart}}, \bibinfo {author} {\bibfnamefont {V.~F.}\ \bibnamefont
  {Mitrovi\ifmmode~\acute{c}\else \'{c}\fi{}}}, \bibinfo {author}
  {\bibfnamefont {M.}~\bibnamefont {Horvati\ifmmode~\acute{c}\else
  \'{c}\fi{}}}, \bibinfo {author} {\bibfnamefont {C.}~\bibnamefont {Berthier}},
  \bibinfo {author} {\bibfnamefont {G.}~\bibnamefont {Lapertot}}, \ and\
  \bibinfo {author} {\bibfnamefont {J.}~\bibnamefont {Flouquet}},\ }\bibfield
  {title} {\enquote {\bibinfo {title} {Field evolution of coexisting
  superconducting and magnetic orders in ${\mathrm{cecoin}}_{5}$},}\ }\href
  {\doibase 10.1103/PhysRevLett.104.087001} {\bibfield  {journal} {\bibinfo
  {journal} {Phys. Rev. Lett.}\ }\textbf {\bibinfo {volume} {104}},\ \bibinfo
  {pages} {087001} (\bibinfo {year} {2010})}\BibitemShut {NoStop}%
\bibitem [{\citenamefont {Kenzelmann}\ \emph {et~al.}(2008)\citenamefont
  {Kenzelmann}, \citenamefont {Strässle}, \citenamefont {Niedermayer},
  \citenamefont {Sigrist}, \citenamefont {Padmanabhan}, \citenamefont
  {Zolliker}, \citenamefont {Bianchi}, \citenamefont {Movshovich},
  \citenamefont {Bauer}, \citenamefont {Sarrao},\ and\ \citenamefont
  {Thompson}}]{kenzelman_science2008}%
  \BibitemOpen
  \bibfield  {author} {\bibinfo {author} {\bibfnamefont {M.}~\bibnamefont
  {Kenzelmann}}, \bibinfo {author} {\bibfnamefont {Th.}\ \bibnamefont
  {Strässle}}, \bibinfo {author} {\bibfnamefont {C.}~\bibnamefont
  {Niedermayer}}, \bibinfo {author} {\bibfnamefont {M.}~\bibnamefont
  {Sigrist}}, \bibinfo {author} {\bibfnamefont {B.}~\bibnamefont
  {Padmanabhan}}, \bibinfo {author} {\bibfnamefont {M.}~\bibnamefont
  {Zolliker}}, \bibinfo {author} {\bibfnamefont {A.~D.}\ \bibnamefont
  {Bianchi}}, \bibinfo {author} {\bibfnamefont {R.}~\bibnamefont {Movshovich}},
  \bibinfo {author} {\bibfnamefont {E.~D.}\ \bibnamefont {Bauer}}, \bibinfo
  {author} {\bibfnamefont {J.~L.}\ \bibnamefont {Sarrao}}, \ and\ \bibinfo
  {author} {\bibfnamefont {J.~D.}\ \bibnamefont {Thompson}},\ }\bibfield
  {title} {\enquote {\bibinfo {title} {Coupled superconducting and magnetic
  order in cecoin$_{5}$},}\ }\href {\doibase 10.1126/science.1161818}
  {\bibfield  {journal} {\bibinfo  {journal} {Science}\ }\textbf {\bibinfo
  {volume} {321}},\ \bibinfo {pages} {1652} (\bibinfo {year}
  {2008})}\BibitemShut {NoStop}%
\bibitem [{\citenamefont {Capan}\ \emph {et~al.}(2004)\citenamefont {Capan},
  \citenamefont {Bianchi}, \citenamefont {Movshovich}, \citenamefont
  {Christianson}, \citenamefont {Malinowski}, \citenamefont {Hundley},
  \citenamefont {Lacerda}, \citenamefont {Pagliuso},\ and\ \citenamefont
  {Sarrao}}]{sarro_prb2004}%
  \BibitemOpen
  \bibfield  {author} {\bibinfo {author} {\bibfnamefont {C.}~\bibnamefont
  {Capan}}, \bibinfo {author} {\bibfnamefont {A.}~\bibnamefont {Bianchi}},
  \bibinfo {author} {\bibfnamefont {R.}~\bibnamefont {Movshovich}}, \bibinfo
  {author} {\bibfnamefont {A.~D.}\ \bibnamefont {Christianson}}, \bibinfo
  {author} {\bibfnamefont {A.}~\bibnamefont {Malinowski}}, \bibinfo {author}
  {\bibfnamefont {M.~F.}\ \bibnamefont {Hundley}}, \bibinfo {author}
  {\bibfnamefont {A.}~\bibnamefont {Lacerda}}, \bibinfo {author} {\bibfnamefont
  {P.~G.}\ \bibnamefont {Pagliuso}}, \ and\ \bibinfo {author} {\bibfnamefont
  {J.~L.}\ \bibnamefont {Sarrao}},\ }\bibfield  {title} {\enquote {\bibinfo
  {title} {Anisotropy of thermal conductivity and possible signature of the
  fulde-ferrell-larkin-ovchinnikov state in
  $\mathrm{Ce}\mathrm{Co}{\mathrm{in}}_{5}$},}\ }\href {\doibase
  10.1103/PhysRevB.70.134513} {\bibfield  {journal} {\bibinfo  {journal} {Phys.
  Rev. B}\ }\textbf {\bibinfo {volume} {70}},\ \bibinfo {pages} {134513}
  (\bibinfo {year} {2004})}\BibitemShut {NoStop}%
\bibitem [{\citenamefont {Martin}\ \emph {et~al.}(2005)\citenamefont {Martin},
  \citenamefont {Agosta}, \citenamefont {Tozer}, \citenamefont {Radovan},
  \citenamefont {Palm}, \citenamefont {Murphy},\ and\ \citenamefont
  {Sarrao}}]{sarro_prb2005}%
  \BibitemOpen
  \bibfield  {author} {\bibinfo {author} {\bibfnamefont {C.}~\bibnamefont
  {Martin}}, \bibinfo {author} {\bibfnamefont {C.~C.}\ \bibnamefont {Agosta}},
  \bibinfo {author} {\bibfnamefont {S.~W.}\ \bibnamefont {Tozer}}, \bibinfo
  {author} {\bibfnamefont {H.~A.}\ \bibnamefont {Radovan}}, \bibinfo {author}
  {\bibfnamefont {E.~C.}\ \bibnamefont {Palm}}, \bibinfo {author}
  {\bibfnamefont {T.~P.}\ \bibnamefont {Murphy}}, \ and\ \bibinfo {author}
  {\bibfnamefont {J.~L.}\ \bibnamefont {Sarrao}},\ }\bibfield  {title}
  {\enquote {\bibinfo {title} {Evidence for the
  fulde-ferrell-larkin-ovchinnikov state in ${\mathrm{cecoin}}_{5}$ from
  penetration depth measurements},}\ }\href {\doibase
  10.1103/PhysRevB.71.020503} {\bibfield  {journal} {\bibinfo  {journal} {Phys.
  Rev. B}\ }\textbf {\bibinfo {volume} {71}},\ \bibinfo {pages} {020503}
  (\bibinfo {year} {2005})}\BibitemShut {NoStop}%
\bibitem [{\citenamefont {Gerber}\ \emph {et~al.}(2014)\citenamefont {Gerber},
  \citenamefont {Bartkowiak}, \citenamefont {Gavilano}, \citenamefont
  {Ressouche}, \citenamefont {Egetenmeyer}, \citenamefont {Niedermayer},
  \citenamefont {Bianchi}, \citenamefont {Movshovich}, \citenamefont {Bauer},
  \citenamefont {Thompson},\ and\ \citenamefont
  {Kenzelmann}}]{gerber_natphys2013}%
  \BibitemOpen
  \bibfield  {author} {\bibinfo {author} {\bibfnamefont {Simon}\ \bibnamefont
  {Gerber}}, \bibinfo {author} {\bibfnamefont {Marek}\ \bibnamefont
  {Bartkowiak}}, \bibinfo {author} {\bibfnamefont {Jorge~L.}\ \bibnamefont
  {Gavilano}}, \bibinfo {author} {\bibfnamefont {Eric}\ \bibnamefont
  {Ressouche}}, \bibinfo {author} {\bibfnamefont {Nikola}\ \bibnamefont
  {Egetenmeyer}}, \bibinfo {author} {\bibfnamefont {Christof}\ \bibnamefont
  {Niedermayer}}, \bibinfo {author} {\bibfnamefont {Andrea~D.}\ \bibnamefont
  {Bianchi}}, \bibinfo {author} {\bibfnamefont {Roman}\ \bibnamefont
  {Movshovich}}, \bibinfo {author} {\bibfnamefont {Eric~D.}\ \bibnamefont
  {Bauer}}, \bibinfo {author} {\bibfnamefont {Joe~D.}\ \bibnamefont
  {Thompson}}, \ and\ \bibinfo {author} {\bibfnamefont {Michel}\ \bibnamefont
  {Kenzelmann}},\ }\bibfield  {title} {\enquote {\bibinfo {title} {Switching of
  magnetic domains reveals spatially inhomogeneous superconductivity},}\ }\href
  {\doibase 10.1038/nphys2833} {\bibfield  {journal} {\bibinfo  {journal}
  {Nature Physics}\ }\textbf {\bibinfo {volume} {10}},\ \bibinfo {pages} {126}
  (\bibinfo {year} {2014})}\BibitemShut {NoStop}%
\bibitem [{\citenamefont {Kumagai}\ \emph {et~al.}(2006)\citenamefont
  {Kumagai}, \citenamefont {Saitoh}, \citenamefont {Oyaizu}, \citenamefont
  {Furukawa}, \citenamefont {Takashima}, \citenamefont {Nohara}, \citenamefont
  {Takagi},\ and\ \citenamefont {Matsuda}}]{matsuda_prl2006}%
  \BibitemOpen
  \bibfield  {author} {\bibinfo {author} {\bibfnamefont {K.}~\bibnamefont
  {Kumagai}}, \bibinfo {author} {\bibfnamefont {M.}~\bibnamefont {Saitoh}},
  \bibinfo {author} {\bibfnamefont {T.}~\bibnamefont {Oyaizu}}, \bibinfo
  {author} {\bibfnamefont {Y.}~\bibnamefont {Furukawa}}, \bibinfo {author}
  {\bibfnamefont {S.}~\bibnamefont {Takashima}}, \bibinfo {author}
  {\bibfnamefont {M.}~\bibnamefont {Nohara}}, \bibinfo {author} {\bibfnamefont
  {H.}~\bibnamefont {Takagi}}, \ and\ \bibinfo {author} {\bibfnamefont
  {Y.}~\bibnamefont {Matsuda}},\ }\bibfield  {title} {\enquote {\bibinfo
  {title} {Fulde-ferrell-larkin-ovchinnikov state in a perpendicular field of
  quasi-two-dimensional ${\mathrm{cecoin}}_{5}$},}\ }\href {\doibase
  10.1103/PhysRevLett.97.227002} {\bibfield  {journal} {\bibinfo  {journal}
  {Phys. Rev. Lett.}\ }\textbf {\bibinfo {volume} {97}},\ \bibinfo {pages}
  {227002} (\bibinfo {year} {2006})}\BibitemShut {NoStop}%
\bibitem [{\citenamefont {Kim}\ \emph {et~al.}(2016)\citenamefont {Kim},
  \citenamefont {Lin}, \citenamefont {Weickert}, \citenamefont {Kenzelmann},
  \citenamefont {Bauer}, \citenamefont {Ronning}, \citenamefont {Thompson},\
  and\ \citenamefont {Movshovich}}]{movshovich_prx2016}%
  \BibitemOpen
  \bibfield  {author} {\bibinfo {author} {\bibfnamefont {Duk~Y.}\ \bibnamefont
  {Kim}}, \bibinfo {author} {\bibfnamefont {Shi-Zeng}\ \bibnamefont {Lin}},
  \bibinfo {author} {\bibfnamefont {Franziska}\ \bibnamefont {Weickert}},
  \bibinfo {author} {\bibfnamefont {Michel}\ \bibnamefont {Kenzelmann}},
  \bibinfo {author} {\bibfnamefont {Eric~D.}\ \bibnamefont {Bauer}}, \bibinfo
  {author} {\bibfnamefont {Filip}\ \bibnamefont {Ronning}}, \bibinfo {author}
  {\bibfnamefont {J.~D.}\ \bibnamefont {Thompson}}, \ and\ \bibinfo {author}
  {\bibfnamefont {Roman}\ \bibnamefont {Movshovich}},\ }\bibfield  {title}
  {\enquote {\bibinfo {title} {Intertwined orders in heavy-fermion
  superconductor ${\mathrm{cecoin}}_{5}$},}\ }\href {\doibase
  10.1103/PhysRevX.6.041059} {\bibfield  {journal} {\bibinfo  {journal} {Phys.
  Rev. X}\ }\textbf {\bibinfo {volume} {6}},\ \bibinfo {pages} {041059}
  (\bibinfo {year} {2016})}\BibitemShut {NoStop}%
\bibitem [{\citenamefont {Lin}\ \emph {et~al.}(2020)\citenamefont {Lin},
  \citenamefont {Kim}, \citenamefont {Bauer}, \citenamefont {Ronning},
  \citenamefont {Thompson},\ and\ \citenamefont
  {Movshovich}}]{movshovic_prl2020}%
  \BibitemOpen
  \bibfield  {author} {\bibinfo {author} {\bibfnamefont {Shi-Zeng}\
  \bibnamefont {Lin}}, \bibinfo {author} {\bibfnamefont {Duk~Y.}\ \bibnamefont
  {Kim}}, \bibinfo {author} {\bibfnamefont {Eric~D.}\ \bibnamefont {Bauer}},
  \bibinfo {author} {\bibfnamefont {Filip}\ \bibnamefont {Ronning}}, \bibinfo
  {author} {\bibfnamefont {J.~D.}\ \bibnamefont {Thompson}}, \ and\ \bibinfo
  {author} {\bibfnamefont {Roman}\ \bibnamefont {Movshovich}},\ }\bibfield
  {title} {\enquote {\bibinfo {title} {Interplay of the spin density wave and a
  possible fulde-ferrell-larkin-ovchinnikov state in ${\mathrm{cecoin}}_{5}$ in
  rotating magnetic field},}\ }\href {\doibase 10.1103/PhysRevLett.124.217001}
  {\bibfield  {journal} {\bibinfo  {journal} {Phys. Rev. Lett.}\ }\textbf
  {\bibinfo {volume} {124}},\ \bibinfo {pages} {217001} (\bibinfo {year}
  {2020})}\BibitemShut {NoStop}%
\bibitem [{\citenamefont {Beyer}\ and\ \citenamefont
  {Wosnitza}(2013)}]{wosnitza_ltp2013}%
  \BibitemOpen
  \bibfield  {author} {\bibinfo {author} {\bibfnamefont {R.}~\bibnamefont
  {Beyer}}\ and\ \bibinfo {author} {\bibfnamefont {J.}~\bibnamefont
  {Wosnitza}},\ }\bibfield  {title} {\enquote {\bibinfo {title} {Emerging
  evidence for fflo states in layered organic superconductors},}\ }\href
  {\doibase 10.1063/1.4794996} {\bibfield  {journal} {\bibinfo  {journal} {Low
  Temperature Physics}\ }\textbf {\bibinfo {volume} {39}},\ \bibinfo {pages}
  {225--231} (\bibinfo {year} {2013})}\BibitemShut {NoStop}%
\bibitem [{\citenamefont {Lortz}\ \emph {et~al.}(2007)\citenamefont {Lortz},
  \citenamefont {Wang}, \citenamefont {Demuer}, \citenamefont {B\"ottger},
  \citenamefont {Bergk}, \citenamefont {Zwicknagl}, \citenamefont {Nakazawa},\
  and\ \citenamefont {Wosnitza}}]{wosnitza_prl2007}%
  \BibitemOpen
  \bibfield  {author} {\bibinfo {author} {\bibfnamefont {R.}~\bibnamefont
  {Lortz}}, \bibinfo {author} {\bibfnamefont {Y.}~\bibnamefont {Wang}},
  \bibinfo {author} {\bibfnamefont {A.}~\bibnamefont {Demuer}}, \bibinfo
  {author} {\bibfnamefont {P.~H.~M.}\ \bibnamefont {B\"ottger}}, \bibinfo
  {author} {\bibfnamefont {B.}~\bibnamefont {Bergk}}, \bibinfo {author}
  {\bibfnamefont {G.}~\bibnamefont {Zwicknagl}}, \bibinfo {author}
  {\bibfnamefont {Y.}~\bibnamefont {Nakazawa}}, \ and\ \bibinfo {author}
  {\bibfnamefont {J.}~\bibnamefont {Wosnitza}},\ }\bibfield  {title} {\enquote
  {\bibinfo {title} {Calorimetric evidence for a
  fulde-ferrell-larkin-ovchinnikov superconducting state in the layered organic
  superconductor
  $\ensuremath{\kappa}\mathrm{\text{\ensuremath{-}}}(\mathrm{BEDT}\mathrm{\text{\ensuremath{-}}}\mathrm{TTF}{)}_{2}\mathrm{Cu}(\mathrm{NCS}{)}_{2}$},}\
  }\href {\doibase 10.1103/PhysRevLett.99.187002} {\bibfield  {journal}
  {\bibinfo  {journal} {Phys. Rev. Lett.}\ }\textbf {\bibinfo {volume} {99}},\
  \bibinfo {pages} {187002} (\bibinfo {year} {2007})}\BibitemShut {NoStop}%
\bibitem [{\citenamefont {Mayaffre}\ \emph {et~al.}(2014)\citenamefont
  {Mayaffre}, \citenamefont {Krämer}, \citenamefont {Horvatić}, \citenamefont
  {Berthier}, \citenamefont {Miyagawa}, \citenamefont {Kanoda},\ and\
  \citenamefont {Mitrović}}]{mitrovic_natphys2014}%
  \BibitemOpen
  \bibfield  {author} {\bibinfo {author} {\bibfnamefont {H.}~\bibnamefont
  {Mayaffre}}, \bibinfo {author} {\bibfnamefont {S.}~\bibnamefont {Krämer}},
  \bibinfo {author} {\bibfnamefont {M.}~\bibnamefont {Horvatić}}, \bibinfo
  {author} {\bibfnamefont {C.}~\bibnamefont {Berthier}}, \bibinfo {author}
  {\bibfnamefont {K.}~\bibnamefont {Miyagawa}}, \bibinfo {author}
  {\bibfnamefont {K.}~\bibnamefont {Kanoda}}, \ and\ \bibinfo {author}
  {\bibfnamefont {V.~F.}\ \bibnamefont {Mitrović}},\ }\bibfield  {title}
  {\enquote {\bibinfo {title} {Evidence of andreev bound states as a hallmark
  of the fflo phase in $\kappa$-(bedt-ttf)$_{2}$cu(ncs)$_{2}$},}\ }\href
  {\doibase 10.1038/nphys3121} {\bibfield  {journal} {\bibinfo  {journal}
  {Nature Physics}\ }\textbf {\bibinfo {volume} {10}},\ \bibinfo {pages} {928}
  (\bibinfo {year} {2014})}\BibitemShut {NoStop}%
\bibitem [{\citenamefont {Wright}\ \emph {et~al.}(2011)\citenamefont {Wright},
  \citenamefont {Green}, \citenamefont {Kuhns}, \citenamefont {Reyes},
  \citenamefont {Brooks}, \citenamefont {Schlueter}, \citenamefont {Kato},
  \citenamefont {Yamamoto}, \citenamefont {Kobayashi},\ and\ \citenamefont
  {Brown}}]{wright_prl2011}%
  \BibitemOpen
  \bibfield  {author} {\bibinfo {author} {\bibfnamefont {J.~A.}\ \bibnamefont
  {Wright}}, \bibinfo {author} {\bibfnamefont {E.}~\bibnamefont {Green}},
  \bibinfo {author} {\bibfnamefont {P.}~\bibnamefont {Kuhns}}, \bibinfo
  {author} {\bibfnamefont {A.}~\bibnamefont {Reyes}}, \bibinfo {author}
  {\bibfnamefont {J.}~\bibnamefont {Brooks}}, \bibinfo {author} {\bibfnamefont
  {J.}~\bibnamefont {Schlueter}}, \bibinfo {author} {\bibfnamefont
  {R.}~\bibnamefont {Kato}}, \bibinfo {author} {\bibfnamefont {H.}~\bibnamefont
  {Yamamoto}}, \bibinfo {author} {\bibfnamefont {M.}~\bibnamefont {Kobayashi}},
  \ and\ \bibinfo {author} {\bibfnamefont {S.~E.}\ \bibnamefont {Brown}},\
  }\bibfield  {title} {\enquote {\bibinfo {title} {Zeeman-driven phase
  transition within the superconducting state of
  $\ensuremath{\kappa}\mathrm{\text{\ensuremath{-}}}(\mathrm{BEDT}\mathrm{\text{\ensuremath{-}}}\mathrm{TTF}{)}_{2}\mathrm{Cu}(\mathrm{NCS}{)}_{2}$},}\
  }\href {\doibase 10.1103/PhysRevLett.107.087002} {\bibfield  {journal}
  {\bibinfo  {journal} {Phys. Rev. Lett.}\ }\textbf {\bibinfo {volume} {107}},\
  \bibinfo {pages} {087002} (\bibinfo {year} {2011})}\BibitemShut {NoStop}%
\bibitem [{\citenamefont {Coniglio}\ \emph {et~al.}(2011)\citenamefont
  {Coniglio}, \citenamefont {Winter}, \citenamefont {Cho}, \citenamefont
  {Agosta}, \citenamefont {Fravel},\ and\ \citenamefont
  {Montgomery}}]{montegomery_prb2011}%
  \BibitemOpen
  \bibfield  {author} {\bibinfo {author} {\bibfnamefont {William~A.}\
  \bibnamefont {Coniglio}}, \bibinfo {author} {\bibfnamefont {Laurel~E.}\
  \bibnamefont {Winter}}, \bibinfo {author} {\bibfnamefont {Kyuil}\
  \bibnamefont {Cho}}, \bibinfo {author} {\bibfnamefont {C.~C.}\ \bibnamefont
  {Agosta}}, \bibinfo {author} {\bibfnamefont {B.}~\bibnamefont {Fravel}}, \
  and\ \bibinfo {author} {\bibfnamefont {L.~K.}\ \bibnamefont {Montgomery}},\
  }\bibfield  {title} {\enquote {\bibinfo {title} {Superconducting phase
  diagram and fflo signature in
  $\ensuremath{\lambda}$-(bets)${}_{2}$gacl${}_{4}$ from rf penetration depth
  measurements},}\ }\href {\doibase 10.1103/PhysRevB.83.224507} {\bibfield
  {journal} {\bibinfo  {journal} {Phys. Rev. B}\ }\textbf {\bibinfo {volume}
  {83}},\ \bibinfo {pages} {224507} (\bibinfo {year} {2011})}\BibitemShut
  {NoStop}%
\bibitem [{\citenamefont {Bergk}\ \emph {et~al.}(2011)\citenamefont {Bergk},
  \citenamefont {Demuer}, \citenamefont {Sheikin}, \citenamefont {Wang},
  \citenamefont {Wosnitza}, \citenamefont {Nakazawa},\ and\ \citenamefont
  {Lortz}}]{lortz_prb2011}%
  \BibitemOpen
  \bibfield  {author} {\bibinfo {author} {\bibfnamefont {B.}~\bibnamefont
  {Bergk}}, \bibinfo {author} {\bibfnamefont {A.}~\bibnamefont {Demuer}},
  \bibinfo {author} {\bibfnamefont {I.}~\bibnamefont {Sheikin}}, \bibinfo
  {author} {\bibfnamefont {Y.}~\bibnamefont {Wang}}, \bibinfo {author}
  {\bibfnamefont {J.}~\bibnamefont {Wosnitza}}, \bibinfo {author}
  {\bibfnamefont {Y.}~\bibnamefont {Nakazawa}}, \ and\ \bibinfo {author}
  {\bibfnamefont {R.}~\bibnamefont {Lortz}},\ }\bibfield  {title} {\enquote
  {\bibinfo {title} {Magnetic torque evidence for the
  fulde-ferrell-larkin-ovchinnikov state in the layered organic superconductor
  $\ensuremath{\kappa}\ensuremath{-}(\mathrm{BEDT}\ensuremath{-}\mathrm{TTF}){}_{2}\mathrm{Cu}(\mathrm{NCS}){}_{2}$},}\
  }\href {\doibase 10.1103/PhysRevB.83.064506} {\bibfield  {journal} {\bibinfo
  {journal} {Phys. Rev. B}\ }\textbf {\bibinfo {volume} {83}},\ \bibinfo
  {pages} {064506} (\bibinfo {year} {2011})}\BibitemShut {NoStop}%
\bibitem [{\citenamefont {Agosta}\ \emph {et~al.}(2012)\citenamefont {Agosta},
  \citenamefont {Jin}, \citenamefont {Coniglio}, \citenamefont {Smith},
  \citenamefont {Cho}, \citenamefont {Stroe}, \citenamefont {Martin},
  \citenamefont {Tozer}, \citenamefont {Murphy}, \citenamefont {Palm},
  \citenamefont {Schlueter},\ and\ \citenamefont {Kurmoo}}]{agosta_prb2012}%
  \BibitemOpen
  \bibfield  {author} {\bibinfo {author} {\bibfnamefont {C.~C.}\ \bibnamefont
  {Agosta}}, \bibinfo {author} {\bibfnamefont {Jing}\ \bibnamefont {Jin}},
  \bibinfo {author} {\bibfnamefont {W.~A.}\ \bibnamefont {Coniglio}}, \bibinfo
  {author} {\bibfnamefont {B.~E.}\ \bibnamefont {Smith}}, \bibinfo {author}
  {\bibfnamefont {K.}~\bibnamefont {Cho}}, \bibinfo {author} {\bibfnamefont
  {I.}~\bibnamefont {Stroe}}, \bibinfo {author} {\bibfnamefont
  {C.}~\bibnamefont {Martin}}, \bibinfo {author} {\bibfnamefont {S.~W.}\
  \bibnamefont {Tozer}}, \bibinfo {author} {\bibfnamefont {T.~P.}\ \bibnamefont
  {Murphy}}, \bibinfo {author} {\bibfnamefont {E.~C.}\ \bibnamefont {Palm}},
  \bibinfo {author} {\bibfnamefont {J.~A.}\ \bibnamefont {Schlueter}}, \ and\
  \bibinfo {author} {\bibfnamefont {M.}~\bibnamefont {Kurmoo}},\ }\bibfield
  {title} {\enquote {\bibinfo {title} {Experimental and semiempirical method to
  determine the pauli-limiting field in quasi-two-dimensional superconductors
  as applied to $\ensuremath{\kappa}$-(bedt-ttf)${}_{2}$cu(ncs)${}_{2}$: Strong
  evidence of a fflo state},}\ }\href {\doibase 10.1103/PhysRevB.85.214514}
  {\bibfield  {journal} {\bibinfo  {journal} {Phys. Rev. B}\ }\textbf {\bibinfo
  {volume} {85}},\ \bibinfo {pages} {214514} (\bibinfo {year}
  {2012})}\BibitemShut {NoStop}%
\bibitem [{\citenamefont {Cho}\ \emph {et~al.}(2009)\citenamefont {Cho},
  \citenamefont {Smith}, \citenamefont {Coniglio}, \citenamefont {Winter},
  \citenamefont {Agosta},\ and\ \citenamefont {Schlueter}}]{schlueter_prb2009}%
  \BibitemOpen
  \bibfield  {author} {\bibinfo {author} {\bibfnamefont {K.}~\bibnamefont
  {Cho}}, \bibinfo {author} {\bibfnamefont {B.~E.}\ \bibnamefont {Smith}},
  \bibinfo {author} {\bibfnamefont {W.~A.}\ \bibnamefont {Coniglio}}, \bibinfo
  {author} {\bibfnamefont {L.~E.}\ \bibnamefont {Winter}}, \bibinfo {author}
  {\bibfnamefont {C.~C.}\ \bibnamefont {Agosta}}, \ and\ \bibinfo {author}
  {\bibfnamefont {J.~A.}\ \bibnamefont {Schlueter}},\ }\bibfield  {title}
  {\enquote {\bibinfo {title} {Upper critical field in the organic
  superconductor
  ${\ensuremath{\beta}}^{\ensuremath{''}}\text{\ensuremath{-}}{(\text{ET})}_{2}{\text{sf}}_{5}{\text{ch}}_{2}{\text{cf}}_{2}{\text{so}}_{3}$:
  Possibility of fulde-ferrell-larkin-ovchinnikov state},}\ }\href {\doibase
  10.1103/PhysRevB.79.220507} {\bibfield  {journal} {\bibinfo  {journal} {Phys.
  Rev. B}\ }\textbf {\bibinfo {volume} {79}},\ \bibinfo {pages} {220507}
  (\bibinfo {year} {2009})}\BibitemShut {NoStop}%
\bibitem [{\citenamefont {Zocco}\ \emph {et~al.}(2013)\citenamefont {Zocco},
  \citenamefont {Grube}, \citenamefont {Eilers}, \citenamefont {Wolf},\ and\
  \citenamefont {L\"ohneysen}}]{lohneysen_prl2013}%
  \BibitemOpen
  \bibfield  {author} {\bibinfo {author} {\bibfnamefont {D.~A.}\ \bibnamefont
  {Zocco}}, \bibinfo {author} {\bibfnamefont {K.}~\bibnamefont {Grube}},
  \bibinfo {author} {\bibfnamefont {F.}~\bibnamefont {Eilers}}, \bibinfo
  {author} {\bibfnamefont {T.}~\bibnamefont {Wolf}}, \ and\ \bibinfo {author}
  {\bibfnamefont {H.~v.}\ \bibnamefont {L\"ohneysen}},\ }\bibfield  {title}
  {\enquote {\bibinfo {title} {Pauli-limited multiband superconductivity in
  ${\mathrm{kfe}}_{2}{\mathrm{as}}_{2}$},}\ }\href {\doibase
  10.1103/PhysRevLett.111.057007} {\bibfield  {journal} {\bibinfo  {journal}
  {Phys. Rev. Lett.}\ }\textbf {\bibinfo {volume} {111}},\ \bibinfo {pages}
  {057007} (\bibinfo {year} {2013})}\BibitemShut {NoStop}%
\bibitem [{\citenamefont {Cho}\ \emph {et~al.}(2011)\citenamefont {Cho},
  \citenamefont {Kim}, \citenamefont {Tanatar}, \citenamefont {Song},
  \citenamefont {Kwon}, \citenamefont {Coniglio}, \citenamefont {Agosta},
  \citenamefont {Gurevich},\ and\ \citenamefont
  {2Prozorov}}]{prozorov_prb2011}%
  \BibitemOpen
  \bibfield  {author} {\bibinfo {author} {\bibfnamefont {K.}~\bibnamefont
  {Cho}}, \bibinfo {author} {\bibfnamefont {H.}~\bibnamefont {Kim}}, \bibinfo
  {author} {\bibfnamefont {M.~A.}\ \bibnamefont {Tanatar}}, \bibinfo {author}
  {\bibfnamefont {Y.~J.}\ \bibnamefont {Song}}, \bibinfo {author}
  {\bibfnamefont {Y.~S.}\ \bibnamefont {Kwon}}, \bibinfo {author}
  {\bibfnamefont {W.~A.}\ \bibnamefont {Coniglio}}, \bibinfo {author}
  {\bibfnamefont {C.~C.}\ \bibnamefont {Agosta}}, \bibinfo {author}
  {\bibfnamefont {A.}~\bibnamefont {Gurevich}}, \ and\ \bibinfo {author}
  {\bibfnamefont {R.}~\bibnamefont {2Prozorov}},\ }\bibfield  {title} {\enquote
  {\bibinfo {title} {Anisotropic upper critical field and possible
  fulde-ferrel-larkin-ovchinnikov state in the stoichiometric pnictide
  superconductor lifeas},}\ }\href {\doibase 10.1103/PhysRevB.83.060502}
  {\bibfield  {journal} {\bibinfo  {journal} {Phys. Rev. B}\ }\textbf {\bibinfo
  {volume} {83}},\ \bibinfo {pages} {060502} (\bibinfo {year}
  {2011})}\BibitemShut {NoStop}%
\bibitem [{\citenamefont {Khim}\ \emph {et~al.}(2011)\citenamefont {Khim},
  \citenamefont {Lee}, \citenamefont {Kim}, \citenamefont {Choi}, \citenamefont
  {Stewart},\ and\ \citenamefont {Kim}}]{kim_prb2011}%
  \BibitemOpen
  \bibfield  {author} {\bibinfo {author} {\bibfnamefont {Seunghyun}\
  \bibnamefont {Khim}}, \bibinfo {author} {\bibfnamefont {Bumsung}\
  \bibnamefont {Lee}}, \bibinfo {author} {\bibfnamefont {Jae~Wook}\
  \bibnamefont {Kim}}, \bibinfo {author} {\bibfnamefont {Eun~Sang}\
  \bibnamefont {Choi}}, \bibinfo {author} {\bibfnamefont {G.~R.}\ \bibnamefont
  {Stewart}}, \ and\ \bibinfo {author} {\bibfnamefont {Kee~Hoon}\ \bibnamefont
  {Kim}},\ }\bibfield  {title} {\enquote {\bibinfo {title} {Pauli-limiting
  effects in the upper critical fields of a clean lifeas single crystal},}\
  }\href {\doibase 10.1103/PhysRevB.84.104502} {\bibfield  {journal} {\bibinfo
  {journal} {Phys. Rev. B}\ }\textbf {\bibinfo {volume} {84}},\ \bibinfo
  {pages} {104502} (\bibinfo {year} {2011})}\BibitemShut {NoStop}%
\bibitem [{\citenamefont {Agterberg}\ \emph {et~al.}(2020)\citenamefont
  {Agterberg}, \citenamefont {Davis}, \citenamefont {Edkins}, \citenamefont
  {Fradkin}, \citenamefont {Van~Harlingen}, \citenamefont {Kivelson},
  \citenamefont {Lee}, \citenamefont {Radzihovsky}, \citenamefont {Tranquada},\
  and\ \citenamefont {Wang}}]{agterberg_annrevcmp2020}%
  \BibitemOpen
  \bibfield  {author} {\bibinfo {author} {\bibfnamefont {Daniel~F.}\
  \bibnamefont {Agterberg}}, \bibinfo {author} {\bibfnamefont {J.C.~Séamus}\
  \bibnamefont {Davis}}, \bibinfo {author} {\bibfnamefont {Stephen~D.}\
  \bibnamefont {Edkins}}, \bibinfo {author} {\bibfnamefont {Eduardo}\
  \bibnamefont {Fradkin}}, \bibinfo {author} {\bibfnamefont {Dale~J.}\
  \bibnamefont {Van~Harlingen}}, \bibinfo {author} {\bibfnamefont {Steven~A.}\
  \bibnamefont {Kivelson}}, \bibinfo {author} {\bibfnamefont {Patrick~A.}\
  \bibnamefont {Lee}}, \bibinfo {author} {\bibfnamefont {Leo}\ \bibnamefont
  {Radzihovsky}}, \bibinfo {author} {\bibfnamefont {John~M.}\ \bibnamefont
  {Tranquada}}, \ and\ \bibinfo {author} {\bibfnamefont {Yuxuan}\ \bibnamefont
  {Wang}},\ }\bibfield  {title} {\enquote {\bibinfo {title} {The physics of
  pair-density waves: Cuprate superconductors and beyond},}\ }\href {\doibase
  https://doi.org/10.1146/annurev-conmatphys-031119-050711} {\bibfield
  {journal} {\bibinfo  {journal} {Annual Review of Condensed Matter Physics}\
  }\textbf {\bibinfo {volume} {11}},\ \bibinfo {pages} {231--270} (\bibinfo
  {year} {2020})}\BibitemShut {NoStop}%
\bibitem [{\citenamefont {Berg}\ \emph
  {et~al.}(2009{\natexlab{a}})\citenamefont {Berg}, \citenamefont {Fradkin},
  \citenamefont {Kivelson},\ and\ \citenamefont
  {Tranquada}}]{tranquada_njp2009}%
  \BibitemOpen
  \bibfield  {author} {\bibinfo {author} {\bibfnamefont {Erez}\ \bibnamefont
  {Berg}}, \bibinfo {author} {\bibfnamefont {Eduardo}\ \bibnamefont {Fradkin}},
  \bibinfo {author} {\bibfnamefont {Steven~A}\ \bibnamefont {Kivelson}}, \ and\
  \bibinfo {author} {\bibfnamefont {John~M}\ \bibnamefont {Tranquada}},\
  }\bibfield  {title} {\enquote {\bibinfo {title} {Striped superconductors: how
  spin, charge and superconducting orders intertwine in the cuprates},}\ }\href
  {\doibase 10.1088/1367-2630/11/11/115004} {\bibfield  {journal} {\bibinfo
  {journal} {New Journal of Physics}\ }\textbf {\bibinfo {volume} {11}},\
  \bibinfo {pages} {115004} (\bibinfo {year} {2009}{\natexlab{a}})}\BibitemShut
  {NoStop}%
\bibitem [{\citenamefont {Berg}\ \emph {et~al.}(2010)\citenamefont {Berg},
  \citenamefont {Fradkin},\ and\ \citenamefont {Kivelson}}]{kivelson_prl2010}%
  \BibitemOpen
  \bibfield  {author} {\bibinfo {author} {\bibfnamefont {Erez}\ \bibnamefont
  {Berg}}, \bibinfo {author} {\bibfnamefont {Eduardo}\ \bibnamefont {Fradkin}},
  \ and\ \bibinfo {author} {\bibfnamefont {Steven~A.}\ \bibnamefont
  {Kivelson}},\ }\bibfield  {title} {\enquote {\bibinfo {title}
  {Pair-density-wave correlations in the kondo-heisenberg model},}\ }\href
  {\doibase 10.1103/PhysRevLett.105.146403} {\bibfield  {journal} {\bibinfo
  {journal} {Phys. Rev. Lett.}\ }\textbf {\bibinfo {volume} {105}},\ \bibinfo
  {pages} {146403} (\bibinfo {year} {2010})}\BibitemShut {NoStop}%
\bibitem [{\citenamefont {Schwemmer}\ \emph {et~al.}(2024)\citenamefont
  {Schwemmer}, \citenamefont {Hohmann}, \citenamefont {D\"urrnagel},
  \citenamefont {Potten}, \citenamefont {Beyer}, \citenamefont {Rachel},
  \citenamefont {Wu}, \citenamefont {Raghu}, \citenamefont {M\"uller},
  \citenamefont {Hanke},\ and\ \citenamefont {Thomale}}]{raghu_prbl2023}%
  \BibitemOpen
  \bibfield  {author} {\bibinfo {author} {\bibfnamefont {Tilman}\ \bibnamefont
  {Schwemmer}}, \bibinfo {author} {\bibfnamefont {Hendrik}\ \bibnamefont
  {Hohmann}}, \bibinfo {author} {\bibfnamefont {Matteo}\ \bibnamefont
  {D\"urrnagel}}, \bibinfo {author} {\bibfnamefont {Janik}\ \bibnamefont
  {Potten}}, \bibinfo {author} {\bibfnamefont {Jacob}\ \bibnamefont {Beyer}},
  \bibinfo {author} {\bibfnamefont {Stephan}\ \bibnamefont {Rachel}}, \bibinfo
  {author} {\bibfnamefont {Yi-Ming}\ \bibnamefont {Wu}}, \bibinfo {author}
  {\bibfnamefont {Srinivas}\ \bibnamefont {Raghu}}, \bibinfo {author}
  {\bibfnamefont {Tobias}\ \bibnamefont {M\"uller}}, \bibinfo {author}
  {\bibfnamefont {Werner}\ \bibnamefont {Hanke}}, \ and\ \bibinfo {author}
  {\bibfnamefont {Ronny}\ \bibnamefont {Thomale}},\ }\bibfield  {title}
  {\enquote {\bibinfo {title} {Sublattice modulated superconductivity in the
  kagome hubbard model},}\ }\href {\doibase 10.1103/PhysRevB.110.024501}
  {\bibfield  {journal} {\bibinfo  {journal} {Phys. Rev. B}\ }\textbf {\bibinfo
  {volume} {110}},\ \bibinfo {pages} {024501} (\bibinfo {year}
  {2024})}\BibitemShut {NoStop}%
\bibitem [{\citenamefont {Agterberg}\ and\ \citenamefont
  {Tsunetsugu}(2008)}]{tsunetsugu_natphys2008}%
  \BibitemOpen
  \bibfield  {author} {\bibinfo {author} {\bibfnamefont {D.~F.}\ \bibnamefont
  {Agterberg}}\ and\ \bibinfo {author} {\bibfnamefont {H.}~\bibnamefont
  {Tsunetsugu}},\ }\bibfield  {title} {\enquote {\bibinfo {title} {Dislocations
  and vortices in pair-density-wave superconductors},}\ }\href {\doibase
  10.1038/nphys999} {\bibfield  {journal} {\bibinfo  {journal} {Nature
  Physics}\ }\textbf {\bibinfo {volume} {4}},\ \bibinfo {pages} {639} (\bibinfo
  {year} {2008})}\BibitemShut {NoStop}%
\bibitem [{\citenamefont {Berg}\ \emph
  {et~al.}(2009{\natexlab{b}})\citenamefont {Berg}, \citenamefont {Fradkin},\
  and\ \citenamefont {Kivelson}}]{kivelson_natphys2009}%
  \BibitemOpen
  \bibfield  {author} {\bibinfo {author} {\bibfnamefont {Erez}\ \bibnamefont
  {Berg}}, \bibinfo {author} {\bibfnamefont {Eduardo}\ \bibnamefont {Fradkin}},
  \ and\ \bibinfo {author} {\bibfnamefont {Steven~A.}\ \bibnamefont
  {Kivelson}},\ }\bibfield  {title} {\enquote {\bibinfo {title} {Charge-4e
  superconductivity from pair-density-wave order in certain high-temperature
  superconductors},}\ }\href {\doibase 10.1038/nphys1389} {\bibfield  {journal}
  {\bibinfo  {journal} {Nature Physics}\ }\textbf {\bibinfo {volume} {5}},\
  \bibinfo {pages} {830} (\bibinfo {year} {2009}{\natexlab{b}})}\BibitemShut
  {NoStop}%
\bibitem [{\citenamefont {Fradkin}\ \emph {et~al.}(2015)\citenamefont
  {Fradkin}, \citenamefont {Kivelson},\ and\ \citenamefont
  {Tranquada}}]{tranquada_rmp2015}%
  \BibitemOpen
  \bibfield  {author} {\bibinfo {author} {\bibfnamefont {Eduardo}\ \bibnamefont
  {Fradkin}}, \bibinfo {author} {\bibfnamefont {Steven~A.}\ \bibnamefont
  {Kivelson}}, \ and\ \bibinfo {author} {\bibfnamefont {John~M.}\ \bibnamefont
  {Tranquada}},\ }\bibfield  {title} {\enquote {\bibinfo {title} {Colloquium:
  Theory of intertwined orders in high temperature superconductors},}\ }\href
  {\doibase 10.1103/RevModPhys.87.457} {\bibfield  {journal} {\bibinfo
  {journal} {Rev. Mod. Phys.}\ }\textbf {\bibinfo {volume} {87}},\ \bibinfo
  {pages} {457--482} (\bibinfo {year} {2015})}\BibitemShut {NoStop}%
\bibitem [{\citenamefont {Loder}\ \emph {et~al.}(2010)\citenamefont {Loder},
  \citenamefont {Kampf},\ and\ \citenamefont {Kopp}}]{kampf_prb2010}%
  \BibitemOpen
  \bibfield  {author} {\bibinfo {author} {\bibfnamefont {Florian}\ \bibnamefont
  {Loder}}, \bibinfo {author} {\bibfnamefont {Arno~P.}\ \bibnamefont {Kampf}},
  \ and\ \bibinfo {author} {\bibfnamefont {Thilo}\ \bibnamefont {Kopp}},\
  }\bibfield  {title} {\enquote {\bibinfo {title} {Superconducting state with a
  finite-momentum pairing mechanism in zero external magnetic field},}\ }\href
  {\doibase 10.1103/PhysRevB.81.020511} {\bibfield  {journal} {\bibinfo
  {journal} {Phys. Rev. B}\ }\textbf {\bibinfo {volume} {81}},\ \bibinfo
  {pages} {020511} (\bibinfo {year} {2010})}\BibitemShut {NoStop}%
\bibitem [{\citenamefont {Tranquada}\ \emph {et~al.}(2008)\citenamefont
  {Tranquada}, \citenamefont {Gu}, \citenamefont {H\"ucker}, \citenamefont
  {Jie}, \citenamefont {Kang}, \citenamefont {Klingeler}, \citenamefont {Li},
  \citenamefont {Tristan}, \citenamefont {Wen}, \citenamefont {Xu},
  \citenamefont {Xu}, \citenamefont {Zhou},\ and\ \citenamefont
  {v.~Zimmermann}}]{zimmermann_prb2008}%
  \BibitemOpen
  \bibfield  {author} {\bibinfo {author} {\bibfnamefont {J.~M.}\ \bibnamefont
  {Tranquada}}, \bibinfo {author} {\bibfnamefont {G.~D.}\ \bibnamefont {Gu}},
  \bibinfo {author} {\bibfnamefont {M.}~\bibnamefont {H\"ucker}}, \bibinfo
  {author} {\bibfnamefont {Q.}~\bibnamefont {Jie}}, \bibinfo {author}
  {\bibfnamefont {H.-J.}\ \bibnamefont {Kang}}, \bibinfo {author}
  {\bibfnamefont {R.}~\bibnamefont {Klingeler}}, \bibinfo {author}
  {\bibfnamefont {Q.}~\bibnamefont {Li}}, \bibinfo {author} {\bibfnamefont
  {N.}~\bibnamefont {Tristan}}, \bibinfo {author} {\bibfnamefont {J.~S.}\
  \bibnamefont {Wen}}, \bibinfo {author} {\bibfnamefont {G.~Y.}\ \bibnamefont
  {Xu}}, \bibinfo {author} {\bibfnamefont {Z.~J.}\ \bibnamefont {Xu}}, \bibinfo
  {author} {\bibfnamefont {J.}~\bibnamefont {Zhou}}, \ and\ \bibinfo {author}
  {\bibfnamefont {M.}~\bibnamefont {v.~Zimmermann}},\ }\bibfield  {title}
  {\enquote {\bibinfo {title} {Evidence for unusual superconducting
  correlations coexisting with stripe order in
  ${\text{la}}_{1.875}{\text{ba}}_{0.125}{\text{cuo}}_{4}$},}\ }\href {\doibase
  10.1103/PhysRevB.78.174529} {\bibfield  {journal} {\bibinfo  {journal} {Phys.
  Rev. B}\ }\textbf {\bibinfo {volume} {78}},\ \bibinfo {pages} {174529}
  (\bibinfo {year} {2008})}\BibitemShut {NoStop}%
\bibitem [{\citenamefont {Jiang}\ \emph {et~al.}(2021)\citenamefont {Jiang},
  \citenamefont {Yin}, \citenamefont {Denner}, \citenamefont {Shumiya},
  \citenamefont {Ortiz}, \citenamefont {Xu}, \citenamefont {Guguchia},
  \citenamefont {He}, \citenamefont {Hossain}, \citenamefont {Liu},
  \citenamefont {Ruff}, \citenamefont {Kautzsch}, \citenamefont {Zhang},
  \citenamefont {Chang}, \citenamefont {Belopolski}, \citenamefont {Zhang},
  \citenamefont {Cochran}, \citenamefont {Multer}, \citenamefont {Litskevich},
  \citenamefont {Cheng}, \citenamefont {Yang}, \citenamefont {Wang},
  \citenamefont {Thomale}, \citenamefont {Neupert}, \citenamefont {Wilson},\
  and\ \citenamefont {Hasan}}]{hasan_natmat2021}%
  \BibitemOpen
  \bibfield  {author} {\bibinfo {author} {\bibfnamefont {Yu-Xiao}\ \bibnamefont
  {Jiang}}, \bibinfo {author} {\bibfnamefont {Jia-Xin}\ \bibnamefont {Yin}},
  \bibinfo {author} {\bibfnamefont {M.~Michael}\ \bibnamefont {Denner}},
  \bibinfo {author} {\bibfnamefont {Nana}\ \bibnamefont {Shumiya}}, \bibinfo
  {author} {\bibfnamefont {Brenden~R.}\ \bibnamefont {Ortiz}}, \bibinfo
  {author} {\bibfnamefont {Gang}\ \bibnamefont {Xu}}, \bibinfo {author}
  {\bibfnamefont {Zurab}\ \bibnamefont {Guguchia}}, \bibinfo {author}
  {\bibfnamefont {Junyi}\ \bibnamefont {He}}, \bibinfo {author} {\bibfnamefont
  {Md~Shafayat}\ \bibnamefont {Hossain}}, \bibinfo {author} {\bibfnamefont
  {Xiaoxiong}\ \bibnamefont {Liu}}, \bibinfo {author} {\bibfnamefont {Jacob}\
  \bibnamefont {Ruff}}, \bibinfo {author} {\bibfnamefont {Linus}\ \bibnamefont
  {Kautzsch}}, \bibinfo {author} {\bibfnamefont {Songtian~S.}\ \bibnamefont
  {Zhang}}, \bibinfo {author} {\bibfnamefont {Guoqing}\ \bibnamefont {Chang}},
  \bibinfo {author} {\bibfnamefont {Ilya}\ \bibnamefont {Belopolski}}, \bibinfo
  {author} {\bibfnamefont {Qi}~\bibnamefont {Zhang}}, \bibinfo {author}
  {\bibfnamefont {Tyler~A.}\ \bibnamefont {Cochran}}, \bibinfo {author}
  {\bibfnamefont {Daniel}\ \bibnamefont {Multer}}, \bibinfo {author}
  {\bibfnamefont {Maksim}\ \bibnamefont {Litskevich}}, \bibinfo {author}
  {\bibfnamefont {Zi-Jia}\ \bibnamefont {Cheng}}, \bibinfo {author}
  {\bibfnamefont {Xian~P.}\ \bibnamefont {Yang}}, \bibinfo {author}
  {\bibfnamefont {Ziqiang}\ \bibnamefont {Wang}}, \bibinfo {author}
  {\bibfnamefont {Ronny}\ \bibnamefont {Thomale}}, \bibinfo {author}
  {\bibfnamefont {Titus}\ \bibnamefont {Neupert}}, \bibinfo {author}
  {\bibfnamefont {Stephen~D.}\ \bibnamefont {Wilson}}, \ and\ \bibinfo {author}
  {\bibfnamefont {M.~Zahid}\ \bibnamefont {Hasan}},\ }\bibfield  {title}
  {\enquote {\bibinfo {title} {Unconventional chiral charge order in kagome
  superconductor kv3sb5},}\ }\href {\doibase 10.1038/s41563-021-01034-y}
  {\bibfield  {journal} {\bibinfo  {journal} {Nature Materials}\ }\textbf
  {\bibinfo {volume} {20}},\ \bibinfo {pages} {1353} (\bibinfo {year}
  {2021})}\BibitemShut {NoStop}%
\bibitem [{\citenamefont {Chen}\ \emph {et~al.}(2021)\citenamefont {Chen},
  \citenamefont {Yang}, \citenamefont {Hu}, \citenamefont {Zhao}, \citenamefont
  {Yuan}, \citenamefont {Xing}, \citenamefont {Qian}, \citenamefont {Huang},
  \citenamefont {Li}, \citenamefont {Ye}, \citenamefont {Ma}, \citenamefont
  {Ni}, \citenamefont {Zhang}, \citenamefont {Yin}, \citenamefont {Gong},
  \citenamefont {Tu}, \citenamefont {Lei}, \citenamefont {Tan}, \citenamefont
  {Zhou}, \citenamefont {Shen}, \citenamefont {Dong}, \citenamefont {Yan},
  \citenamefont {Wang},\ and\ \citenamefont {Gao}}]{gao_nature2021}%
  \BibitemOpen
  \bibfield  {author} {\bibinfo {author} {\bibfnamefont {Hui}\ \bibnamefont
  {Chen}}, \bibinfo {author} {\bibfnamefont {Haitao}\ \bibnamefont {Yang}},
  \bibinfo {author} {\bibfnamefont {Bin}\ \bibnamefont {Hu}}, \bibinfo {author}
  {\bibfnamefont {Zhen}\ \bibnamefont {Zhao}}, \bibinfo {author} {\bibfnamefont
  {Jie}\ \bibnamefont {Yuan}}, \bibinfo {author} {\bibfnamefont {Yuqing}\
  \bibnamefont {Xing}}, \bibinfo {author} {\bibfnamefont {Guojian}\
  \bibnamefont {Qian}}, \bibinfo {author} {\bibfnamefont {Zihao}\ \bibnamefont
  {Huang}}, \bibinfo {author} {\bibfnamefont {Geng}\ \bibnamefont {Li}},
  \bibinfo {author} {\bibfnamefont {Yuhan}\ \bibnamefont {Ye}}, \bibinfo
  {author} {\bibfnamefont {Sheng}\ \bibnamefont {Ma}}, \bibinfo {author}
  {\bibfnamefont {Shunli}\ \bibnamefont {Ni}}, \bibinfo {author} {\bibfnamefont
  {Hua}\ \bibnamefont {Zhang}}, \bibinfo {author} {\bibfnamefont {Qiangwei}\
  \bibnamefont {Yin}}, \bibinfo {author} {\bibfnamefont {Chunsheng}\
  \bibnamefont {Gong}}, \bibinfo {author} {\bibfnamefont {Zhijun}\ \bibnamefont
  {Tu}}, \bibinfo {author} {\bibfnamefont {Hechang}\ \bibnamefont {Lei}},
  \bibinfo {author} {\bibfnamefont {Hengxin}\ \bibnamefont {Tan}}, \bibinfo
  {author} {\bibfnamefont {Sen}\ \bibnamefont {Zhou}}, \bibinfo {author}
  {\bibfnamefont {Chengmin}\ \bibnamefont {Shen}}, \bibinfo {author}
  {\bibfnamefont {Xiaoli}\ \bibnamefont {Dong}}, \bibinfo {author}
  {\bibfnamefont {Binghai}\ \bibnamefont {Yan}}, \bibinfo {author}
  {\bibfnamefont {Ziqiang}\ \bibnamefont {Wang}}, \ and\ \bibinfo {author}
  {\bibfnamefont {Hong-Jun}\ \bibnamefont {Gao}},\ }\bibfield  {title}
  {\enquote {\bibinfo {title} {Roton pair density wave in a strong-coupling
  kagome superconductor},}\ }\href {\doibase 10.1038/s41586-021-03983-5}
  {\bibfield  {journal} {\bibinfo  {journal} {Nature}\ }\textbf {\bibinfo
  {volume} {599}},\ \bibinfo {pages} {222} (\bibinfo {year}
  {2021})}\BibitemShut {NoStop}%
\bibitem [{\citenamefont {Gu}\ \emph {et~al.}(2023)\citenamefont {Gu},
  \citenamefont {Carroll}, \citenamefont {Wang}, \citenamefont {Ran},
  \citenamefont {Broyles}, \citenamefont {Siddiquee}, \citenamefont {Butch},
  \citenamefont {Saha}, \citenamefont {Paglione}, \citenamefont {Davis},\ and\
  \citenamefont {Liu}}]{liu_nature2023}%
  \BibitemOpen
  \bibfield  {author} {\bibinfo {author} {\bibfnamefont {Qiangqiang}\
  \bibnamefont {Gu}}, \bibinfo {author} {\bibfnamefont {Joseph~P.}\
  \bibnamefont {Carroll}}, \bibinfo {author} {\bibfnamefont {Shuqiu}\
  \bibnamefont {Wang}}, \bibinfo {author} {\bibfnamefont {Sheng}\ \bibnamefont
  {Ran}}, \bibinfo {author} {\bibfnamefont {Christopher}\ \bibnamefont
  {Broyles}}, \bibinfo {author} {\bibfnamefont {Hasan}\ \bibnamefont
  {Siddiquee}}, \bibinfo {author} {\bibfnamefont {Nicholas~P.}\ \bibnamefont
  {Butch}}, \bibinfo {author} {\bibfnamefont {Shanta~R.}\ \bibnamefont {Saha}},
  \bibinfo {author} {\bibfnamefont {Johnpierre}\ \bibnamefont {Paglione}},
  \bibinfo {author} {\bibfnamefont {J.~C.~Séamus}\ \bibnamefont {Davis}}, \
  and\ \bibinfo {author} {\bibfnamefont {Xiaolong}\ \bibnamefont {Liu}},\
  }\bibfield  {title} {\enquote {\bibinfo {title} {Detection of a pair density
  wave state in ute2},}\ }\href {\doibase 10.1038/s41586-023-05919-7}
  {\bibfield  {journal} {\bibinfo  {journal} {Nature}\ }\textbf {\bibinfo
  {volume} {618}},\ \bibinfo {pages} {921} (\bibinfo {year}
  {2023})}\BibitemShut {NoStop}%
\bibitem [{\citenamefont {Aishwarya}\ \emph {et~al.}(2023)\citenamefont
  {Aishwarya}, \citenamefont {May-Mann}, \citenamefont {Raghavan},
  \citenamefont {Nie}, \citenamefont {Romanelli}, \citenamefont {Ran},
  \citenamefont {Saha}, \citenamefont {Paglione}, \citenamefont {Butch},
  \citenamefont {Fradkin},\ and\ \citenamefont
  {Madhavan}}]{madhavan_nature2023}%
  \BibitemOpen
  \bibfield  {author} {\bibinfo {author} {\bibfnamefont {Anuva}\ \bibnamefont
  {Aishwarya}}, \bibinfo {author} {\bibfnamefont {Julian}\ \bibnamefont
  {May-Mann}}, \bibinfo {author} {\bibfnamefont {Arjun}\ \bibnamefont
  {Raghavan}}, \bibinfo {author} {\bibfnamefont {Laimei}\ \bibnamefont {Nie}},
  \bibinfo {author} {\bibfnamefont {Marisa}\ \bibnamefont {Romanelli}},
  \bibinfo {author} {\bibfnamefont {Sheng}\ \bibnamefont {Ran}}, \bibinfo
  {author} {\bibfnamefont {Shanta~R.}\ \bibnamefont {Saha}}, \bibinfo {author}
  {\bibfnamefont {Johnpierre}\ \bibnamefont {Paglione}}, \bibinfo {author}
  {\bibfnamefont {Nicholas~P.}\ \bibnamefont {Butch}}, \bibinfo {author}
  {\bibfnamefont {Eduardo}\ \bibnamefont {Fradkin}}, \ and\ \bibinfo {author}
  {\bibfnamefont {Vidya}\ \bibnamefont {Madhavan}},\ }\bibfield  {title}
  {\enquote {\bibinfo {title} {Magnetic-field-sensitive charge density waves in
  the superconductor ute2},}\ }\href {\doibase 10.1038/s41586-023-06005-8}
  {\bibfield  {journal} {\bibinfo  {journal} {Nature}\ }\textbf {\bibinfo
  {volume} {618}},\ \bibinfo {pages} {928} (\bibinfo {year}
  {2023})}\BibitemShut {NoStop}%
\bibitem [{\citenamefont {Zhao}\ \emph {et~al.}(2023)\citenamefont {Zhao},
  \citenamefont {Blackwell}, \citenamefont {Thinel}, \citenamefont {Handa},
  \citenamefont {Ishida}, \citenamefont {Zhu}, \citenamefont {Iyo},
  \citenamefont {Eisaki}, \citenamefont {Pasupathy},\ and\ \citenamefont
  {Fujita}}]{fujita_nature2023}%
  \BibitemOpen
  \bibfield  {author} {\bibinfo {author} {\bibfnamefont {He}~\bibnamefont
  {Zhao}}, \bibinfo {author} {\bibfnamefont {Raymond}\ \bibnamefont
  {Blackwell}}, \bibinfo {author} {\bibfnamefont {Morgan}\ \bibnamefont
  {Thinel}}, \bibinfo {author} {\bibfnamefont {Taketo}\ \bibnamefont {Handa}},
  \bibinfo {author} {\bibfnamefont {Shigeyuki}\ \bibnamefont {Ishida}},
  \bibinfo {author} {\bibfnamefont {Xiaoyang}\ \bibnamefont {Zhu}}, \bibinfo
  {author} {\bibfnamefont {Akira}\ \bibnamefont {Iyo}}, \bibinfo {author}
  {\bibfnamefont {Hiroshi}\ \bibnamefont {Eisaki}}, \bibinfo {author}
  {\bibfnamefont {Abhay~N.}\ \bibnamefont {Pasupathy}}, \ and\ \bibinfo
  {author} {\bibfnamefont {Kazuhiro}\ \bibnamefont {Fujita}},\ }\bibfield
  {title} {\enquote {\bibinfo {title} {Smectic pair-density-wave order in
  eurbfe4as4},}\ }\href {\doibase 10.1038/s41586-023-06103-7} {\bibfield
  {journal} {\bibinfo  {journal} {Nature}\ }\textbf {\bibinfo {volume} {618}},\
  \bibinfo {pages} {940} (\bibinfo {year} {2023})}\BibitemShut {NoStop}%
\bibitem [{\citenamefont {Hu}\ \emph {et~al.}(2025)\citenamefont {Hu},
  \citenamefont {Liu},\ and\ \citenamefont {Liu}}]{liu_prb2025}%
  \BibitemOpen
  \bibfield  {author} {\bibinfo {author} {\bibfnamefont {Hui}\ \bibnamefont
  {Hu}}, \bibinfo {author} {\bibfnamefont {Zhao}\ \bibnamefont {Liu}}, \ and\
  \bibinfo {author} {\bibfnamefont {Xia-Ji}\ \bibnamefont {Liu}},\ }\bibfield
  {title} {\enquote {\bibinfo {title} {Unconventional superconductivity of an
  altermagnetic metal: Polarized bcs and inhomogeneous fflo states},}\ }\href
  {\doibase 10.1103/whm1-76jc} {\bibfield  {journal} {\bibinfo  {journal}
  {Phys. Rev. B}\ }\textbf {\bibinfo {volume} {112}},\ \bibinfo {pages}
  {184501} (\bibinfo {year} {2025})}\BibitemShut {NoStop}%
\bibitem [{\citenamefont {Hong}\ \emph {et~al.}(2025)\citenamefont {Hong},
  \citenamefont {Park},\ and\ \citenamefont {Kim}}]{kim_prb2025}%
  \BibitemOpen
  \bibfield  {author} {\bibinfo {author} {\bibfnamefont {SeungBeom}\
  \bibnamefont {Hong}}, \bibinfo {author} {\bibfnamefont {Moon~Jip}\
  \bibnamefont {Park}}, \ and\ \bibinfo {author} {\bibfnamefont {Kyoung-Min}\
  \bibnamefont {Kim}},\ }\bibfield  {title} {\enquote {\bibinfo {title}
  {Unconventional $p$-wave and finite-momentum superconductivity induced by
  altermagnetism through the formation of bogoliubov fermi surface},}\ }\href
  {\doibase 10.1103/PhysRevB.111.054501} {\bibfield  {journal} {\bibinfo
  {journal} {Phys. Rev. B}\ }\textbf {\bibinfo {volume} {111}},\ \bibinfo
  {pages} {054501} (\bibinfo {year} {2025})}\BibitemShut {NoStop}%
\bibitem [{\citenamefont {Wei}\ \emph {et~al.}(2024)\citenamefont {Wei},
  \citenamefont {Xiang}, \citenamefont {Xu}, \citenamefont {Zhang},
  \citenamefont {Tang},\ and\ \citenamefont {Wang}}]{wang_prbl2024}%
  \BibitemOpen
  \bibfield  {author} {\bibinfo {author} {\bibfnamefont {Miaomiao}\
  \bibnamefont {Wei}}, \bibinfo {author} {\bibfnamefont {Longjun}\ \bibnamefont
  {Xiang}}, \bibinfo {author} {\bibfnamefont {Fuming}\ \bibnamefont {Xu}},
  \bibinfo {author} {\bibfnamefont {Lei}\ \bibnamefont {Zhang}}, \bibinfo
  {author} {\bibfnamefont {Gaomin}\ \bibnamefont {Tang}}, \ and\ \bibinfo
  {author} {\bibfnamefont {Jian}\ \bibnamefont {Wang}},\ }\bibfield  {title}
  {\enquote {\bibinfo {title} {Gapless superconducting state and mirage gap in
  altermagnets},}\ }\href {\doibase 10.1103/PhysRevB.109.L201404} {\bibfield
  {journal} {\bibinfo  {journal} {Phys. Rev. B}\ }\textbf {\bibinfo {volume}
  {109}},\ \bibinfo {pages} {L201404} (\bibinfo {year} {2024})}\BibitemShut
  {NoStop}%
\bibitem [{\citenamefont {Ouassou}\ \emph {et~al.}(2023)\citenamefont
  {Ouassou}, \citenamefont {Brataas},\ and\ \citenamefont
  {Linder}}]{linder_prl2023}%
  \BibitemOpen
  \bibfield  {author} {\bibinfo {author} {\bibfnamefont {Jabir~Ali}\
  \bibnamefont {Ouassou}}, \bibinfo {author} {\bibfnamefont {Arne}\
  \bibnamefont {Brataas}}, \ and\ \bibinfo {author} {\bibfnamefont {Jacob}\
  \bibnamefont {Linder}},\ }\bibfield  {title} {\enquote {\bibinfo {title} {dc
  josephson effect in altermagnets},}\ }\href {\doibase
  10.1103/PhysRevLett.131.076003} {\bibfield  {journal} {\bibinfo  {journal}
  {Phys. Rev. Lett.}\ }\textbf {\bibinfo {volume} {131}},\ \bibinfo {pages}
  {076003} (\bibinfo {year} {2023})}\BibitemShut {NoStop}%
\bibitem [{\citenamefont {Papaj}(2023)}]{papaj_prbl2023}%
  \BibitemOpen
  \bibfield  {author} {\bibinfo {author} {\bibfnamefont {Micha\l{}}\
  \bibnamefont {Papaj}},\ }\bibfield  {title} {\enquote {\bibinfo {title}
  {Andreev reflection at the altermagnet-superconductor interface},}\ }\href
  {\doibase 10.1103/PhysRevB.108.L060508} {\bibfield  {journal} {\bibinfo
  {journal} {Phys. Rev. B}\ }\textbf {\bibinfo {volume} {108}},\ \bibinfo
  {pages} {L060508} (\bibinfo {year} {2023})}\BibitemShut {NoStop}%
\bibitem [{\citenamefont {Hubbard}(1959)}]{hs1}%
  \BibitemOpen
  \bibfield  {author} {\bibinfo {author} {\bibfnamefont {J.}~\bibnamefont
  {Hubbard}},\ }\bibfield  {title} {\enquote {\bibinfo {title} {Calculation of
  partition functions},}\ }\href {\doibase 10.1103/PhysRevLett.3.77} {\bibfield
   {journal} {\bibinfo  {journal} {Phys. Rev. Lett.}\ }\textbf {\bibinfo
  {volume} {3}},\ \bibinfo {pages} {77--78} (\bibinfo {year}
  {1959})}\BibitemShut {NoStop}%
\bibitem [{\citenamefont {Schulz}(1990)}]{hs2}%
  \BibitemOpen
  \bibfield  {author} {\bibinfo {author} {\bibfnamefont {H.~J.}\ \bibnamefont
  {Schulz}},\ }\bibfield  {title} {\enquote {\bibinfo {title} {Effective action
  for strongly correlated fermions from functional integrals},}\ }\href
  {\doibase 10.1103/PhysRevLett.65.2462} {\bibfield  {journal} {\bibinfo
  {journal} {Phys. Rev. Lett.}\ }\textbf {\bibinfo {volume} {65}},\ \bibinfo
  {pages} {2462--2465} (\bibinfo {year} {1990})}\BibitemShut {NoStop}%
\bibitem [{\citenamefont {Fratini}\ and\ \citenamefont
  {Ciuchi}(2021)}]{ciuchi_scipost2021}%
  \BibitemOpen
  \bibfield  {author} {\bibinfo {author} {\bibfnamefont {Simone}\ \bibnamefont
  {Fratini}}\ and\ \bibinfo {author} {\bibfnamefont {Sergio}\ \bibnamefont
  {Ciuchi}},\ }\bibfield  {title} {\enquote {\bibinfo {title} {Displaced drude
  peak and bad metal from the interaction with slow fluctuations.}}\ }\href
  {\doibase 10.21468/SciPostPhys.11.2.039} {\bibfield  {journal} {\bibinfo
  {journal} {SciPost Phys.}\ }\textbf {\bibinfo {volume} {11}},\ \bibinfo
  {pages} {039} (\bibinfo {year} {2021})}\BibitemShut {NoStop}%
\bibitem [{\citenamefont {Ciuchi}\ and\ \citenamefont
  {Fratini}(2023)}]{fratini_prb2023}%
  \BibitemOpen
  \bibfield  {author} {\bibinfo {author} {\bibfnamefont {Sergio}\ \bibnamefont
  {Ciuchi}}\ and\ \bibinfo {author} {\bibfnamefont {Simone}\ \bibnamefont
  {Fratini}},\ }\bibfield  {title} {\enquote {\bibinfo {title} {Strange metal
  behavior from incoherent carriers scattered by local moments},}\ }\href
  {\doibase 10.1103/PhysRevB.108.235173} {\bibfield  {journal} {\bibinfo
  {journal} {Phys. Rev. B}\ }\textbf {\bibinfo {volume} {108}},\ \bibinfo
  {pages} {235173} (\bibinfo {year} {2023})}\BibitemShut {NoStop}%
\bibitem [{\citenamefont {Murthy}\ \emph {et~al.}(2023)\citenamefont {Murthy},
  \citenamefont {Pandey}, \citenamefont {Esterlis},\ and\ \citenamefont
  {Kivelson}}]{kivelson_pans2023}%
  \BibitemOpen
  \bibfield  {author} {\bibinfo {author} {\bibfnamefont {Chaitanya}\
  \bibnamefont {Murthy}}, \bibinfo {author} {\bibfnamefont {Akshat}\
  \bibnamefont {Pandey}}, \bibinfo {author} {\bibfnamefont {Ilya}\ \bibnamefont
  {Esterlis}}, \ and\ \bibinfo {author} {\bibfnamefont {Steven~A.}\
  \bibnamefont {Kivelson}},\ }\bibfield  {title} {\enquote {\bibinfo {title} {A
  stability bound on the t-linear resistivity of conventional metals},}\ }\href
  {\doibase 10.1073/pnas.2216241120} {\bibfield  {journal} {\bibinfo  {journal}
  {Proceedings of the National Academy of Sciences}\ }\textbf {\bibinfo
  {volume} {120}},\ \bibinfo {pages} {e2216241120} (\bibinfo {year}
  {2023})}\BibitemShut {NoStop}%
\bibitem [{\citenamefont {Kunwar}\ and\ \citenamefont
  {Karmakar}(2026)}]{karmakar_prml2025}%
  \BibitemOpen
  \bibfield  {author} {\bibinfo {author} {\bibfnamefont {Shashikant~Singh}\
  \bibnamefont {Kunwar}}\ and\ \bibinfo {author} {\bibfnamefont {Madhuparna}\
  \bibnamefont {Karmakar}},\ }\bibfield  {title} {\enquote {\bibinfo {title}
  {Straintronics across lieb-kagome interconversion and variable transport
  scaling exponents},}\ }\href {\doibase 10.1103/8v9g-fwdm} {\bibfield
  {journal} {\bibinfo  {journal} {Phys. Rev. Mater.}\ }\textbf {\bibinfo
  {volume} {10}},\ \bibinfo {pages} {L011002} (\bibinfo {year}
  {2026})}\BibitemShut {NoStop}%
\bibitem [{\citenamefont {Ok}\ \emph {et~al.}(2020)\citenamefont {Ok},
  \citenamefont {Kwon}, \citenamefont {Kohama}, \citenamefont {You},
  \citenamefont {Park}, \citenamefont {Kim}, \citenamefont {Jo}, \citenamefont
  {Choi}, \citenamefont {Kindo}, \citenamefont {Kang}, \citenamefont {Kim},
  \citenamefont {Moon}, \citenamefont {Gurevich},\ and\ \citenamefont
  {Kim}}]{ok_prb2020}%
  \BibitemOpen
  \bibfield  {author} {\bibinfo {author} {\bibfnamefont {Jong~Mok}\
  \bibnamefont {Ok}}, \bibinfo {author} {\bibfnamefont {Chang~Il}\ \bibnamefont
  {Kwon}}, \bibinfo {author} {\bibfnamefont {Yoshimitsu}\ \bibnamefont
  {Kohama}}, \bibinfo {author} {\bibfnamefont {Jung~Sang}\ \bibnamefont {You}},
  \bibinfo {author} {\bibfnamefont {Sun~Kyu}\ \bibnamefont {Park}}, \bibinfo
  {author} {\bibfnamefont {Ji-hye}\ \bibnamefont {Kim}}, \bibinfo {author}
  {\bibfnamefont {Y.~J.}\ \bibnamefont {Jo}}, \bibinfo {author} {\bibfnamefont
  {E.~S.}\ \bibnamefont {Choi}}, \bibinfo {author} {\bibfnamefont {Koichi}\
  \bibnamefont {Kindo}}, \bibinfo {author} {\bibfnamefont {Woun}\ \bibnamefont
  {Kang}}, \bibinfo {author} {\bibfnamefont {Ki-Seok}\ \bibnamefont {Kim}},
  \bibinfo {author} {\bibfnamefont {E.~G.}\ \bibnamefont {Moon}}, \bibinfo
  {author} {\bibfnamefont {A.}~\bibnamefont {Gurevich}}, \ and\ \bibinfo
  {author} {\bibfnamefont {Jun~Sung}\ \bibnamefont {Kim}},\ }\bibfield  {title}
  {\enquote {\bibinfo {title} {Observation of in-plane magnetic field induced
  phase transitions in fese},}\ }\href {\doibase 10.1103/PhysRevB.101.224509}
  {\bibfield  {journal} {\bibinfo  {journal} {Phys. Rev. B}\ }\textbf {\bibinfo
  {volume} {101}},\ \bibinfo {pages} {224509} (\bibinfo {year}
  {2020})}\BibitemShut {NoStop}%
\bibitem [{\citenamefont {Kasahara}\ \emph {et~al.}(2021)\citenamefont
  {Kasahara}, \citenamefont {Suzuki}, \citenamefont {Machida}, \citenamefont
  {Sato}, \citenamefont {Ukai}, \citenamefont {Murayama}, \citenamefont
  {Suetsugu}, \citenamefont {Kasahara}, \citenamefont {Shibauchi},
  \citenamefont {Hanaguri},\ and\ \citenamefont {Matsuda}}]{kasahara_prl2021}%
  \BibitemOpen
  \bibfield  {author} {\bibinfo {author} {\bibfnamefont {S.}~\bibnamefont
  {Kasahara}}, \bibinfo {author} {\bibfnamefont {H.}~\bibnamefont {Suzuki}},
  \bibinfo {author} {\bibfnamefont {T.}~\bibnamefont {Machida}}, \bibinfo
  {author} {\bibfnamefont {Y.}~\bibnamefont {Sato}}, \bibinfo {author}
  {\bibfnamefont {Y.}~\bibnamefont {Ukai}}, \bibinfo {author} {\bibfnamefont
  {H.}~\bibnamefont {Murayama}}, \bibinfo {author} {\bibfnamefont
  {S.}~\bibnamefont {Suetsugu}}, \bibinfo {author} {\bibfnamefont
  {Y.}~\bibnamefont {Kasahara}}, \bibinfo {author} {\bibfnamefont
  {T.}~\bibnamefont {Shibauchi}}, \bibinfo {author} {\bibfnamefont
  {T.}~\bibnamefont {Hanaguri}}, \ and\ \bibinfo {author} {\bibfnamefont
  {Y.}~\bibnamefont {Matsuda}},\ }\bibfield  {title} {\enquote {\bibinfo
  {title} {Quasiparticle nodal plane in the fulde-ferrell-larkin-ovchinnikov
  state of fese},}\ }\href {\doibase 10.1103/PhysRevLett.127.257001} {\bibfield
   {journal} {\bibinfo  {journal} {Phys. Rev. Lett.}\ }\textbf {\bibinfo
  {volume} {127}},\ \bibinfo {pages} {257001} (\bibinfo {year}
  {2021})}\BibitemShut {NoStop}%
\bibitem [{\citenamefont {Zhang}\ \emph {et~al.}(2023)\citenamefont {Zhang},
  \citenamefont {Liu}, \citenamefont {Wang}, \citenamefont {Tsang},
  \citenamefont {Wang}, \citenamefont {Lam}, \citenamefont {Wang},
  \citenamefont {Xie}, \citenamefont {Zhou}, \citenamefont {Zhao},
  \citenamefont {Wang}, \citenamefont {Tallon}, \citenamefont {Lai},\ and\
  \citenamefont {Goh}}]{goh_nanolett2023}%
  \BibitemOpen
  \bibfield  {author} {\bibinfo {author} {\bibfnamefont {Wei}\ \bibnamefont
  {Zhang}}, \bibinfo {author} {\bibfnamefont {Xinyou}\ \bibnamefont {Liu}},
  \bibinfo {author} {\bibfnamefont {Lingfei}\ \bibnamefont {Wang}}, \bibinfo
  {author} {\bibfnamefont {Chun~Wai}\ \bibnamefont {Tsang}}, \bibinfo {author}
  {\bibfnamefont {Zheyu}\ \bibnamefont {Wang}}, \bibinfo {author}
  {\bibfnamefont {Siu~Tung}\ \bibnamefont {Lam}}, \bibinfo {author}
  {\bibfnamefont {Wenyan}\ \bibnamefont {Wang}}, \bibinfo {author}
  {\bibfnamefont {Jianyu}\ \bibnamefont {Xie}}, \bibinfo {author}
  {\bibfnamefont {Xuefeng}\ \bibnamefont {Zhou}}, \bibinfo {author}
  {\bibfnamefont {Yusheng}\ \bibnamefont {Zhao}}, \bibinfo {author}
  {\bibfnamefont {Shanmin}\ \bibnamefont {Wang}}, \bibinfo {author}
  {\bibfnamefont {Jeff}\ \bibnamefont {Tallon}}, \bibinfo {author}
  {\bibfnamefont {Kwing~To}\ \bibnamefont {Lai}}, \ and\ \bibinfo {author}
  {\bibfnamefont {Swee~K.}\ \bibnamefont {Goh}},\ }\bibfield  {title} {\enquote
  {\bibinfo {title} {Nodeless superconductivity in kagome metal csv3sb5 with
  and without time reversal symmetry breaking},}\ }\href {\doibase
  10.1021/acs.nanolett.2c04103} {\bibfield  {journal} {\bibinfo  {journal}
  {Nano Letters}\ }\textbf {\bibinfo {volume} {23}},\ \bibinfo {pages} {872}
  (\bibinfo {year} {2023})}\BibitemShut {NoStop}%
\bibitem [{\citenamefont {Swain}\ \emph {et~al.}(2022)\citenamefont {Swain},
  \citenamefont {Karmakar},\ and\ \citenamefont {Majumdar}}]{karmakar_spinliq}%
  \BibitemOpen
  \bibfield  {author} {\bibinfo {author} {\bibfnamefont {Nyayabanta}\
  \bibnamefont {Swain}}, \bibinfo {author} {\bibfnamefont {Madhuparna}\
  \bibnamefont {Karmakar}}, \ and\ \bibinfo {author} {\bibfnamefont {Pinaki}\
  \bibnamefont {Majumdar}},\ }\bibfield  {title} {\enquote {\bibinfo {title}
  {Spin-orbital liquids and insulator-metal transitions on the pyrochlore
  lattice},}\ }\href {\doibase 10.1103/PhysRevB.106.245114} {\bibfield
  {journal} {\bibinfo  {journal} {Phys. Rev. B}\ }\textbf {\bibinfo {volume}
  {106}},\ \bibinfo {pages} {245114} (\bibinfo {year} {2022})}\BibitemShut
  {NoStop}%
\bibitem [{\citenamefont {Karmakar}\ and\ \citenamefont
  {Swain}(2022)}]{karmakar_tri}%
  \BibitemOpen
  \bibfield  {author} {\bibinfo {author} {\bibfnamefont {Madhuparna}\
  \bibnamefont {Karmakar}}\ and\ \bibinfo {author} {\bibfnamefont {Nyayabanta}\
  \bibnamefont {Swain}},\ }\bibfield  {title} {\enquote {\bibinfo {title}
  {Transport and spectroscopic signatures of a disorder-stabilized metal in
  two-dimensional frustrated mott insulators},}\ }\href {\doibase
  10.1103/PhysRevB.105.195146} {\bibfield  {journal} {\bibinfo  {journal}
  {Phys. Rev. B}\ }\textbf {\bibinfo {volume} {105}},\ \bibinfo {pages}
  {195146} (\bibinfo {year} {2022})}\BibitemShut {NoStop}%
\bibitem [{\citenamefont {Swain}\ and\ \citenamefont
  {Karmakar}(2020)}]{lieb_strain}%
  \BibitemOpen
  \bibfield  {author} {\bibinfo {author} {\bibfnamefont {Nyayabanta}\
  \bibnamefont {Swain}}\ and\ \bibinfo {author} {\bibfnamefont {Madhuparna}\
  \bibnamefont {Karmakar}},\ }\bibfield  {title} {\enquote {\bibinfo {title}
  {Strain-induced superconductor-insulator transition on a lieb lattice},}\
  }\href {\doibase 10.1103/PhysRevResearch.2.023136} {\bibfield  {journal}
  {\bibinfo  {journal} {Phys. Rev. Res.}\ }\textbf {\bibinfo {volume} {2}},\
  \bibinfo {pages} {023136} (\bibinfo {year} {2020})}\BibitemShut {NoStop}%
\bibitem [{\citenamefont {Kunwar}\ and\ \citenamefont
  {Karmakar}(2024)}]{shashi_kagome2024}%
  \BibitemOpen
  \bibfield  {author} {\bibinfo {author} {\bibfnamefont {Shashikant~Singh}\
  \bibnamefont {Kunwar}}\ and\ \bibinfo {author} {\bibfnamefont {Madhuparna}\
  \bibnamefont {Karmakar}},\ }\bibfield  {title} {\enquote {\bibinfo {title}
  {Kagome hubbard model away from the strong coupling limit: Flat band
  localization and non fermi liquid signatures},}\ }\href
  {https://arxiv.org/abs/2404.05787} {\  (\bibinfo {year} {2024})},\ \Eprint
  {http://arxiv.org/abs/2404.05787} {arXiv:2404.05787 [cond-mat.str-el]}
  \BibitemShut {NoStop}%
\bibitem [{\citenamefont {Kannan}\ \emph {et~al.}(2026)\citenamefont {Kannan},
  \citenamefont {Savla},\ and\ \citenamefont {Karmakar}}]{santhosh_dalm2026}%
  \BibitemOpen
  \bibfield  {author} {\bibinfo {author} {\bibfnamefont {Santhosh}\
  \bibnamefont {Kannan}}, \bibinfo {author} {\bibfnamefont {Jainam}\
  \bibnamefont {Savla}}, \ and\ \bibinfo {author} {\bibfnamefont {Madhuparna}\
  \bibnamefont {Karmakar}},\ }\bibfield  {title} {\enquote {\bibinfo {title}
  {Metal-insulator transition and thermal scales in $d$-wave altermagnet},}\
  }\href {https://arxiv.org/abs/2603.02707} {\  (\bibinfo {year} {2026})},\
  \Eprint {http://arxiv.org/abs/2603.02707} {arXiv:2603.02707
  [cond-mat.str-el]} \BibitemShut {NoStop}%
\end{thebibliography}%

\end{document}